# The Brussels Effect and Artificial Intelligence:

## How EU regulation will impact the global AI market

Charlotte Siegmann* and Markus Anderljung* | August 2022

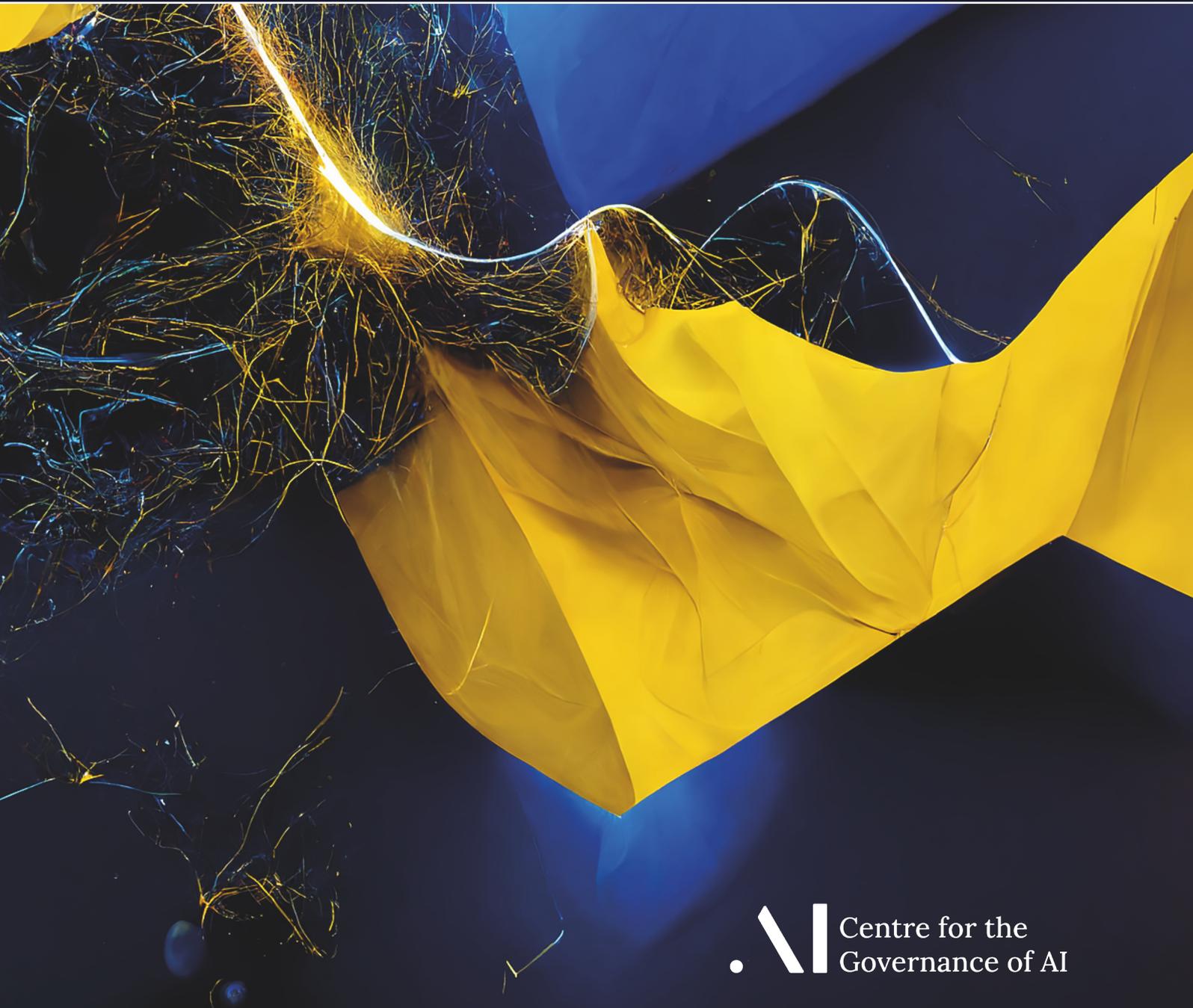

Centre for the Governance of AI

# ABSTRACT

The European Union is likely to introduce among the first, most stringent, and most comprehensive AI regulatory regimes of the world's major jurisdictions. In this report, we ask whether the EU's upcoming regulation for AI will diffuse globally, producing a so-called "Brussels Effect". Building on and extending Anu Bradford's work, we outline the mechanisms by which such regulatory diffusion may occur. We consider both the possibility that the EU's AI regulation will incentivise changes in products offered in non-EU countries (a de facto Brussels Effect) and the possibility it will influence regulation adopted by other jurisdictions (a de jure Brussels Effect). Focusing on the proposed EU AI Act, we tentatively conclude that both de facto and de jure Brussels effects are likely for parts of the EU regulatory regime. A de facto effect is particularly likely to arise in large US tech companies with AI systems that the AI Act terms "high-risk". We argue that the upcoming regulation might be particularly important in offering the first and most influential operationalisation of what it means to develop and deploy trustworthy or human-centred AI. If the EU regime is likely to see significant diffusion, ensuring it is well-designed becomes a matter of global importance.

*This report was jointly authored by Charlotte Siegmann and Markus Anderljung with equal contributions. Author order randomized. Charlotte Siegmann is a Predoc Fellow in Economics at the Global Priorities Institute at the University of Oxford. Markus Anderljung is Head of Policy at the Centre for the Governance of AI.

# Executive Summary

In 2019, two of the most powerful European politicians, Ursula von der Leyen and Angela Merkel, called for the European Union to create a GDPR for AI.[3] The General Data Protection Regulation[4] (GDPR) is one of the most influential pieces of European Union (EU) legislation in the last decade. It not only changed business practices within the EU, but also caused a "Brussels Effect" abroad. It incentivised changes in products offered in several non-EU countries (a *de facto* Brussels Effect) and influenced regulation adopted by other jurisdictions (a *de jure* Brussels Effect).

This report argues that upcoming EU regulation of AI is poised to have a similarly global impact. Focusing on the proposed AI Act and proposed updates to liability regimes, we argue that:

- Both de facto and de jure Brussels Effects are likely for parts of the EU's AI regulation.

- The Brussels Effect will likely be more significant than the "Washington Effect" or the "Beijing Effect".

If there is a significant AI Brussels Effect, this could lead to stricter AI regulation globally. The details of EU AI regulation could also influence how "trustworthy AI" is conceived across the world, shaping research agendas aimed at ensuring the safety and fairness of AI systems. Ultimately, the likelihood of a Brussels Effect increases the importance of helping shape the EU AI regulatory regime: getting the regulation right would become a matter of global importance.

## Findings

**For parts of the EU's AI regulation, a de facto Brussels Effect is likely**

We expect multinational companies to offer some EU-compliant AI products outside the EU. Once a company has decided to produce an EU-compliant product for the EU market, it will sometimes be more profitable to offer that product in some other jurisdictions or even globally rather than offering a separate non-compliant version outside the EU.

Drawing and building on Anu Bradford's work, we highlight several factors that make this behaviour ("non-differentiation") more likely.

- The EU has relatively **favourable market properties**. In particular, the market for AI-based products is large and heavily serviced by multinational firms. The EU market size incentivises firms to develop and offer EU-compliant products, rather than simply abandoning the EU market. The fact that these firms also service non-EU markets opens the door to a de facto effect.

---

[3] Directorate-General for Neighbourhood and Enlargement Negotiations, *"Speech by President-Elect von Der Leyen in the European Parliament Plenary on the Occasion of the Presentation of Her College of Commissioners and Their Programme,"* European Neighbourhood Policy and Enlargement Negotiations, November 27, 2019

[4] European Parliament, "Regulation (EU) 2016/679 of the European Parliament and of the Council of 27 April 2016 on the Protection of Natural Persons with Regard to the Processing of Personal Data and on the Free Movement of Such Data, and Repealing Directive 95/46/EC (General Data Protection Regulation) (Text with EEA Relevance)."

- The EU's AI regulation is likely to be especially **stringent**. If it were not more stringent than other jurisdictions' regulations on at least some dimensions, then there would be no room for it to have an effect abroad.

- The EU has high **regulatory capacity**. The EU's ability to produce well-crafted regulation decreases the chance that its AI regulations will be either difficult to enforce or overly cumbersome to comply with. Customers may also see EU compliance as a sign of the trustworthiness of the product, further incentivising firms to offer EU-complaint products in other jurisdictions.

- Demand for some affected AI products is likely to be fairly **inelastic**. Compliance with EU AI rules may raise the cost or decrease the quality of AI products by reducing product functionality. If demand were too elastic in response to such changes in cost and quality, then this could shrink the size of the EU market and make multinational firms more willing to abandon it. The incentive to offer non-EU-compliant products outside the EU would also increase.

- The **cost of differentiation** for some, but not all, AI products is likely to be high. Creating both compliant and non-compliant versions of a product may require developers to practise "early forking" (i.e. changing a fundamental feature early on in the development process) and maintain two separate technology stacks in parallel. If a company has already decided to develop a compliant version of the product, then simply offering this same version outside the EU may allow them to cut development costs without comparably large **costs of non-differentiation**, e.g. the costs of offering EU-compliant products globally.

The proposed AI Act would introduce new standards and conformity assessment requirements for "high-risk" AI products sold in the EU, estimated at 5–15% of the EU AI market. We anticipate a de facto effect for some high-risk AI products and for some categories of requirements, but not others, owing to variation in how strongly the above factors apply. A de facto effect is particularly likely, for instance, for medical devices, some worker management systems, certain legal technology, and a subset of biometric categorisation systems. A de facto effect may be particularly likely for requirements concerning risk management, record-keeping, transparency, accuracy, robustness, and cybersecurity. A de facto effect is less likely for products whose markets tend to be more regionalised, such as creditworthiness assessment systems and various government applications.

Although so-called "foundation models" are not classed as high-risk in the EU Commission's AI Act proposal, they may also experience a de facto effect. Foundation models are general purpose, pre-trained AI systems that can be used to create a wide range of AI products. Developers of these models may wish to ensure that AI products derived from their models will satisfy certain EU requirements by default. In addition, the AI Act proposal may also be amended to introduce specific requirements on general purpose systems and foundation models.

The AI Act proposal also introduces prohibitions on certain uses of AI systems. There is a small chance that prohibitions on the use of "subliminal techniques" could have implications for the design of recommender systems. If so, companies may choose to offer EU-compliant recommender systems in other jurisdictions. Other prohibitions (such as on the real-time use of facial recognition for law enforcement) also have a small chance of influencing non-EU products by shaping norms.

The proposal also requires people to be made aware if they are engaging with certain AI systems, e.g. content-generating systems (such as authentic-seeming images and chatbot conversations) or remote biometric surveillance. There is a modest chance that these requirements will lead companies to also e.g. display tags indicating some piece of content is AI-generated in other jurisdictions, since removing the tags could come to be seen as dishonest behaviour.

**A de jure Brussels Effect is also likely for parts of the EU's AI regulation**

We expect that other jurisdictions will adopt some EU-inspired AI regulation. This could happen for several different reasons:

- Foreign jurisdictions may expect EU-like regulation to be high quality and consistent with their own regulatory goals.

- The EU may promote its blueprint through participation in international institutions and negotiations.

- A de facto Brussels Effect with regard to a jurisdiction increases its incentive to adopt EU-like regulations, for instance by reducing the *additional* burden that would be placed on companies that serve both markets.

- The EU may actively incentivise the adoption of EU-like regulations, for instance through trade rules.

We think de jure diffusion is particularly likely for jurisdictions with significant trade relations with the EU, as introducing requirements incompatible with the AI Act's requirements for "high-risk" systems would impose frictions to trade. We also think there is a significant chance that these requirements will produce a de jure effect by becoming the international gold standard for the responsible development and deployment of AI.

A de jure effect is more likely for China than for the US, as China has chosen to adopt many EU-inspired laws in the past. However, China is unlikely to include individual protections from state uses of AI. Further, China has already adopted some new AI regulation, somewhat reducing the opportunity for a de jure effect.

**We are more likely to see a Brussels Effect than a Washington or Beijing Effect**

The US is unlikely to implement more stringent legislation than the EU, making a Washington Effect unlikely. Beijing will struggle to create a de facto Beijing Effect as companies often already offer products specifically for the Chinese market, though there could be a de facto Beijing Effect through Chinese firm exports. There is some chance that we see a de jure Beijing Effect with regard to countries that share the Chinese Communist Party's regulatory goals.

## Implications

**EU policymakers** and other **actors with an interest in AI regulation** should take especially great care to ensure the EU's regulatory regime addresses risks from AI, since the regime may diffuse across the world. It is especially important, for instance, to ensure that EU AI regulation is future-proof and can be adapted to a world of increasingly transformative AI capabilities.

**Policymakers worldwide** should expect their jurisdictions to experience a partial de facto Brussels Effect. As a result, they – and non-EU AI companies – might want to increase participation in the EU regulatory process. They may also face incentives to ensure that their regulation is compatible with the EU's regime.

The **global AI field** should invest in certain research topics – including explainability, fairness, transparency, robustness, and human oversight – to help guide the EU's regulatory efforts. The proposed regulation should be seen as a rallying cry to engage with policymakers and produce the research needed to support the development and enforcement of useful standards.

Finally, **regulators and standard setters beyond the EU legislative process** should take note. Higher prospects of an AI Brussels Effect might suggest that other rules and standards for AI could diffuse globally. This includes Californian AI regulation affecting US federal regulation – a "California Effect" – and standards set by organisations such as the ISO, NIST, and the IEEE having a global effect.

## STRUCTURE OF THE REPORT

We have aimed to make the report modular. We encourage readers to skip to the sections they expect to find most informative.

**The introduction** summarises the EU's upcoming regulatory regime for AI, as well as the rest of the report.

**Section 2** concerns the de facto Brussels Effect: whether firms outside the EU will voluntarily comply with EU AI regulation. It outlines the core mechanisms of de facto diffusion and assesses its likelihood, for various kinds of AI systems and requirements in the proposed AI Act.

**Section 3** concerns the de jure Brussels Effect: whether other jurisdictions will adopt EU-like regulation. It outlines the core mechanisms of de jure diffusion and assesses its likelihood.

**The appendix** details three relevant case studies of the Brussels Effect: the EU's regulatory regime for data privacy, its product liability regime, and its product safety scheme (including CE marking).

**Table 1** summarises the EU Commission's proposed AI Act.

**Table 2** summarises our conclusions on the likelihood that various parts of the proposed AI Act will produce a Brussels Effect.

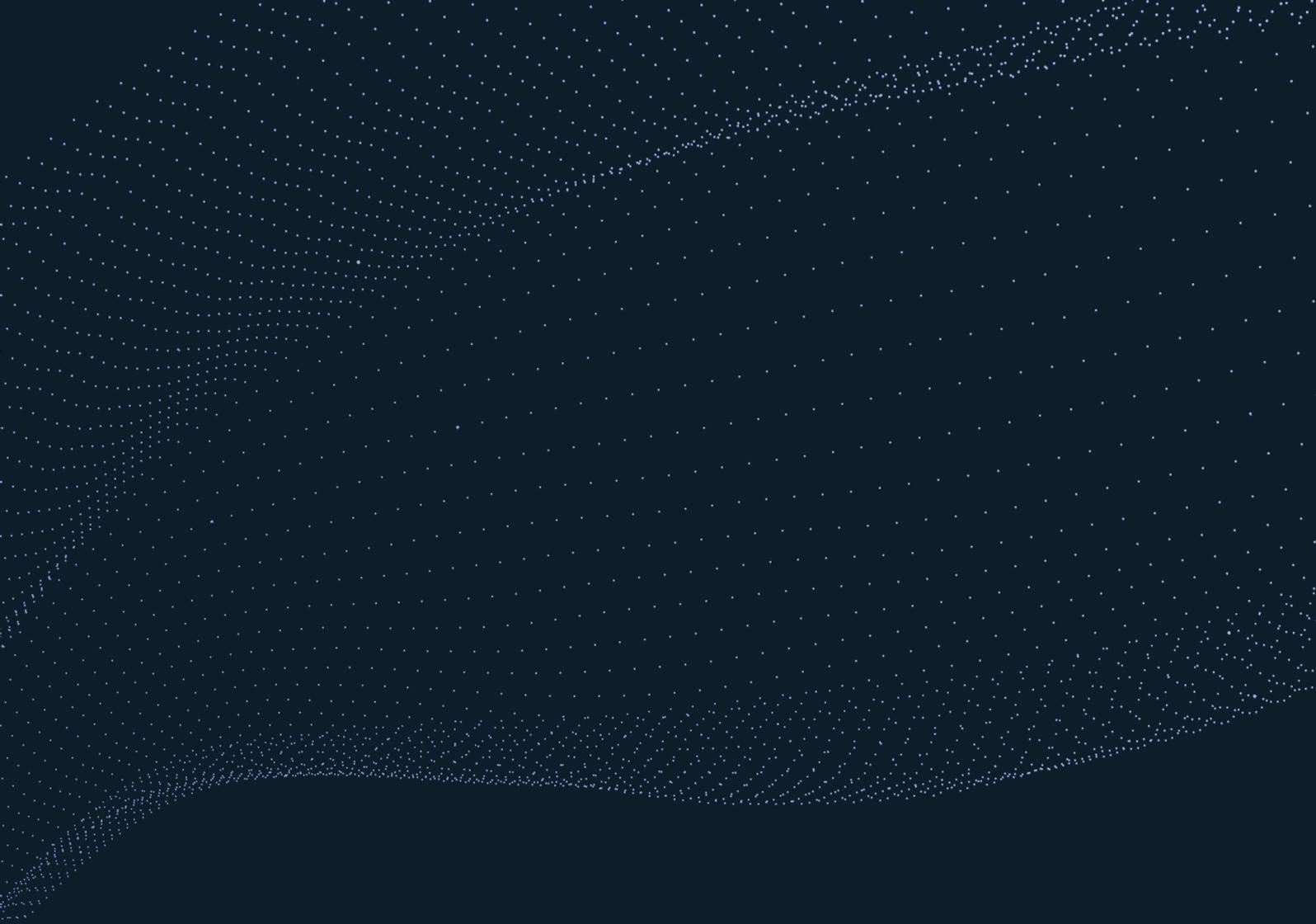

## ACKNOWLEDGEMENTS

For valuable comments, input, and discussion, we thank Joslyn Barnhart, Haydn Belfield, Alexandra Belias, Mathias Bonde, Miles Brundage, Will Carter, Allan Dafoe, Tom Davidson, Jeff Ding, Noemi Dreksler, Gillian Hadfield, Shin-Shin Hua, Henry Josephson, Jade Leung, Darius Meissner, Nicolas Moës, Ben Mueller, Zach Robinson, Daniel Schiff, Toby Shevlane, Charlotte Stix, Emma Bluemke, and Robert Trager. We also thank audiences at the CHAI all-hands meeting and the European Governance Research Network bookclub in September 2021, as well as attendees of various Centre for the Governance of AI work-in-progress sessions. We also thank Wes Cowley for copy editing, Maria Valente De Almeida Sineiro Vau for fact-checking, José Luis León Medina for referencing assistance, and Aleksandra Knežević for typesetting. Special thanks to Stefan Torges and Ben Garfinkel.

The cover image was generated using Midjourney, a machine learning image generation tool, covered by a Creative Commons License. Prompt engineered by Noemi Dreksler.

# Abbreviations

**AI**
artificial intelligence

**AIA**
EU AI Act

**B2B**
business-to-business

**B2C**
business-to-consumer

**CE**
Conformité Européenne (the European conformity marking for "safe" products)

**CEN**
European Committee for Standardization

**CENELEC**
European Committee for Electrotechnical Standardization

**CoE**
Council of Europe

**DMA**
Digital Market Act

**DPC**
Data Protection Commission

**DPD**
Data Protection Directive

**DSA**
Digital Services Act

**EC**
European Commission

**EMAS**
Eco-Management and Auditing Scheme

**EU**
European Union

**GAFAM**
Google, Apple, Facebook, Amazon, and Microsoft

**GAN**
generative adversarial network

**GDPR**
General Data Protection Regulation

**GMO**
genetically modified organism

**HLEG**
High-level expert group

**ICT**
information and communications technology

**ISO**
International Organization for Standardization

**MRA**
Mutual Recognition Agreement on Conformity Marking

**MSA**
market surveillance authority

**OMB**
Office of Management and Budget (a US government agency)

**PET**
privacy-enhancing technologies

**PLD**
Product Liability Directive

**PNR**
passenger name record

**PSD**
Product Safety Directive

**QMS**
quality management system

**REACH**
Registration, Evaluation, Authorisation and Restriction of Chemicals regulation

**RoHS**
Restriction of Hazardous Substances directive

**SME**
small-to-medium enterprise

# Contents

# 1. Introduction

## 1.1. The EU's Upcoming Regulatory Regime for AI

Over the past two years, the European Commission ("the Commission" below) has proposed a number of updates and additions to EU regulation[5] that will likely have significant impact on the AI industry.[6] These include the AI Act[7] (the primary focus of this report), updates to the EU liability regime, the Digital Market Act (DMA), and the Digital Services Act (DSA). The Commission has not yet proposed liability updates. It may take 1–2 years before the AI Act has been finalised in negotiations between the Parliament and Council. The Digital Markets Act and the Digital Services Act are both expected to be formally adopted in the summer of 2022.

### 1.1.1. The Digital Market Act and the Digital Service Act

In December 2020, the Commission presented their proposed Digital Market Act and Digital Services Act.[8] These acts jointly seek to rein in the power of big tech companies and make digital markets more competitive.[9] They are expected to be formally adopted in the summer of 2022, following adoption by the Parliament in July 2022.[10]

The DMA would prohibit some practices of "gatekeeper" companies, such as self-preferencing their products in search results, and would restrict gatekeepers' ability to reuse personal data across platforms. Most big tech companies are expected to be considered gatekeepers.[11] Without removing or amending existing EU competition law, the DMA adds new rules which consider certain actions unfair ex ante, before the fact. This is a break with the current competition law regime which requires investigations into whether there has been a breach of competition law after some potential breach has been committed. The DMA will also require companies to report upcoming mergers and acquisitions to the Commission, though it does not give the Commission new powers to block them.

---

[5] We speak of regulation to mean all regulatory instruments. Regulation is also one legal instrument of the EU, which is directly translated into national law. In contrast, directives are legislative acts that set out the goals that all member states must achieve, while preserving the freedom of member states to decide how to achieve those goals best. The AI Act is a proposed piece of regulation; other future legislation could be in the form of a directive, e.g. for AI liability rules.

[6] A more thorough overview and history can be found in Mark Dempsey et al., *"Transnational Digital Governance and Its Impact on Artificial Intelligence,"* in The Oxford Handbook of AI Governance, ed. Justin Bullock et al. (Oxford University Press, May 19, 2022)

[7] European Commission, *"Proposal for a Regulation of the European Parliament and of the Council Laying Down Harmonised Rules on Artificial Intelligence (Artificial Intelligence Act) and Amending Certain Union Legislative Acts COM/2021/206 Final,"* CELEX number: 52021PC0206, April 21, 2020

[8] European Commission, *"Proposal for a Regulation of the European Parliament and of the Council on Contestable and Fair Markets in the Digital Sector (Digital Markets Act) COM/2020/842 Final,"* CELEX number: 52020PC0842, December 15, 2020; European Commission, *"Proposal for a Regulation of the European Parliament and of the Council on a Single Market for Digital Services (Digital Services Act) and Amending Directive 2000/31/EC COM/2020/825 Final,"* CELEX number: 52020PC0825, Dec,15,2020.

[9] Aline Blankertz and Julian Jaursch, *"What the European DSA and DMA Proposals Mean for Online Platforms,"* Brookings, January 14, 2021.

[10] Both proposals were adopted by the European Parliament in July 2022. They are expected to be formally adopted by Council and published in the EU Official Journal. European Parliament, *"Digital Services: Landmark Rules Adopted for a Safer, Open Online Environment,"* May 7, 2022; European Parliament, "Digital Markets Act: EP Committee Endorses Agreement with Council," May 16, 2022; European Parliament, *"Digital Services Act: Agreement for a Transparent and Safe Online Environment,"* April 23, 2022.

[11] "'Gatekeeper' platforms with a turnover of at least €6.5bn; activities in at least 3 EU countries; at least 45 million monthly active end-users and 10,000 yearly active business users (both in the EU); having met these thresholds in the last three years. Alternatively, an investigation can determine applicability." Blankertz and Jaursch, *"What the European DSA and DMA Proposals Mean for Online Platforms."*

The DSA focuses specifically on content moderation on large platforms, mainly those with more than 45 million EU users.[12] It will continue to be the case that platforms are not sanctioned for having illegal content on their websites, but there are new obligations to try to find such content and to remove it if found. The DSA will also include provisions requiring companies to disclose some details of how their content moderation algorithms work, how they decide what content to remove, and how advertisers are targeting users.

Though we focus on the AI Act and changes to liability in this report, as these are directly aimed at regulating AI systems, the impacts of the DMA and DSA on the global AI industry may be significant. Those acts could more significantly impact how big technology companies deploy AI systems in Europe than the AI Act will. We encourage others to explore whether the DMA and DSA would lead to a de facto and/or de jure Brussels Effect as some have suggested.[13]

We hypothesise that much of the DMA and DSA will not have a strong *de facto* Brussels Effect, as the costs of differentiation, e.g. implementing different pricing strategies in different jurisdictions, might be low and because the benefits of behaviour in breach of the proposed legislation may be significant. However, there may be a significant de facto effect with regard to mergers and acquisitions, as the Commission has powers to block global mergers if the merging parties have sufficient turnover in the EU.[14] The disclosure requirements introduced by the DSA[15] could exhibit a de facto effect, so long as it does not require disclosure of information that could be particularly detrimental to the company. For example, if Google were to release their model for search in full, that could make it possible to exploit the algorithm using search engine optimization to place one's website high in the search results without that website being what the user is looking for. In such a situation, Google might be forced to keep separate algorithms for EU and non-EU markets. There might also be a significant de jure effect considering increasing interest among US legislators in updating US antitrust laws, including proposals with new ex ante regulatory requirements for big tech companies, similar to the DMA.[16] Further, the UK is currently considering adopting a Digital Markets Bill, which shares many features of the DMA.[17]

**1.1.2. The AI Act**

In April 2021, the EU Commission published its AI Act (AIA) proposal.[18] It may take 1–2 years before the bill has been finalised in negotiations between the Parliament and Council. The AI Act takes a risk-based approach, classifying AI systems as creating unacceptable, high, limited, or minimal risk. The level of risk is judged by the likelihood that the system may harm specific individuals,[19] potentially violating their fundamental rights. The requirements imposed on systems are related to the level of risk, ranging from prohibitions to the voluntary adoption of codes of conduct. The AIA proposes prohibitions on AI applications that pose "unacceptable risks", including "real-time" remote biometric identification systems used by governments. It requires conformity assessments for "high-risk" AI systems, such as some AI systems deployed in worker management, critical infrastructure operation, border control, remote biometric identification, medical devices, machinery, and other areas.[20] Certain limited-risk AI systems need to comply with transparency rules, requiring that users are made aware e.g. if they are engaging with AI-generated content that may appear authentic such as chatbots or deepfakes. All other AI systems, termed "minimal risk", face no additional obligations, though providers are encouraged to follow voluntary codes of conduct. We summarise the proposed AI Act in Table 1.

---

[12] European Commission, *"Proposal for a Regulation of the European Parliament and of the Council on a Single Market for Digital Services (Digital Services Act) and Amending Directive 2000/31/EC COM/2020/825 Final."*

[13] Anu Bradford, *"The Brussels Effect Comes for Big Tech,"* December 17, 2020; Alex Engler, *"The EU AI Act Will Have Global Impact, but a Limited Brussels Effect,"* Brookings, June 8, 2022.

[14] Anu Bradford, *The Brussels Effect: How the European Union Rules the World* (Oxford University Press, 2020).

[15] See European Commission, *"Proposal for a Regulation of the European Parliament and of the Council on a Single Market for Digital Services (Digital Services Act) and Amending Directive 2000/31/EC COM/2020/825 Final"*, articles 13, 23, 24.

[16] Senate Republican Policy Committee, *"Big Tech Gets Bigger, Calls for Antitrust Changes Get Louder,"* Senate RPC, November 18, 2021.

[17] DCMS and BEIS, *"A New pro-Competition Regime for Digital Markets - Government Response to Consultation, Command Paper: CP 657,"* May 6, 2022.

[18] AI Act.

[19] That is, it does not seek to mitigate small harms that afflict a large number of people.

The draft legislation builds on years of policy efforts in the EU, including the Commission's AI Whitepaper in February 2020 and the High-level expert group's AI Ethics Guidelines in April 2019.[21] The proposed AI Act is expected to enter into force in a few years after being negotiated and amended by the European Parliament and the Council of the European Union.

The proposed AIA prohibits the following uses of AI: (i) systems that deploy "subliminal techniques" or use vulnerabilities of a specific group[22] to materially distort their behaviour such that they cause harm or are likely to do so to themselves or other persons, (ii) the use of "social scores" by public authorities or on their behalf, and (iii) the use of 'real-time' remote biometric identification systems in publicly accessible spaces for the purpose of law enforcement" with a small number of exceptions.[23] There is significant uncertainty about how to interpret the ban on subliminal techniques, e.g. when a group's vulnerability has been used, and what level of harm to an individual is required.[24] Thus, it is not clear whether recommender systems and algorithms used in social media news feeds could be prohibited under the regulation.[25] Such systems could avoid the prohibition because companies are not held liable for harm caused by content on their platforms that has been posted by others, but this is not yet clear. Further, ambiguity on whether e.g. the Google search algorithm would be considered manipulative would likely impose large costs to tech companies, as these companies have already pointed out.[26]

The Commission's proposal classifies some AI systems as high-risk.[27] Producers of such systems are obligated to go through a conformity assessment to ensure they comply with certain standards before they are put on the EU market. Systems identified as high-risk are firstly those that are safety components in or constitute products in domains that are already covered by 12 EU product safety regulations and that require third party conformity assessments. The full list is available in Annex II, Section A, and most notably includes medical devices (including those for in vitro diagnostics), toys, and machinery. For these products, the AI Act proposes that existing product safety regulation be updated such that it ensures compliance also with the AI Act, to reduce regulatory complexity.[28] There is also a list of seven additional product safety regulations listed in Annex II, Section B, covering e.g. aviation and cars, where the AI Act introduces no new requirements for producers.[29] However, in the recitals accompanying the AI Act, the Commission suggests that "the ex-ante essential requirements for high-risk AI systems set out in this proposal will have to be taken into account when adopting relevant implementing or delegated legislation under those acts."[30]

---

[20] See AI Act, annex II. When we refer to "high-risk" AI systems throughout this report, we simply refer to the Commission's definition.

[21] European Commission, *Ethics Guidelines for Trustworthy AI* (Publications Office of the European Union, 2019).

[22] Due to "their age, physical or mental disability."

[23] AI Act, title II, art. 5.

[24] Michael Veale and Frederik Zuiderveen Borgesius, *"Demystifying the Draft EU Artificial Intelligence Act — Analysing the Good, the Bad, and the Unclear Elements of the Proposed Approach,"* Computer Law Review International 22, no. 4 (August 1, 2021): 97–112. It has also been criticised for excluding forms of manipulation. Dan Taylor, *"Op-Ed: The EU's Artificial Intelligence Act Does Little to Protect Democracy,"* Tech.eu, March 14, 2022.

[25] Facebook, *"Response to the European Commission's Proposed AI Act,"* August 6, 2021; Will Douglas Heaven, *"This Has Just Become a Big Week for AI Regulation,"* MIT Technology Review, April 21, 2021.

[26] Facebook, *"Response to the European Commission's Proposed AI Act"*; Google, "Consultation on the EU AI Act Proposal," July 15, 2021.

[27] AI Act, annex III(1).

[28] In addition to this, the Commission started a process of renewing the EU's General Product Safety Regulation in June 2021. European Parliament, *"General Product Safety Regulation,"* Legislative Train Schedule European Parliament, June 23, 2022; European Parliament, *"2021/0170(COD),"* 2021.

The AI Act lists additional high-risk uses of AI in Annex III. This list includes remote biometric identification and categorisation, admission or grading in education, management and operation of critical infrastructure, law enforcement, and certain aspects of employment and worker management. The category of AI used for "employment, worker management, and access to self-employment opportunities," appears particularly sizable and fast-growing:[31] it likely includes nearly all gig economy companies, ranging from new ride-hailing companies (e.g. Uber and Bolt) to the collection of errant e-scooters (e.g. Bolt, Bird, and Lime), delivery companies (e.g. Deliveroo, Foodora, and Just Eat), and various other freelancing platforms (e.g. Fiverr, Amazon Mechanical Turk, and TaskRabbit). Further, it will likely apply to the growing industry of software for staff scheduling and hiring, often referred to as workforce management.

Further, there will be many systems in the financial sector that determine "[a]ccess to and enjoyment of essential private services … services and benefits." Critical infrastructure systems include road traffic, gas, water, heating, and electricity. Remote biometric identification does not seem to cover facial recognition systems used in place of signatures,[32] though it will likely apply to automatic tagging of photos by e.g. Google or Facebook. High-risk systems also include a number of government uses of AI, including certain uses in law enforcement, border control and migration, the courts, and social benefit allocation.[33] For more details, see Table 1.

---

[29] AI Act, art. 2 §2. Though Article 84 will still apply, which refers to the EU Commission's responsibilities to review and evaluate the AI Act at certain intervals, Article 84 §7 suggests that such a review could result in the Commission recommending legislation be introduced to have the rest of the AI Act's requirements apply to these Old Approach product safety regulations.

[30] AI Act, recitals, 1.2.

[31] For example, the ride-hailing and food-delivery markets both garnered 150 million users in Europe in 2021 and are expected to see significant growth in the coming few years. The food-delivery app market is estimated to grow by 10% annually. David Curry, *"Taxi App Revenue and Usage Statistics (2022),"* Business of Apps, November 10, 2020; David Curry, *"Food Delivery App Revenue and Usage Statistics (2022),"* Business of Apps, October 29, 2020; Deloitte LLP, *"Delivering Growth,"* Deloitte United Kingdom, November 26, 2019, 8.

[32] European Commission, *"Speech by Executive Vice-President Vestager at the Press Conference on Fostering a European Approach to Artificial Intelligence,"* April 21, 2021.

[33] For more, see AI Act, annex III.

| CATEGORY | SCOPE | REQUIREMENTS | SANCTIONS |
| --- | --- | --- | --- |
| **Unacceptable Risk: Prohibited (Title II)** | • Subliminal techniques or exploiting vulnerabilities of specific populations which cause harm<br>• "Social scores" used by public authorities or on their behalf<br>• Real-time remote biometrics in public spaces used by law enforcement (with some exceptions) | These uses are prohibited. | Fines up to 6% of global revenue or 30mn euros, whichever is higher |
| **High-Risk Systems: Conformity Assessment (Title III)** | **Annex II:**<br>• AI systems that are products or safety components of products covered by 12 product safety regulation regimes and that require third party conformity assessments, including medical devices (including for in vitro diagnostics), toys, and machinery.<br>**Annex III:**<br>• Remote biometric identification and categorisation of natural persons (e.g. a system classifying the number of people of different skin tones walking down a street)<br>• Management and operation of critical infrastructure (road traffic and the supply of water, gas, heating, and electricity)<br>• Education and vocational training, where systems are used for e.g. admission and grading<br>• Employment, worker management, and access to self-employment opportunities, including systems that make or inform decisions about hiring, firing, and task allocation<br>• Access to and enjoyment of essential private services and public services and benefits<br>• Specific uses of law enforcement<br>• Specific uses in migration, asylum, and border control management<br>• Administration of justice and democratic processes, in particular when used to research and establish facts or applying the law to some facts | Providers of high-risk systems must perform a conformity assessment to make sure that they are compliant with requirements including:<br>• Risk management system<br>• Data requirements<br>• Technical documentation<br>• Record-keeping<br>• Transparency on the system's functioning<br>• Human oversight<br>• Accuracy, robustness, and cybersecurity<br>• Post-market monitoring | Fines up to 4% of global revenue or 20mn euros, whichever is higher, for everything except the data requirements, where the same fines apply as for the prohibited systems |
| **Limited Risk: Transparency Obligations (Title IV)** | • AI systems interacting with natural persons<br>• Emotion recognition systems or biometric categorisation systems<br>• AI system that generates or manipulates image, audio, or video content that appears real | Notify the user that they are engaging with an AI system | Fines up to 4% of global revenue or 20mn euros, whichever is higher |
| **Minimal Risk: Voluntary Codes of Conduct (Title IX)** | All AI systems that are not either prohibited or high-risk | Providers can choose to comply with voluntary codes of conduct. The Commission and Member States will encourage the creation and voluntary compliance with these codes. | Not applicable as there are no requirements. |

Table 1: A summary of the EU Commission's proposed AI Act.[34]

---

[34] Inspired by the graphic in Eve Gaumond, *"Artificial Intelligence Act: What Is the European Approach for AI?,"* Lawfare, June 4, 2021.

The list of high-risk AI systems can be updated over time. The EU Commission can add additional uses to the list in Annex II, so long as they are under the eight categories outlined in the annex (e.g. education, law enforcement, and biometric identification) and the use poses similar risks to the uses currently on the list.[35]

In the Commission's proposed AI Act, producers of high-risk AI systems have to comply with specific standards and procedures before putting the products on the EU market, after which they must add the CE mark to their product. Producers of high-risk systems would be required to have a risk management system that includes identifying and analysing risks, post-market monitoring, implementing suitable risk management measures, and communicating residual risks to users. Moreover, producers are required to eliminate or reduce risks through adequate product design and development. In addition, they need to conform to requirements for data governance, technical documentation, and record-keeping. Producers should also integrate human oversight into their products, such as with human-machine interface tools, for example to ensure that individuals overseeing the system "fully understand the capacities and limitations of the high-risk AI system and be able to duly monitor its operation".[36] Finally, the AI Act proposes requirements for accuracy, robustness, and cybersecurity of AI systems. This includes both resilience to errors and to attempts by unauthorised parties to alter the system's use or performance by exploiting vulnerabilities, including via data poisoning, adversarial examples, or model flaws.[37]

For "limited-risk" systems, the Commission's proposed AI Act includes provisions requiring deployers to inform users if their system (i) interacts with humans, (ii) is used to detect emotions or determine association with (social) categories based on biometric data, or (iii) generates or manipulates content, e.g. deepfakes or chatbots.[38]

All other AI systems, termed "minimal risk", face no additional obligations, though providers are encouraged to follow voluntary codes of conduct. The proposed AI Act tasks the Commission and member states with encouraging and facilitating the drawing up of voluntary codes of conduct.

The AI Act includes clauses to promote compliance. Member states are, according to the AI Act, obligated to designate or create market surveillance authorities (MSAs) to oversee and ensure the implementation of the regulation, with significant powers to request information from providers of AI systems. Non-compliance with the AI Act would come with significant fines. Breaching the prohibitions or the data governance requirements for high-risk systems can produce fines of up to 30 million euros or 6% of global annual turnover, whichever is higher. Non-compliance with all other requirements in the AI Act may have the actor incur up to 20 million euros or 4% of global annual turnover, whichever is higher.[39]

**Regulatory Costs of the AI Act**

Though it is exceedingly difficult to predict the costs imposed by new regulation, there have been attempts to estimate them. These regulatory costs consist of compliance costs, which are those associated with meeting the requirements, and verification costs, those associated with being able to evidence compliance.

---

[35] AI Act, art. 73.
[36] Other notable clauses state "Human oversight shall aim at preventing or minimising the risks to health, safety or fundamental rights that may emerge when a high-risk AI system is used in accordance with its intended purpose or under conditions of reasonably foreseeable misuse, in particular when such risks persist notwithstanding the application of other requirements set out in this Chapter" as well as the individual overseeing the system's functioning being "able to intervene on the operation of the high-risk AI system or interrupt the system through a 'stop' button or a similar procedure."
[37] AI Act, title III, chapter 2.
[38] However, Veale and Borgesius argue that the transparency obligation may be unenforceable. Market surveillance authorities will struggle to find the undisclosed deepfakes, especially if there are limited routes for citizens to file complaints. Veale and Borgesius, *"Demystifying the Draft EU Artificial Intelligence Act — Analysing the Good, the Bad, and the Unclear Elements of the Proposed Approach."* See also AI Act, title IV.
[39] AI Act, art. 71

While the EU hopes to reduce the regulatory costs of operating AI systems in the EU, the AI Act, in the form suggested by the EU Commission, could be costly. For example, many commentators have pointed to the impracticability of Article 10§3, which states that "training, validation and testing data sets shall be relevant, representative, free of errors and complete."[40] Meeting such a requirement could be incredibly costly as it is nearly impossible to ensure that a dataset is free of errors or complete. No dataset is perfect. However, there are indications that this requirement will be different in the final bill. The recitals that accompanied and contextualised the EU Commission's draft AI Act include a weaker, more practicable version of the statute: "Training, validation and testing data sets *should be sufficiently relevant,* representative and free of errors and complete *in view of the intended purpose of the system"* [our emphasis].[41] Further, the French presidency of the EU Council proposed changes to the AI Act in early 2022, wherein datasets would need only be e.g. free of errors "to the best extent possible."[42]

The Commission's impact assessment estimates that the AI Act would impose additional regulatory costs of 6–10% to investments in developing high-risk AI systems (including the cost of verifying compliance),[43] suggesting that prices for EU products may rise by the same amount. The Commission says this represents the "theoretical maximum costs" imposed on high-risk systems, as it assumes that none of the requirements are already being complied with.[44] Further, these costs could fall over time because a significant proportion of the compliance and verifications costs will only be paid once.[45] They could also be reduced further if the AI Act reduces the net regulatory complexity of deploying AI systems, which are already regulated by existing rules that may be overlapping or otherwise inappropriate for AI systems.

The total cost could also be higher than the Commission's estimate. First, these estimates only focus on the cost imposed on high-risk AI systems, excluding regulatory costs as a result of voluntary codes of conduct and transparency requirements for e.g. chatbots. However, one might reason that the voluntary codes will only be followed should it look like a sound business decision and that the transparency requirements will impose small costs. Second, the study commissioned by the Commission before the draft AI Act was released finds higher regulatory costs of up to 17% of high-risk systems' development costs.[46]

### 1.1.3. Updated Liability Rules

In addition to the AI Act, the Commission seeks to adopt liability rules for AI products. In 2021, the Commission stated its intention to propose regulation either via an update of the Product Liability Directive (PLD) or by separately harmonising aspects of the national civil liability framework regarding certain AI systems in the first quarter of 2022.[47]

---

[40] AI Act, art. 10 (3). Microsoft, Google, Facebook, and DeepMind (part of Google) maintained in their submissions to the EU AI Act consultation that in certain cases this data requirement is unnecessary, and in others, impossible.

[41] AI Act, recitals, 44

[42] La Présidence Française du Conseil de l'Union européenne, *"Proposition de Règlement Du Parlement Européen et Du Conseil établissant Des Règles Harmonisées Concernant L'intelligence Artificielle (législation Sur L'intelligence Artificielle) et Modifiant Certains Actes Législatifs de l'Union - Texte de Compromis de La Présidence - Articles 16-29."*

[43] European Commission, *"Commission Staff Working Document Impact Assessment Accompanying the Proposal for a Regulation of the European Parliament and of the Council Laying Down Harmonised Rules on Artificial Intelligence (Artificial Intelligence Act) and Amending Certain Union Legislative Acts SWD/2021/84 Final,"* CELEX number: 52021SC0084, April 21, 2021, 67–70.

[44] European Commission, 66.

[45] See European Commission, *"Commission Staff Working Document Impact Assessment Accompanying the Proposal for a Regulation of the European Parliament and of the Council Laying Down Harmonised Rules on Artificial Intelligence (Artificial Intelligence Act) and Amending Certain Union Legislative Acts SWD/2021/84 Final."*

[46] Andrea Renda et al., *"Study to Support an Impact Assessment of Regulatory Requirements for Artificial Intelligence in Europe Final Report (D5)"* (Luxembourg: European Commission, April 2021), Chapter 4.

[47] See European Commission, *"Commission Staff Working Document Impact Assessment Accompanying the Proposal for a Regulation of the European Parliament and of the Council Laying Down Harmonised Rules on Artificial Intelligence (Artificial Intelligence Act) and Amending Certain Union Legislative Acts SWD/2021/84 Final."* The Inception Impact Assessment from June 2021 also clearly communicates these aims.

## 1.2. Will We See a Brussels Effect for the EU AI Regulatory Regime?

Having described the contours of the upcoming EU regulation of AI above, we now summarise the mechanisms that may lead to de facto and de jure Brussels Effects, and their plausibility for upcoming EU AI regulation.

Some clarifications could be helpful at this point. Throughout the report, we use "Brussels Effect" to simply refer to regulatory diffusion from the European Union. We do not limit our discussion to diffusion that occurs solely due to market forces. We also do not treat regulatory diffusion as an all-or-nothing phenomenon – we allow for degrees of diffusion. Further, we focus primarily on the EU Commission's April 2021 proposed AI Act. For the most part, we do not consider whether proposed amendments from the EU Parliament and Council to the Commission's draft may differ in their propensity for a Brussels Effect. Further, we do not look closely at the chance of a Brussels Effect from the recently passed Digital Services Act and Digital Markets Act. We encourage others to pursue that work, in particular as these bills could have a large impact on how some of the world's most widely interacted with AI systems are developed and deployed.[48]

We also contribute to the conceptual understanding of the drivers of regulatory diffusion. While similar factors of the Brussels Effect have been introduced in Bradford (2020), we either generalise or disentangle each of Bradford's factors into 2–4 components.[49]

### 1.2.1. De Facto Brussels Effect

A de facto Brussels Effect occurs when companies voluntarily comply with EU regulation in non-EU jurisdictions without those jurisdictions requiring it. As with all jurisdictions, when the EU introduces new rules, multinational companies face two decisions. First, they must decide whether to remain in the EU market. New regulation could sufficiently reduce the market size and profit margins to make operating in the EU market unprofitable. Second, assuming firms stay in the EU market, they must decide whether to comply with the new regulation internationally or offer two different products: one EU-compliant and one non-EU-compliant.[50] We use the term "differentiation" to refer to offering different products for different jurisdictions, and "non-differentiation" for offering an EU-compliant product outside the EU. A de facto Brussels Effect has occurred if firms stay in the EU market and sell EU-compliant products worldwide (see Figure 1).[51] Note that throughout the report, we refer to products and services simply as products.

This section summarises section 2 of this report, describing the mechanisms by which a de facto Brussels Effect can occur and our high-level conclusions on its plausibility with regard to the EU's forthcoming regulatory regime.

---

[48] As discussed in Engler, *"The EU AI Act Will Have Global Impact, but a Limited Brussels Effect."*
[49] See also the introduction of section 2 for more detail.
[50] Though note that these two decisions will in reality be made at the same time. Firms need not prefer both differentiation and non-differentiation to leaving in order to choose to stay in the market.
[51] Technically, multinational companies are also a requirement for a de facto Brussels Effect (§2.1.2). If all firms in the industry only sell nationally, which is, for instance, the predominant case in the metal industry, a de facto Brussels Effect will never occur. For the particular example of the metal industry and regulation – wherein the local industry did not exhibit a de facto Brussels Effect. David Hanson, *CE Marking, Product Standards and World Trade* (Edward Elgar, 2005),

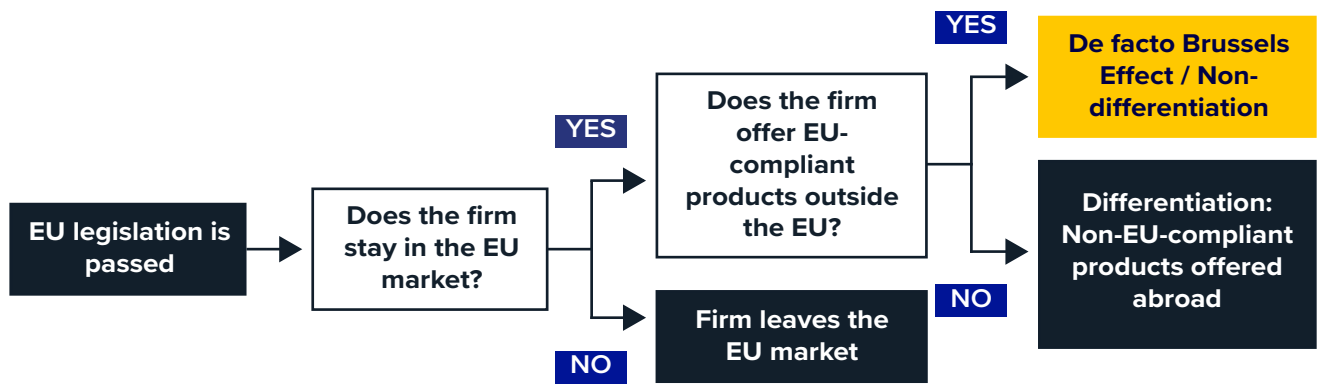

Figure 1: De facto Brussels Effect decision tree. A de facto Brussels Effect occurs when (i) the EU puts in place legislation that is more stringent than other jurisdictions, (ii) the company decides to stay in the EU market, and (iii) decides to adopt the regulation outside the EU.

*Building on*[52] Anu Bradford's 2020 book *The Brussels Effect*[53] and 2012 paper by the same name,[54] we consider five determining factors of the de facto Brussels Effect:

(i) *Favourable market properties* – The larger the absolute EU market, the more likely companies will stay in the market despite the legislation. The larger the relative EU market, the more likely companies will sell EU-compliant rather than non-EU-compliant products outside the EU. The more the market is oligopolistic and consists of multinational firms, the more likely is de facto regulatory diffusion. (See section 2.1.)

(ii) *Stringency* – The EU regulation must be more stringent than other jurisdictions' regulations, at least on some dimensions, for the de facto Brussel Effect to occur. (See section 2.2.)

(iii) *Regulatory capacity* – This concerns a jurisdiction's expertise and capacity to produce well-crafted legislation, ideally earlier than other jurisdictions, and to sanction non-compliance. Well-crafted legislation lowers the regulatory costs and increases the likelihood that companies comply with the regulation and that customers value EU-compliant products, while ideally meeting the same regulatory goals. We refer to the sum of compliance and verification costs as regulatory costs, where compliance costs are those associated with meeting the requirements of the regulation and verification costs are those associated with showing and documenting that this is the case. (See section 2.3.)

(iv) *Inelasticity within and outside the EU* – Demand and supply, both within and outside the EU, need to be relatively inelastic, such that the market size does not shrink in response to the regulation, e.g. due to negative changes to price, cost, or quality resulting from the new regulation. Low elasticity within the EU in response to the new rules increases the chance of companies remaining in the EU, while low elasticity outside the EU in response to EU-compliant products increases the chance of non-differentiation. (See section 2.4.)

(v) *Costs of differentiation* – The costs of differentiation being higher than those of non-differentiation increases the likelihood of a de facto effect. It can be more costly to choose differentiation – maintaining both EU-compliant and non-EU-compliant products – as it might come with higher fixed and variable regulatory costs, and duplication costs associated with maintaining two separate products rather than one. (See section 2.5.)

We argue in section 2.6 that a de facto Brussels Effect for at least some of the EU's AI

---

[52] We summarise how our proposed framework differs from Bradford's in section 2.
[53] Bradford, *The Brussels Effect: How the European Union Rules the World*.
[54] Anu Bradford, *"The Brussels Effect,"* Northwestern University Law Review 107 (Northwestern University School of Law, 2012).

INTRODUCTIONregulation is likely. High-risk systems deployed by multinational companies seem particularly likely to see a de facto effect. These systems include those used in products covered by existing EU product safety regulation such as machinery and medical technology.[55] They could also include systems used for worker management and hiring, remote biometric identification systems, and legal tech, especially if compliance with EU requirements becomes seen as a strong signal of product trustworthiness. Foundation models could see a de facto effect if it turns out to be difficult to comply with the regulation without making fundamental changes to those systems and if the market for high-risk systems grows significantly. A de facto effect is more likely for some requirements of high-risk systems, such as those regarding setting up risk management systems, record-keeping, and making the system's functioning sufficiently transparent, as well as accuracy, robustness, and cybersecurity requirements.

Transparency requirements, such as for e.g. AI systems producing content that may appear authentic, could see a de facto effect, as extending compliance beyond the EU would likely be cheap and because customers might appreciate the transparency. On the other hand, the cost of differentiation is likely to be low. The AI Act's prohibitions are unlikely to see a de facto effect, though they could have a weak effect by changing norms in other jurisdictions about the acceptability of such systems.

### 1.2.2. De Jure Brussels Effect

There is a de jure Brussels Effect if foreign jurisdictions adopt rules influenced by EU regulation. We analyse four channels for the diffusion of EU AI regulation.

1. *Blueprint Adoption Channel* – Foreign jurisdictions adopt the EU regulation voluntarily as they believe the legislation will meet their regulatory goals. This may be because of imitation before the results of the regulation are known, e.g. because EU regulations are usually well-crafted due to the EU's regulatory expertise and capacity, or it could be a result of learning, with non-EU jurisdictions adopting the regulation once positive results are seen. (See section 3.1.)

2. *Multilateralism Channel* – The EU promotes its regulation in multilateral and bilateral negotiations and institutions. For instance, EU product safety standards are regularly promoted in and influence work at the International Organization for Standardization (ISO).[56] (See section 3.2.)

3. *De Facto Channel* – Subjected to a de facto Brussels Effect, multinational companies may be put at a disadvantage in non-EU markets compared to national companies operating only in the non-EU market. Therefore, multinational companies are incentivised to lobby legislators in non-EU jurisdictions to adopt EU-equivalent standards. For such jurisdictions, the cost of adopting such standards is also lower, as some companies are already complying with them. (See section 3.3.)

4. *Conditionality Channel* – EU trade requirements, extraterritoriality (that is, when the legal power of a jurisdiction is extended beyond its territorial boundaries), and economic pressure encourage other countries to adopt EU-equivalent regulation. (See section 3.4.)

The *Blueprint Adoption Channel* is plausible for AI because of the EU's first mover advantage, the Commission's active promotion of their AI regulation,[57] and the diffusion of the EU's AI policy narrative over the last three years.[58] This channel seems most likely to impact jurisdic-tions smaller than the US and China for which compatibility with EU regulation is particularly important and where there are no large domestic AI companies to oppose the measures. A de jure Brussels Effect reaching the US federal level seems much less likely. Historically, there have been few instances of a de jure Brussels Effect reaching the US via this channel.[59] However, it does seem plausible that we will see some regulatory diffusion to US states – notably California, which has adopted data protection laws similar to the GDPR – which could affect future regulation at the federal level.

---

[55] Listed in AI Act, annex II.

[56] Reasons cited are the first mover advantage. Moreover, the outsized influence in health and safety standards might also come from the more hierarchical regulatory structure of the EU compared to the US. Abraham L. Newman and Elliot Posner, *"Putting the EU in Its Place: Policy Strategies and the Global Regulatory Context,"* Journal of European Public Policy 22, no. 9 (October 21, 2015): 1316–1335. See for instance: Deborah Hairston, "Hunting for Harmony in Pharmaceutical Standards," Chemical Engineering 104, no. 20 (1997); Annika Björkdahl et al., eds., *Importing EU Norms Conceptual Framework and Empirical Findings, vol. 8,* United Nations University Series on Regionalism 8 (Springer International Publishing, 2015), 122.

[57] This includes the "International Alliance on Trustworthy AI". However, it might be too early to evaluate the extent of the Commission's promotion.

[58] See AccessNow report: Daniel Leufer and Laureline Lemoine, *"Europe's Approach to Artificial Intelligence: How AI Strategy Is Evolving"* (accessnow, December 2020). Note however that it is difficult to distinguish between the EU regulation causing other jurisdictions to adopt EU-esque regulation from the EU and the other jurisdictions simply responding to the same regulatory need.

A de jure Brussels Effect reaching China via this channel is plausible due to the country's extensive history of adapting regulatory blueprints from the EU and United States, though it might be less likely because the Chinese government is starting to adopt regulation for certain AI applications. Chinese legal documents often reference EU regulation. For instance, data protection legislation; the RoHS directive, which manages hazardous substances; the labelling schemes for genetically modified foods; energy regulation; and the chemical regulation REACH have been used as blueprints for Chinese law.[60] In 2021, China adopted the Personal Information Protection Law, which provides GDPR-like protections for citizens against private corporations.[61] However, Chinese regulators have recently charged ahead in some domains, regulating AI sooner and more stringently than the EU is likely to with regard to recommender systems and potentially systems that generate content.[62] Further, de jure regulatory diffusion to China is likely to be limited to regulation of private companies, and is unlikely to infringe on the Chinese state's uses of AI e.g. for surveillance of its citizens.

Perhaps the most important de jure effect via the Blueprint Channel would be if the AI Act sets the global gold standard for what requirements a responsible developer and deployer of risky AI systems ought to fulfil. These requirements seem likely to inspire other jurisdictions, even if they choose to define the riskiness of systems differently.

Upcoming EU AI regulation also comprises updates to the liability regime. A de jure Brussels Effect of the Product Liability Directive (PLD) (1985) reached more than a dozen countries through the Blueprint Adoption Channel (appendix 4.2). At the same time, the PLD did not lead to significant litigation cases in the EU. Hence, the de jure Brussels Effect of product liability could see limited real-world effects.

The *Multilateralism Channel* is plausible since the EU has historically influenced standard setting bodies, such as the ISO. The ISO sets and seeks to set AI product safety standards.[63] Moreover, bilateral coordination on technology policy, such as via the US-EU Trade and Technology Council,[64] make such de jure regulatory diffusion more likely. Taken together, the US and the EU constitute more than 50% of the AI market's spending.[65] Hence, they could more effectively push for their respective AI regulatory agendas if they cooperate effectively. On the other hand, there has been an increase in engagement with international standard setting efforts for AI from China and the US.[66]

---

The *De Facto Channel* of the de jure Brussels Effect is contingent on a de facto Brussels Effect. If this condition is fulfilled, one would expect multinational AI companies to lobby other jurisdictions to pass EU-like AI regulation, as the AI industry is relatively large and has an oligopolistic structure. For example, since the GDPR's passage, we have seen some big tech companies arguing that the US needs a federal equivalent. However, it is unclear how successful such lobbying efforts would be. While the De Facto Channel is common for US and Californian regulation,[67] it has only been demonstrated for a single EU regulation: the Eco-Management and Auditing Scheme (EMAS).[68]

Perhaps the most likely route by which this channel could lead to an effect on US federal AI regulation is if US states start adopting EU-like regulation, which in turn incentivises US companies to lobby the government to adopt similar regulation, as has been seen with a lot of environmental regulation in the US.[69]

The *Conditionality Channel* is currently implausible because EU AI legislation would likely not comprise a high degree of extraterritoriality, e.g. through equivalency clauses.[70]

### 1.2.3. Will There Be a Brussels Effect for the AI Act?

In this report, we suggest that there is likely to be a de facto Brussels Effect for parts of the AI Act. Prohibitions are generally unlikely to create a de facto Brussels Effect, as they aim to remove certain products from the EU market. However, there is some chance that the prohibitions on manipulation, e.g. subliminal techniques, could produce a Brussels Effect if recommendation algorithms used by social media companies risk being classified as manipulative or if such bans increase the reputational costs to offer EU-prohibited products abroad. Transparency obligations will likely only produce a very weak de facto effect as compliance might only require surface-level changes to the product for EU customers, such as adding a disclaimer at the start of a conversation with a chatbot. However, a de facto effect might occur if such disclaimers become seen as a signal of a high-quality, trustworthy product. For high-risk systems, the requirements that have low variable costs, that increase perceived product quality, and that may require early forking of systems are more likely to see a de facto Brussels Effect. Specifically, we are most likely to see a de facto effect with regard to products under certain existing product safety regulation, worker management systems (e.g. those used by the gig economy and logistics companies), potentially for general or foundational AI systems, and in remote biometric identification and categorisation systems and legal technology.

It is even more difficult to assess the likelihood of a de jure effect. We are unclear about the international impacts of the prohibitions and transparency obligations. However, we do think that perhaps the most important impact of the AI Act will be in the design of the conformity assessments, which may set the gold standard for regulation and standards in the EU and beyond. Our conclusions are summarised in Table 2 and explained in greater detail in sections 3 and 4.

---

67 See section 3.3 of this report. Birdsall and Wheeler discuss the de facto to de jure regulatory diffusion leading to de jure diffusion of US pollution standards to South American and other developing countries. Nancy Birdsall and David Wheeler, "Trade Policy and Industrial Pollution in Latin America: Where Are the Pollution Havens?," *Journal of Environment & Development* 2, no. 1 (January 1993): 137–49. Perkins and Neumayer find evidence for the hypothesis that the countries that have more transnational corporations and more imports are more likely to have stricter automobile emission standards. Richard Perkins and Eric Neumayer, "Does the 'California Effect' Operate across Borders? Trading- and Investing-up in Automobile Emission Standards," *Journal of European Public Policy* 19, no. 2 (March 1, 2012): 217–37. For other US environmental standards influencing non-US countries see: Elizabeth R. DeSombre, "The Experience of the Montreal Protocol: Particularly Remarkable, and Remarkably Particular," *UCLA Journal of Environmental Law and Policy* 19, no. 1 (2000). For Mexico and Brazil in particular see: Ronie Garcia-Johnson, *Exporting Environmentalism: U.S. Multinational Chemical Corporations in Brazil and Mexico, Global Environmental Accord: Strategies for Sustainability and Institutional Innovation* (MIT Press, 2000).

68 European Commission, "EMAS – Environment," June 14, 2016; Walter Mattli and Ngaire Woods, "In Whose Benefit? Explaining Regulatory Change in Global Politics," in The Politics of Global Regulation (Princeton University Press, 2009), 1–43. Also see the discussion in Bradford, "The Brussels Effect."

69 David Vogel, Trading Up: *Consumer and Environmental Regulation in a Global Economy* (Harvard University Press, 1995); David Vogel, California Greenin': *How the Golden State Became an Environmental Leader,* Princeton Studies in American Politics: Historical, International, and Comparative Perspectives (Princeton University Press, 2018).

70 Note that the AI Act exhibits some extraterritoriality. A producer would fall under the scope of the regulation if they were producing an AI system whose output is used in the EU, as would a user of an AI system whose output is used in the EU. Graham Greenleaf, "The 'Brussels Effect' of the EU's 'AI Act' on Data Privacy Outside Europe," *171 Privacy Laws & Business International Report 1*, June 7, 2021.

| | REQUIREMENT | DE FACTO | DE JURE |
|---|---|---|---|
| **PROHIBITIONS** | Manipulation | Perhaps, if e.g. recommendation algorithms are considered plausibly manipulative and if foundational adjustments to such are needed to avoid manipulative behaviour. | Unclear; depends on to what extent the EU exports its narrative. |
| | Social credit scores | Likely not, as the requirements apply primarily to governments. However, it could increase the reputational cost of offering such products in other jurisdictions. | Unclear; depends on to what extent the EU exports its narrative. |
| | Real-time biometrics | Likely not, as the requirements apply primarily to governments. However, it could increase the reputational cost of offering such products in other jurisdictions. | Unclear; depends on to what extent the EU exports its narrative. |
| **CONFORMITY** | Products already covered by some product safety rules, including medical devices, toys, and machinery | Likely to see a de facto effect as long as the requirements are new. | The requirements laid out in the regulation might have a de jure Brussels Effect. This could plausibly be the most important impact of the regulation. |
| | Largely regional or government uses of AI, including in critical infrastructure, education, the financial sector, and law enforcement | Likely not, as these uses are regionalised. Could change if the provision of these systems becomes globalised or if the EU requirements become seen as the gold standard. | |
| | Worker management, including hiring, firing, and task allocation | Plausibly, though it largely depends on the extent to which the EU requirements are perceived as a net quality benefit and how regionalised the industry is. | |
| | Remote biometric identification / categorisation systems and "legal tech" | Perhaps, if the EU's requirements become seen as the gold standard in these more contentious applications of AI. | |
| | General AI systems and foundation models, which could be used in high-risk applications | Likely for some, though it depends on e.g. how large the market for high-risk AI uses becomes, whether general purpose AI systems will be covered by the AI Act, and whether compliance requires early forking, e.g. differences in the systems' pre-training. | |
| | Transparency obligations | Likely for some, though strength will depend on the extent to which disclosures are seen as quality signals. | Unclear; depends on to what extent the EU exports its narrative and the extent to which California's Bot Disclosure Act is more causally responsible for the diffusion. |

Table 2: A summary of our conclusions on the likelihood of a Brussels Effect from various parts of the proposed AI Act. Deeper blues indicate that we think a Brussels Effect is likely.

## 1.3. What About China and the US?

The EU has successfully exported different regulatory standards to less geopolitically powerful jurisdictions in the past, including states in Africa, Oceania, Latin America, and Asia. These countries have less regulatory capacity and international bargaining power relative to the EU. However, to evaluate the impact of EU regulation on the global AI industry, it is essential to know whether EU regulation diffuses to the advanced and prosperous nations that dominate the AI industry and AI sales market, especially China and the United States.

The United States will most likely not adopt more stringent AI regulations than the EU (for more information, see section 2.2.). Since about the 1990s, US regulation on harms from business to citizens has become less stringent than EU regulation.[71] However, there could be a de jure effect spreading to US states such as California, which has already adopted GDPR-like regulation. Should a sufficient number of US states adopt EU-esque regulation, it could diffuse to the federal level via a de facto channel.

China may adopt AI regulation that is more stringent than that of the EU. Indeed, in early 2022 it adopted regulation with regard to recommender systems and proposed regulation for AI systems that generate content in early 2022. This regulation shares many features with the EU's Digital Services Act, Digital Markets Act, and the forthcoming AI Act but goes beyond the EU in some respects. Overall, we expect the Chinese Communist Party to regulate its technology sector more severely than the EU, but it will be unwilling to rein in government use of AI. We may still see a de jure Brussels Effect with regard to domains where the EU is regulating first or by Chinese regulators using the EU requirements for high-risk systems as a blueprint.

Until 2020, the US was the EU's biggest trading partner before being overtaken by China.[72] This means that many firms are operating in both the EU market and the US or Chinese market, making a de facto Brussels Effect possible.

Because of the significant EU-US trade in digital technology, with many multinationals serving both markets,[73] the probability of a de facto Brussels Effect for the United States increases. Historically, various EU legislative efforts have exhibited some measure of a de facto Brussels Effect on the US, including the Data Protection Directive (DPD), the GDPR, the chemical regulation REACH, toy safety standards, and the EU Code of Conduct regarding online hate speech.[74] However, the US has demonstrated that it can selectively resist or avoid EU regulation. For instance, the Safe Harbor Agreement helped the US avoid some GDPR requirements (for more information, see the appendix section 4.1).[75]

A de facto AI Brussels Effect reaching China is less likely since the dominant Chinese technology companies mostly do not operate outside China[76] and the ones that do tend to already have differentiated products.[77]

---

[71] Interestingly, from the 1960s to the 1990s, the US adopted more stringent regulation than the EU. Vogel, *The Politics of Precaution: Regulating Health, Safety, and Environmental Risks in Europe and the United States* pp. 4-6.

[72] Mario Damen, *"The European Union and Its Trade Partners,"* Fact Sheets on the European Union (European Parliament, September 2021). However, in the case of digital technology trade, the US and the EU are aiming to foster their bilateral trade. See the Tech Alliance. Global Times, *"China Replaces US to Become Largest Trade Partner of EU,"* December 4, 2020; European Commission, "EU-US Launch Trade and Technology Council to Lead Values-Based Global Digital Transformation."

[73] "In 2019, U.S. exports of information and communications technology (ICT) services to the EU was $31 billion, with potentially ICT-enabled services adding another $196 billion." Rachel F. Fefer, "EU Digital Policy and International Trade," R46732 (Congressional Research Service, March 25, 2021).

[74] For more, see Bradford, *The Brussels Effect: How the European Union Rules the World.*

[75] Engler also discusses the effect of the AI Act on the US. Engler, *"The EU AI Act Will Have Global Impact, but a Limited Brussels Effect."*

[76] Chinese exports to the EU consist more of physical products than software.

[77] Huawei and Lenovo operate outside China, and there are Western companies (like Apple) that operate in the Chinese market. Chinese ByteDance offers TikTok outside of China, with supposedly separated businesses and technology.

## 1.4. Analogues to EU AI Regulation

To assess a future AI Brussels Effect, it is useful to consider not only its determinants and dynamics but also relevant case studies. We do so in the appendix.

In addition to the AI Act, the upcoming EU AI regulation will comprise changes to the liability regime and the product safety regime. Many of the 29 pieces of EU sectoral product safety legislation have exhibited substantial de jure Brussels Effects, reaching Oceania, Africa, South America, and Asia, including China. A de facto Brussels Effect also reached the United States among many other countries. AI product safety standards and the general product safety regime share many characteristics. For instance, it is plausible that future EU AI regulation will impact the relevant ISO standards, as other EU product safety standards have in the past.

The regulatory diffusion of EU data protection legislation may also help understand the prospects of an AI Brussels Effect because data protection legislation regulates parts of AI development and deployment. The 1995 Data Protection Directive (DPD) experienced a significant de jure Brussels Effect,[78] though some authors attribute this diffusion of norms similar to those in the DPD to the Council of Europe's Convention 108, which preceded and influenced the DPD. The 2018 General Data Protection Regulation (GDPR) has shown a robust de facto Brussels Effect. For instance, 58% of popular websites offer US subjects both the right to erasure (GDPR Article 17) and the right to portability (GDPR Article 20).[79] In addition and despite its recentness, six countries, including Japan, Argentina, and New Zealand, have already adopted similar rules – a de jure Brussels Effect.[80] Regulatory diffusion of the DPD and GDPR may have benefited from extraterritorial demands and high costs for differentiation. For more details, see the appendix.

---

[78] For a more detailed discussion see section 3.1. As one reference, see: Graham Greenleaf, *"The Influence of European Data Privacy Standards Outside Europe: Implications for Globalization of Convention 108,"* International Data Privacy Law 2, no. 2 (April 4, 2012): 68–92.
[79] Jeremy Colvin, *"Unchecked Ambiguity and the Globalization of User Privacy Controls Under the GDPR,"* ed. Jonathan Mayer (Senior Theses, Princeton University, 2019).
[80] Greenleaf, *"The 'Brussels Effect' of the EU's 'AI Act' on Data Privacy Outside Europe."*

# 2. Determinants of the De Facto Brussels Effect

When new more stringent legislation is introduced in a jurisdiction, multinational actors are faced with two choices. First, they need to decide whether it is worth remaining in the market. Second, if they choose to stay in the market, they must decide whether to comply with the new regulation globally or offer two or more products, one compliant with the jurisdiction's requirements and at least one non-compliant version. With regard to EU regulation, if companies choose to stay in the market *and* sell EU-compliant products outside the EU, we have a de facto Brussels Effect, as illustrated in Figure 1.

Whether firms stay in the EU depends to a large degree on the market size after the relevant regulation taking effect, which depends on the EU market size before the regulation (§2.1), how compliance is likely to affect product quality and costs, and how much buyers and sellers are expected to react to accompanying price and product changes (§2.4).

If a company chooses to remain in the EU market, their next choice is whether to offer their EU-compliant product outside the EU or not. We consider the factors which make it likely profitable for companies to offer one EU-compliant product globally ("non-differentiation") rather than to differentiate their products into one EU-compliant product and at least one non-EU-compliant product ("differentiation"). In short, we assume non-differentiation will be chosen when it is deemed more profitable to sell EU-compliant products, rather than non-EU-compliant products, outside the EU.

More specifically, the following inequality must hold if a company is to choose non-differentiation, creating a de facto Brussels Effect:

| | Non-differentiation profits outside the EU | | Differentiation profits outside the EU |
|---|---|---|---|
| REVENUES | Revenue from selling EU compliant products outside the EU | ≥ | Revenue from selling non-EU-compliant products outside the EU |
| | − | | − |
| COSTS | Variable compliance cost of producing an EU-compliant product for non-EU markets | | Additional regulatory costs of non-EU-compliant products (fixed + variable costs) + Duplication costs associated with |

**Figure 2:** The conditions under which non-differentiation is more profitable and a de facto Brussels Effect would be produced.

In choosing non-differentiation, the company would not have to pay additional fixed compliance costs, as that cost has already been borne in choosing to stay in the EU market. On the other hand, it might see smaller revenues if the EU-compliant product is less desirable to non-EU customers (§2.4), and it will have to pay the variable compliance costs associated with offering an EU-compliant product outside the EU (§2.5.1).

In choosing differentiation, the company's profit outside the EU is equal to its revenue from selling the non-EU-compliant product minus the variable compliance cost in producing the non-EU-compliant product (§2.5.3) and the fixed compliance costs from complying with regulation in other jurisdictions. In addition, it may have to bear duplication costs associated with needing to maintain two separate production processes (§2.5.2). The company choosing differentiation may also need to pay additional verification costs in non-EU jurisdictions. Even though such costs would be borne in choosing non-differentiation,[81] they would if anything be lower in the non-differentiation case, as verification efforts (e.g. documentation) for the EU market could be more easily reused in other jurisdictions if the system remains the same.

As a simplification, we assume that if a product is EU-compliant, it is automatically compliant with all non-EU regulation. This would not be the case if other jurisdictions introduced regulation incompatible with EU rules, undermining de facto diffusion. Another simplification is that we do not consider in detail differences in verification costs outside the EU between non-differentiation and differentiation.

In this section, we discuss five factors which make a de facto Brussels Effect more likely. Roughly speaking, we discuss the role of three actors: regulators (§§2.2 and 2.3), the market and consumer behaviour (§§2.1 and 2.4), and the firms' production processes (§2.5).

Firstly, market properties (§2.1) such as market size, market concentration, and globalisation influence the chance of a de facto Brussels Effect. Some properties, such as the EU's relative market size, make it more likely that firms stay in the EU. The bigger the EU's relative and absolute market size, the more likely companies are to stay in the market. The more globalised the market structure, the more likely it is that firms offer products outside and within the EU, creating the preconditions for a de facto Brussels Effect. Further, the more oligopolistic the market structure (see §2.1.2), the more likely it is that companies choose non-differentiation, as they can coordinate their compliance strategies, e.g. by choosing to all offer non-differentiated products, thereby not being put at a disadvantage compared to their competitors.[82]

Secondly, a requirement for the de facto Brussels Effect is that EU regulation must be more stringent than that of other jurisdictions (§2.2). Higher stringency with regard to all regulatory dimensions across all other jurisdictions is not necessary, but without any higher stringency there cannot be de facto diffusion.

Thirdly, the more regulatory capacity (§2.3), such as regulatory expertise (see §2.3.1), is brought to bear on the design of the EU regulation, the more likely companies are to stay in the EU and the smaller the costs of non-differentiation.[83] This is because better-crafted regulation might decrease the cost of complying with EU regulation (particularly if it decreases variable compliance costs) while ensuring minimal (or positive) impacts on revenue within and outside the EU. Well-crafted regulation may also be required to ensure that EU-compliant products will be compatible with the laws of other jurisdictions.

---

[81] Unless the jurisdiction put in place e.g. unilateral recognition of CE marked products.

[82] The EU's Code of Conduct on countering illegal hate speech online could illustrate such oligopolistic coordination. The big tech companies, including Google and Facebook, implemented the new Code of Conduct worldwide. European Commission, *"The EU Code of Conduct on Countering Illegal Hate Speech Online: The Robust Response Provided by the European Union,"* accessed July 11, 2022; Bradford, *The Brussels Effect: How the European Union Rules the World,* chap. 6.

[83] Compliance cost is the cost of meeting the requirements, and verification cost refers to the cost of being able to verify and evidence that this is the case. We refer to the sum of these two costs as regulatory costs.

Otherwise, a Brussels Effect with regard to those jurisdictions would be undermined. We also discuss verification costs in section 2.5.3. Further, competent enforcement of the regulation might be necessary for a de facto Brussels Effect. Competent enforcement can reduce regulatory costs by reducing regulatory uncertainty and ensuring that firms are compliant. Without enforcement, firms could choose not to comply with the EU rules even within the EU, undermining the opportunity for de facto diffusion.

Fourthly, demand and supply, within and outside the EU, must be relatively inelastic (§2.4) in that the market size does not change much in response to a given change in regulatory costs or in product quality. Low elasticity within the EU in response to the new EU rules increases the chance of companies remaining in the EU, whereas low elasticity outside the EU increases the chance of non-differentiation. We discuss four determinants of inelasticity. First, if buyers prefer compliant companies and products, companies are more willing to pay the regulatory costs and are less responsive to regulation. Second, if EU buyers can, without great effort, move their consumption of AI products out of the EU, then the demand is more elastic. Third, the more substitutes or alternatives for a comparable price are available, the greater the likelihood that buyers substitute AI products with alternatives – increasing the elasticity of non-EU and EU demand. Fourth, there are supply-side effects, where firms might e.g. start taking longer to place their products on the EU market. Firms' investments into the EU market being inelastic increases the chance of a de facto Brussels Effect. For the EU, we are unlikely to see immediate effects on EU consumption of AI products – EU end consumers are unlikely to e.g. move their consumption of AI products out of the EU – but the increased regulatory burdens could decrease EU consumption over time via supply-side effects. Outside the EU, we argue, demand is likely inelastic in the regions and domains where EU compliance is seen as a quality signal or if EU norms have diffused.

Lastly, we consider the regulatory cost associated with applying the EU standards globally (in proportion to the market size (§2.1) and the existing production costs), i.e. the cost of non-differentiation, compared to that associated with producing non-EU compliant products, the cost of differentiation (§2.5). This section is focused on how the production process and costs change when the EU-compliant product is also sold outside the EU. We argue that some of the crucial factors determining non-differentiation production cost in the AI industry are (i) whether compliance requires early forking of an AI system – i.e. changes to the foundational parts of the system – which often results in higher duplication costs, (ii) the variable cost accrued by offering EU-compliant products globally, and (iii) the extent to which there is existing product differentiation (reducing the costs of differentiation).

This report contributes to the literature studying the drivers of regulatory diffusion. We break down Bradford's[84] second and fourth determinant of the de facto Brussels Effect, regulatory capacity and inelasticity, into four and three components respectively, and generalise both concepts to include further considerations. Our first determinant discusses favourable market properties, whereas Bradford only discusses one such market property: market size. Bradford describes the fifth component as "compliance indivisibility." We attempt to make this criterion more precise by having it refer to the difference in cost between non-differentiation and differentiation, and we offer a breakdown of these two costs. Different authors, such as Bradford , usually discuss two channels of the de jure Brussels Effect. We suggest four different channels, overall presenting a hopefully more comprehensive picture.

## 2.1. Favourable Market Properties

The AI industry's market properties are generally conducive to a de facto Brussels Effect: the EU AI market is large in both absolute and relative terms (at least 15% of the global AI market), and multinational companies dominate the global AI industry. However, many of the AI applications that will have the highest regulatory burdens imposed by the AI Act – high-risk AI systems – are in less globalised industries. For example, many of the high-risk uses of AI are in government services. Moreover, the EU AI relative market size may be reduced in the future if the AI Act proves very costly.

---

[84] Bradford, *The Brussels Effect: How the European Union Rules the World.*

### 2.1.1. Market Size

The larger the *absolute* EU market size, the more incentivised companies will be to stay in the market when new legislation is introduced. A firm might leave the EU market in response to stringent regulation, but if the absolute market size of the EU market is large, the foregone profits of leaving the EU market are also larger.

As the *relative size* of the EU market increases, the more likely companies are to sell EU-compliant rather than non-EU-compliant products outside the EU.[85] The profits of non-differentiation, i.e. of selling and producing EU-compliant products worldwide, increase with the absolute size of the market outside the EU. A bigger market outside the EU allows firms to absorb additional fixed costs associated with complying with non-EU rules or duplication of production processes in exchange for potential lower variable costs or higher consumption of non-EU-compliant products. Hence, the likelihood of the Brussels Effect increases with the absolute market size within the EU and decreases with the absolute market size outside the EU. In other words, the likelihood of the Brussels Effect will increase with the absolute and relative market size of the EU.[86]

What is the size of the EU AI market? There are no highly rigorous estimates of the EU's AI market size, as the industry is growing quickly and there are disagreements about what counts as AI and how much of current AI spending is on R&D. Therefore, we believe it best to use multiple methods to estimate it. We can start by looking at AI spending – investments made in developing and deploying AI – as a proxy of the EU AI market. The International Data Corporation estimates that European[87] AI spending was approximately 17 billion US dollars in 2021 and is projected to grow by 27% on average per year from 2022 to 2025.[88] Given estimates of global AI spending at $85 billion in 2021, the EU's share of global spending is around 20%. This might be an under-estimate if we expect the EU's share of AI spending to go up as the technology matures and if the US has a higher share of investment in development than of consumption of AI products. The EU Commission's AI Act Impact Assessment used another method: assuming that the EU AI market share is similar to its market share in software, they estimate the EU AI market at approximately 22% of the global AI market.[89] Another method would be to assume that the EU AI market will be at least proportional to its global GDP share. Hence, the relative EU AI market size may be at least 15% because this is the EU's share of global GDP in 2021.[90] Moreover, projections assume that Europe's position will not significantly change over the following years.[91] Taking these together, we believe the EU's AI market share is likely to be no lower than 15% of the global market. This is a sizable market, which may well produce pressures in favour of a de facto Brussels Effect.

---

[85] Chad Damro, *"Market Power Europe,"* Journal of European Public Policy 19, no. 5 (June 1, 2012): 682–99; Daniel W. Drezner, *"Globalization, Harmonization, and Competition: The Different Pathways to Policy Convergence,"* Journal of European Public Policy 12, no. 5 (October 1, 2005): 841–59; Vogel, *Trading Up: Consumer and Environmental Regulation in a Global Economy.*

[86] If there are only small-to-medium enterprises (SMEs) and 60% of their profits are in the EU, then the relative market size is large (the absolute size of a firm's customer base is small). Because of the small customer base for the firm, they might not be prepared to pay the fixed costs of regulatory adaptation and would instead focus on the customer base outside the EU.

[87] One should note that in the following we use data from the European continent which includes countries, such as Norway and Switzerland, that are not part of the EU.

[88] IDC, *"European Spending on Artificial Intelligence Will Reach $22 Billion in 2022, Supported by Strong Investments Across Banking and Manufacturing, Says IDC,"* IDC: The premier global market intelligence company, October 7, 2021.

[89] European Commission, *"Commission Staff Working Document Impact Assessment Accompanying the Proposal for a Regulation of the European Parliament and of the Council Laying Down Harmonised Rules on Artificial Intelligence (Artificial Intelligence Act) and Amending Certain Union Legislative Acts SWD/2021/84 Final."*

[90] IMF, *"European Union: Share in Global Gross Domestic Product Based on Purchasing-Power-Parity from 2017 to 2027,"* April 2022, Statista,

[91] IDC, *"Worldwide Artificial Intelligence Spending Guide,"* IDC: The premier global market intelligence company, accessed July 5, 2022.

Note that none of these estimates take into account that EU regulation could reduce or increase[92] the supply and demand in the EU market. We discuss these dynamics and the expected effect in section 2.4 on inelasticity.

The AI Act does not regulate all AI systems, however. What is the expected share of the global market of high-risk AI systems, as defined by the proposed AI Act? The Commission's impact assessment of the AI Act estimates that only 5–15% of AI systems on the EU market will be considered high-risk.[93] It seems likely that the largest market segments using high-risk AI systems will include AI systems used for recruitment, determining access to self-employment opportunities, and task allocation, likely affecting many gig economy companies; multiple uses in the financial services sector; and sectors already covered by some existing product safety regulation, such as the use of AI in medical devices, toys, and machinery.

### 2.1.2. Oligopolistic Competition and Multinational Companies

In addition to sufficiently high absolute and relative market size, the market must be adequately globalised and oligopolistic to produce a de facto effect.[94] Without companies straddling multiple jurisdictions, there is no possibility of a de facto Brussels Effect. If all companies produced and sold goods in a single country or region, no company would bring regulatory norms to other jurisdictions. This condition is exemplified by a comparison between European metal and chemical regulation. As the chemical market is highly globalised, EU regulation, such as REACH,[95] exhibited a strong de facto Brussels Effect. In contrast, the metal industry is predominantly regional. As almost no international firm could spread the EU blueprint to consumers elsewhere, the metal regulation did not exhibit a Brussels Effect.[96]

Multinational firms dominate the AI industry, making the AI market structure conducive to a de facto Brussels Effect. This interconnectedness is illustrated by the fact that foreign markets were strongly affected by the GDPR.[97] Based on a survey, PricewaterhouseCoopers estimated that 68% of American companies were expected to spend $1–10 million on GDPR compliance, and 9% of American companies would spend more than $10 million.[98] For more details, see the appendix section 4.1.

However, some of the industries and applications classed as high-risk AI systems fail to or only partly fulfil this criterion.[99] Many high-risk systems in Annex III of the proposed AI Act are likely to largely be deployed by EU governments — e.g. for border control, certain uses in education, public benefit allocation, law enforcement, management of critical infrastructure, and administration of justice[100] — who will prefer their AI systems be developed in the EU.[101] Financial services companies are also likely to have AI systems used "to evaluate the creditworthiness of natural persons or establish their credit score" covered by the AI Act.[102] However, due to the differences in national regulation, the financial services industry already sees significant regionalisation in the business-to-consumer market, with few companies providing credit checks internationally.

---

[92] Tatjana Evas, *"European Framework on Ethical Aspects of Artificial Intelligence, Robotics and Related Technologies: European Added Value Assessment: Study"* (European Parliamentary Research Service, 2020).

[93] European Commission, *"Commission Staff Working Document Impact Assessment Accompanying the Proposal for a Regulation of the European Parliament and of the Council Laying Down Harmonised Rules on Artificial Intelligence (Artificial Intelligence Act) and Amending Certain Union Legislative Acts SWD/2021/84 Final."*

[94] This condition has not been discussed in Bradford, *The Brussels Effect: How the European Union Rules the World*; Björkdahl et al., *Importing EU Norms Conceptual Framework and Empirical Findings*. Fini discusses that New Zealand industries are likely to adhere to the EU norms if they are exporting a significant part of the goods to the EU, as has happened in the New Zealand wine industry. Melissa Fini, *"The EU as Force to 'Do Good': The EU's Wider Influence on Environmental Matters,"* Australian and New Zealand Journal of European Studies 3, no. 1 (May 5, 2011).

[95] REACH: European Parliament, *"Regulation (EC) No 1907/2006 of the European Parliament and of the Council of 18 December 2006 Concerning the Registration, Evaluation, Authorisation and Restriction of Chemicals (REACH), Establishing a European Chemicals Agency, Amending Directive 1999/45/EC and Repealing Council Regulation (EEC) No 793/93 and Commission Regulation (EC) No 1488/94 as Well as Council Directive 76/769/EEC and Commission Directives 91/155/EEC, 93/67/EEC, 93/105/EC and 2000/21/EC,"* CELEX number: 32006R1907, Official Journal of the European Union L 396 49 (December 2006).

[96] Concerning REACH, the chemical regulation, see Hanson, *CE Marking, Product Standards and World Trade*; Bradford, *The Brussels Effect: How the European Union Rules the World*.

[97] Though, it is important to stress that data-processing activities are much more common and distinct from the usage of AI systems.

[98] He Li, Lu Yu, and Wu He, *"The Impact of GDPR on Global Technology Development,"* Journal of Global Information Technology Management 22, no. 1 (January 2, 2019): 1–6; PwC, *"Pulse Survey: US Companies Ramping Up General Data Protection Regulation (GDPR) Budgets,"* GDPR Series (PwC, 2017).

[99] See AI Act, annex II and III for a list of the high-risk AI applications.

[100] See Table 1 for more details.

Even if an industry is regionalised, there can still be a de facto Brussels Effect if the provision of AI products is globalised and if the regulation would affect that provision. This could for example be the case if regionalised industries relied heavily on foundation models provided by big multinational technology companies and those models need to be adjusted to meet the EU's requirements. Whether this will be the case depends partly on how general systems (for example large language models like OpenAI's GPT-3) that are adapted to a more specific domain are handled by the AI Act. The EU Council recently released a proposal where the responsibility to ensure conformity of a high-risk AI system would only go to the actor that deploys it in a high-risk domain, even if they use a general system to do so.[103] We discuss this more in section 2.6.

Other high-risk uses covered by the AI Act are more likely to be highly globalised. In particular, the AI Act classifies a range of products already covered by various product safety regulations – notably medical devices, toys, and machinery – as high-risk.[104] Some of these industries are highly globalised with a small number of multinational companies dominating the market.

Further, a more oligopolistic market is more likely to see a de facto Brussels Effect. Companies are more prepared to pay the fixed costs of regulatory compliance if they have larger EU revenues. Further, as an oligopolistic market includes fewer firms, the customer base of every single firm will be greater. In addition, companies in an oligopolistic market may find it easier to converge on the same compliance strategy, all of them choosing non-differentiation, and may face a greater need to maintain a positive reputation. Therefore, the more we can expect the AI industry to be dominated by big oligopolistic companies like IBM, Amazon, Google, Facebook, and Apple,[105] as well as companies in the medical devices industry, the more we can expect these firms to stay in the EU market and pay the regulatory costs. Whether this is the case will partly depend on the extent to which big technology companies will be the main developers and sellers of AI systems globally or if the market becomes less concentrated as it matures.[106]

**2.1.3. Territorial Scope**

A broader territorial scope of regulation, that is, a further jurisdictional reach of a regulation, makes the de facto Brussels Effect more likely because it effectively increases the size of the affected market. The territorial scope of a regulation is very broad if the regulation affects companies even though they are only selling, producing, or are registered outside the EU. A broad territorial scope effectively increases the global proportion of the

---

products to which the EU regulation applies, making it more likely that a company uses the EU regulation as its internal global policy. Illustrations of such a broad territorial scope are the General Data Protection Regulation (GDPR) and Data Protection Directive (DPD). These regulations apply to any organisation, institute, and website which interacts with European residents, offering goods or services or monitoring behaviour.[107] Similarly, EU Competition Law effectively rules extraterritorially.[108] All else being equal, extraterritoriality makes a de facto Brussels Effect more likely.

Because the proposed AI Act does not have a very broad territorial scope – in contrast to, for instance, data protection or competition legislation – the conditions for regulatory diffusion are not optimal. The EU AI Act includes some extraterritoriality,[109] although not to the same extent as the GDPR. Firms fall under the scope of the regulation if they are the users[110] or producers of an AI system whose output is used in the EU.[111] Exports from the EU are not covered. For high-risk uses of AI, an EU importer must make sure that the non-EU-produced product has gone through the required conformity assessment,[112] introducing a measure of extraterritoriality.[113] For example, a non-EU company providing a recruitment assessment tool for EU companies using machine learning falls under the EU AI Act. Similarly, a non-EU company offering an AI-based assessment of medical risks for an EU insurance company falls under Annex III(5) of the proposed EU AI Act.[114] Moreover, many past EU regulations have exhibited a de facto Brussels Effect with similar degrees of extraterritoriality as the proposed AI Act.

One example of this is the numerous product safety regulations under the New Legislative Framework, as analysed in the appendix (§4.3).

It is possible for other jurisdictions to increase the de facto territorial scope of the AI Act if compliance with the EU requirements allows access to their market. For example, New Zealand has incorporated the EU's CE mark in its national regulation, allowing CE-marked products onto the EU market without additional checks.[115]

## 2.2. Regulatory Stringency

A requirement for the de facto Brussels Effect is that EU regulation be more stringent than the regulation in other jurisdictions.[116] The forthcoming EU AI regulation will likely be more stringent than that of other large jurisdictions such as the US and potentially China.

Among the jurisdictions in which a multinational company operates, the one with the most stringent regulation is more likely to shape the company's global internal policy, if that regulation is compatible across jurisdictions.[117] EU regulation, however, does not have to be most stringent on all possible regulatory dimensions for a de facto Brussels Effect to occur. It must only have non-overlapping obligations.

The EU will likely create a more stringent regulatory regime for AI than the US will. EU public opinion and regulatory culture are significantly more prone to produce stringent risk regulation. This has not always been the case. The US had more stringent risk regulation between the 1960s and the 1990s, after which EU regulation started becoming more stringent.[118] David Vogel describes this pattern[119] and seeks to explain it. Firstly, he claims it stems from an increase in public demand for more stringent risk regulation in the EU and a decrease in the US, partially as a consequence of the success of regulation pursued in the 1960s to 1990s. For example, differences in public opinion regarding food safety and data privacy have driven laxer rules in the United States and stricter rules in the EU, which might also happen for AI regulation.[120] Secondly, risk regulation has become politically polarised in the US since the 1990s, while this has not occurred in the EU. Republican President Nixon created the US Environmental Protection Agency in 1970, while Donald Trump called for its abolition during his presidency, indicating an increased polarisation in environmental risk regulation.[121] Thirdly, Vogel suggests, while the US has adopted regulatory principles and approaches that make risk regulation less likely, requiring formal risk assessments based on claims with high levels of scientific certainty, the EU has done the opposite in e.g. enshrining the precautionary principle in the 1992 Maastricht Treaty.[122]

---

[107] GDPR, art. 3.
[108] Bradford, *The Brussels Effect: How the European Union Rules the World,* chap. 4.
[109] Greenleaf, *"The 'Brussels Effect' of the EU's 'AI Act' on Data Privacy Outside Europe,"* 3.
[110] A private non-professional activity cannot be a user according to the EU AI Act.
[111] *Greenleaf,* 3..
[112] AI Act, art. 26.
[113] *Greenleaf,* 3.
[114] Both examples are from Greenleaf, 4. For the second example, Greenleaf notes that even though this is not explicitly listed, it should fall under "essential private services and benefits".
[115] See §4.3 and W. John Hopkins and Henrietta S. McNeill, *"Exporting Hard Law Through Soft Norms: New Zealand's Reception of European Standards,"* in Importing EU Norms: Conceptual Framework and Empirical Findings, ed. Annika Björkdahl et al. (Cham: Springer International Publishing, 2015), chap. 8; Fini, *"The EU as Force to 'Do Good': The EU's Wider Influence on Environmental Matters."*
[116] Bradford, *The Brussels Effect: How the European Union Rules the World,* chap. 4.
[117] This is only true under the assumption that non-differentiation is profit-maximising.

EU citizens seem more favourably inclined towards regulation of AI technology than do their US counterparts. When asked in a 2019 poll whether companies like Google, Apple, Facebook, or Amazon have been sufficiently regulated by the EU in the past 5 years, 64% of respondents said big tech companies had been regulated insufficiently.[123] In a similar 2019 US Gallup poll, 48% of Americans favoured more regulation of big tech companies.[124] Given the EU's higher levels of existing regulatory burden for big tech, these results suggest preferences for significantly more regulation in the EU than in the US. Other survey data shows a less clear picture. For example, a 2019 survey of US public opinion found that 82% of respondents agreed with the statement that "Robots and artificial intelligence are technologies that require careful management", while a 2017 Eurobarometer survey found that 88% of EU respondents agreed with the statement.[125]

The United States regulatory discourse on AI differs from the European discourse in that it focuses less on product safety or fundamental rights, is more national security focused, and is expected to be less stringent.[126] In 2020, the Office of Management and Budget (OMB) published guidelines for federal agencies concerning AI regulation. While the OMB does not have the authority to propose new legislation, its framing and interest in AI governance are very different from those of the 2020 EU AI White Paper.[127] While the EU AI White Paper discusses competitiveness, trustworthiness, and safety, the OMB AI memorandum is framed around breaking down barriers to innovation and the adoption of AI. The memorandum states that "[a]gencies must avoid a precautionary approach that holds AI systems to such an impossibly high standard that society cannot enjoy their benefits."[128] Furthermore, digital companies, particu-larly Google, Amazon, Facebook, Apple, and Microsoft, are more influential in United States politics than in EU politics.[129] While some contest that US and EU policy do not differ significantly in their precaution across all policy domains,[130] the difference does seem significant with regard to product safety regulation.[131] We therefore expect less stringent AI regulation in the United States than in the EU.

---

[118] Though some contest this point. See e.g. James Hammit et al., *The Reality of Precaution, 1st Edition* (Routledge, 2010), chap. 15.

[119] Vogel, *The Politics of Precaution: Regulating Health, Safety, and Environmental Risks in Europe and the United States,* 4–6.

[120] Bradford, *The Brussels Effect: How the European Union Rules the World*, chap. 5.

[121] Arthur Neslen, *"Donald Trump 'Taking Steps to Abolish Environmental Protection Agency,'"* The Guardian, February 2, 2017

[122] Vogel, *The Politics of Precaution: Regulating Health, Safety, and Environmental Risks in Europe and the United States,* 34–36.

[123] Jean-Daniel Lévy and Pierre-Hadrien Bartoli, *"Copyrights & Tech Giants: What Are the Expectations in Europe?"* (harris interactive, February 2019).

[124] Lydia Saad, *"Americans Split on More Regulation of Big Tech,"* August 21, 2019

[125] Baobao Zhang and Allan Dafoe, "Artificial Intelligence: American Attitudes and Trends" (Centre for the Governance of AI, Future of Humanity Institute, University of Oxford, January 2019) sec. 2; Eurobarometer, *"Attitudes towards the Impact of Digitisation and Automation on Daily Life"* (European Commission, May 2017).

[126] This assessment relies, among others, on a comparison of EU AI Whitepaper and the Office of Management and Budget's AI draft memorandum and the respective submissions to the consultation process. The appearance of keywords such as safety, rights, trust or investment and their connotations differs between the two jurisdictions. European Commission, *"On Artificial Intelligence - A European Approach to Excellence and Trust COM/2020/65 Final,"* CELEX number: 52020DC0065, February 19, 2020; Russell T. Vought to Heads of Executive Departments and Agencies, *"Draft Memorandum for the Heads of Executive Departments and Agencies, Guidance for Regulation of Artificial Intelligence Applications,"* January 7, 2019; European Commission, *"White Paper on Artificial Intelligence - a European Approach,"* European Commission, accessed July 12, 2022; Regulations.gov, *"Draft Memorandum to the Heads of Executive Departments and Agencies, Guidance for Regulation of Artificial Intelligence Applications,"* Regulations.gov, accessed July 21, 2022.

[127] Russell T. Vought to Heads of Executive Departments and Agencies, *"Memorandum for the Heads of Executive Departments and Agencies, Guidance for Regulation of Artificial Intelligence Applications,"* November 17, 2020. And the Trump Administration criticised the EU for their potentially strict rules: David Shepardson, *"Trump Administration Seeks to Limit 'Overreach' of Regulation of Artificial Intelligence,"* Insurance Journal, January 8, 2020.

The situation with regard to AI-powered facial recognition is less clear but may nonetheless indicate differences in regulatory culture between the EU and other jurisdictions such as the United States. The proposed AI Act includes a ban on "real-time" biometric identification for law enforcement purposes with certain exceptions, such as particularly serious crimes.[132] Belgium has found facial recognition applications unlawful.[133] In the EU AI White Paper consultation, 55% of all citizens and 29% of civil society called for a ban of remote biometric identification systems in publicly accessible spaces. Out of all respondents, 77% responded that remote biometric systems should be banned (28%), only allowed conditional on certain requirements being met (29%), or only allowed in certain cases (20%), with 17% of respondents not expressing an opinion.[134] In part, there have been similar tendencies in the United States. The states of Oregon and New Hampshire have enacted bans on using facial recognition technologies in law enforcement body cameras. California introduced a three-year moratorium on the same uses in January 2020.[135] Further, the facial recognition debate has become charged by the Black Lives Matter protests of 2020. However, as of June 2020, 59% of Americans still favoured facial recognition technology for law enforcement.[136]

China, on the other hand, may adopt more stringent regulation than the EU, but would only be likely to do so for private sector uses of AI. For a more detailed assessment, see the discussion in section 2.3.4. The Chinese Communist Party (CCP) is unlikely to limit its ability to use AI technology for e.g. surveillance and censorship. Even if China adopted more stringent regulation than the EU, we should not necessarily expect a de facto "Beijing Effect". Firms may seek to avoid risks to their reputation from potentially being regarded as co-operating with autocracies. For instance, firms do not wish to be viewed as "complicit in state censorship in the most speech-restricting nation".[137] More importantly, as discussed in section 2.5, many globalised companies already offer different products in China than in the rest of the world.

## 2.3. Regulatory Capacity

The EU's generally high regulatory capacity – which includes expertise, coherence within and between relevant policy institutions, and sanctioning authority – increases the chance of discovery and sanctions of infractions and ensures regulation is well-crafted, though its capacity with regard to AI may be weaker.[138] Moreover, being the first jurisdiction to regulate a particular issue increases the chances for a de facto Brussels Effect; there is a first mover advantage. For AI, the EU will likely be the first large jurisdiction to comprehensively regulate the technology, it has sufficient sanctioning authority, and it has set up new AI policy bodies to gain more expertise. However, some argue the expertise that current regulators' have in AI may be limited.[139]

---

[128] Vought to Heads of Executive Departments and Agencies, *"Draft Memorandum for the Heads of Executive Departments and Agencies, Guidance for Regulation of Artificial Intelligence Applications,"* January 7, 2019.

[129] One indicator would be the difference in lobby spending. In the EU, the GAFAM companies report combined annual spending of around 22.5 million euros. In the US in 2020, GAFAM spent 63.53 million US dollars (56 million euros using 2020 exchange rate). For the EU: Transparency International EU, *"Integrity Watch - EU Lobbyists,"* Transparency International EU, accessed July 12, 2022. For the US: Senate Office of Public Records, *"Lobbying Expenses of Amazon in the United States from 2009 to 2020,"* 2021, Statista, ; Senate Office of Public Records, *"Lobbying Expenses of Apple in the United States from 2009 to 2020,"* January 2021, Statista; Senate Office of Public Records, *"Lobbying Expenses of Microsoft in the United States from 2009 to 2020,"* January 2021, Statista; Senate Office of Public Records, *"Lobbying Expenses of Alphabet Inc in the United States from 2015 to 2021,"* October 2021, Statista; Senate Office of Public Records, *"Lobbying Expenses of Facebook in the United States from 2009 to 2020,"* April 2021, Statista.

[130] Hammit et al., *The Reality of Precaution*.

[131] Vogel, *The Politics of Precaution: Regulating Health, Safety, and Environmental Risks in Europe and the United States.*

[132] However, others have argued that such a practice would already be incompatible with the GDPR from 2018. Veale and Borgesius, *"Demystifying the Draft EU Artificial Intelligence Act — Analysing the Good, the Bad, and the Unclear Elements of the Proposed Approach."* It is also noteworthy that this ban will still be changed by the European Parliament and the Council of the European Union.

[133] Nicolás Elena Sánchez, *"Pandemic Speeds Calls for Ban on Facial Recognition,"* EUobserver, May 18, 2021.

[134] European Commission, "Public Consultation on the AI White Paper: Final Report," November 2020, 11.

[135] Haley Samsel, *"California Becomes Third State to Ban Facial Recognition Software in Police Body Cameras,"* Security Today, October 10, 2019; Leufer and Lemoine, *"Europe's Approach to Artificial Intelligence: How AI Strategy Is Evolving."*

[136] Katharina Buchholz, *"Americans Accept Facial Recognition for Public Safety,"* Statista, June 10, 2020.

[137] Bradford, *The Brussels Effect: How the European Union Rules the World*, chap. 5.

### 2.3.1. Regulatory Expertise

Regulatory expertise means that relevant authorities have knowledge and resources relevant to the regulatory domain. Regulatory expertise often reduces compliance costs while still achieving the same regulatory aims, making regulation more effective.

Usually, the EU is regarded as having high regulatory expertise, though its expertise regarding AI is harder to judge. For instance, many EU civil servants have technical or economic PhDs.[140] Further, European regulatory agencies that enforce the EU product safety rules are led by experts.[141] For AI in particular, an assessment of the skills of policymakers and institutional expertise on the national and European level is complicated as the issue is relatively novel and there is no existing agency on the subject.[142] The Commission sought to address this by establishing technical expert groups, such as the High-level expert group on artificial intelligence[143] and the Expert Group on Liability and Emerging Technologies, while already having regulatory expertise on product safety testing.[144]

However, at the same time, the Commission has been accused of lacking an evidence-based AI policy plan.[145] In places, the Commission's AI Act draft seems to show a lack of understanding of AI technology. For instance, the act requires "[t]raining, validation and testing data sets [to] […] be relevant, representative, free of errors and complete."[146] On common sense interpretations of these requirements, it seems technically near impossible to ensure datasets are free of errors and complete.[147] However, it is worth noting that the recitals accompanying the Commission's proposal and the French presidency of the EU Council's proposal both include weaker, more achievable versions of the requirement.[148]

Lower regulatory expertise can increase the regulatory costs for the relevant industry and unnecessarily reduce product quality by disallowing too many practices. This may in turn lead to buyers substituting AI products with alternatives and otherwise reducing their consumption of AI products. In response, the EU market size (§2.1.1) would be reduced, making it less profitable to not differentiate the EU and non-EU products, as described in section 2.4. This reduces the likelihood of a de facto Brussels Effect.[149]

---

### 2.3.2. Regulatory Coherence

Regulatory coherence concerns the degree to which the demands of regulatory targets are clear and consistent.[150] The proposed EU regulation seems well set-up to ensure such coherence by (i) aiming to establish EU-level rules for AI, instead of going through a period with national governments adopting their own policies, and (ii) clearly identifying the relevant actors responsible for supervision and enforcement.

As a collective of 27 member states, the European Union at times has greater difficulty finding common solutions to regulatory problems compared to other jurisdictions such as China and the US. This can hinder a de facto Brussels Effect. Importantly, it can also undermine the free movement of goods within the EU and the EU single market, one of the union's core objectives. Thus, to achieve this goal, the EU has put significant effort into harmonising regulation since the 1990s.[151]

Coherence in aims and intentions is high for AI regulation. In April 2018, EU member states committed to a joint approach in a Declaration of Cooperation on Artificial Intelligence.[152] The draft EU AI Act published in April 2021 is a maximum harmonisation instrument,[153] meaning that once the act is passed, national law cannot exceed the EU-level rules.

Maximum harmonisation and the resulting coherence make a de facto Brussels Effect more likely.[154] For instance, food safety standards do not exhibit a Brussels Effect, partly because the rules differ between EU member states, effectively shrinking the EU market covered by the regulation and also because the EU is a preference outlier.[155] Even if AI regulation did not achieve maximal harmonisation and coherence in a first law, coherence can also develop over time. The case of data protection regulation is illustrative, where the first efforts (notably in Germany) were national, followed by an EU directive in 1995 (which states could implement in different ways). Fully harmonised EU-level regulation came with GDPR taking effect in 2018.[156]

Further, all else being equal, the Brussels Effect of AI regulation will be greater if specific and known regulatory bodies are clearly made responsible for the issue and for shaping and enforcing market rules. To do so, the Commission, in the EU AI Act proposed in April 2021, seeks to set up a European Artificial Intelligence Board.[157] New national market surveillance authorities (MSAs) will be set up and specifically tasked with enforcing the AI Act.[158]

### 2.3.3. Sanctioning Authority

The sanctioning authority of a regulator, such as the Commission, has two parts. First, it consists of creating laws with sufficient sanctioning clauses. Second, the Commission must have the legal institutions and resources to identify and sanction violations. The current AI Act proposal includes significant sanctioning powers, including the ability to levy heavy fines. It is less clear whether there will be sufficient resources to identify and sanction violations.

---

[150] Bach and Newman, *"The European Regulatory State and Global Public Policy: Micro-Institutions, Macro-Influence."*

[151] Bradford, *The Brussels Effect: How the European Union Rules the World*, 36–37; Dempsey et al., *"Transnational Digital Governance and Its Impact on Artificial Intelligence."*

[152] European Commission, *"EU Member States Sign up to Cooperate on Artificial Intelligence,"* Shaping Europe's digital future, April 10, 2018,

[153] In contrast to a minimum harmonisation instrument, a maximum harmonisation instrument prohibits member states from passing national law which exceeds the principles of the EU regulation. Veale and Borgesius, *"Demystifying the Draft EU Artificial Intelligence Act — Analysing the Good, the Bad, and the Unclear Elements of the Proposed Approach."*

[154] It is worth noting that this harmonisation likely means that some countries will implement less strict regulation than they would have if it were not for the EU-level rules.

[155] Alasdair R. Young, *"Europe as a Global Regulator? The Limits of EU Influence in International Food Safety Standards,"* Journal of European Public Policy 21, no. 6 (2014): 904–22, Björkdahl et al., *Importing EU Norms Conceptual Framework and Empirical Findings*, vol. 8, chap. 8ff. Three-quarters of Bradford's de facto Brussels Effect examples are regulations, the EU legislation that is directly implemented into national law, as opposed to directives for policy fields which are not fully harmonised. Bradford, *The Brussels Effect: How the European Union Rules the World*.

[156] For more, see appendix §4.1.

[157] Veale and Borgesius, *"Demystifying the Draft EU Artificial Intelligence Act — Analysing the Good, the Bad, and the Unclear Elements of the Proposed Approach."*

[158] One for each member state. They are expected to be staffed with between 1 and 25 FTEs. AI systems already covered by existing product safety regulation will continue to be covered by their current notified bodies and market surveillance authorities.

The EU tends to back up regulation with sufficient sanctioning authority and capacity. For instance, financial penalties in the case of a violation of EU competition law can amount to 10% of the company's annual turnover. The GDPR allows for fines of up to 4% of the company's annual turnover.[159]

What about sanctioning capacity? The enforcement of GDPR offers a helpful case study. In the first year of the GDPR after it came into force in May 2018, an estimated 91 fines were issued.[160] As of the end of 2021, there has been a total of 990 fines and penalties – a significant increase in fines per year.[161] Half a year after the GDPR's implementation, Google was fined 50 million euros;[162] in July 2021, the Luxembourg National Commission for Data Protection issued Amazon a 746 million euro fine;[163] and in December 2021, the French National Data Protection Commission issued total fines of 200 million euros to Google and its subsidiaries[164] as well as 60 million euros to Facebook.[165]

The budgets of the national Data Protection Commissions (DPCs), the enforcement agencies responsible for the GDPR, have increased since the regulation's introduction. The DPC in Dublin has significant responsibility for enforcing the GDPR for Amazon, Facebook, and Google.[166] In 2016, it had an annual budget of 9 million euros, which increased to 23 million in 2022, with another two million added per year since the introduction of the GDPR.[167] However, it is still criticised for being too slow, in particular with regard to cross-border cases. According to a 2021 Irish Council for Civil Liberties report, "[a]lmost all (98%) major GDPR cases referred to Ireland remain unresolved."[168]

Similar to the GDPR, the EU product safety framework has significant sanctioning capacity. Consider, for illustration, the case of toy safety standards. While the legal toy safety standards in the United States and Europe are similar, there have been ten times as many recalls of Chinese children's toys in the EU than in the United States.[169] Such sanctioning capacity at existing product safety enforcers is important to AI Act enforcement, as they will continue to be responsible for product safety even when products introduce AI systems.

The proposed AI Act allows for large penalties, which can be up to 6% of global turnover or 30 million euros – whichever is higher – for breaches of the Title II prohibitions of e.g. social scoring or of the Title III data quality requirements for high-risk systems. For other rules of the proposed AI Act, the maximums are lower: up to 20 million euros or 4% of global turnover (whichever is higher) for non-compliance with other obligations in the law and up to 10 million euros or 2% of global turnover (whichever is higher) for providing incorrect, incomplete, or misleading information to the relevant authorities.[170]

---

[159] European Commission, *"Fines for Breaking EU Competition Law,"* November 2011; *"GDPR: Fines / Penalties,"* GDPR, accessed July 12, 2022.

[160] Catherine Barrett, *"Emerging Trends from the First Year of EU GDPR Enforcement,"* Data, Spring 2020 16, no. 3 (2020): 22–25,

[161] CMS, *"GDPR Enforcement Tracker,"* accessed July 13, 2022.

[162] O. Tambou, *"France · Lessons from the First Post-GDPR Fines of the CNIL against Google LLC."* European Data Protection Law Review 5, no. 1 (2019): 80–84,

[163] Which they intend to defend themselves against. Amazon.com, Inc., *"Form 10-Q,"* Washington, D.C., June 30, 2021.

[164] CNIL, *"Cookies: la CNIL sanctionne GOOGLE à hauteur de 150 millions d'euros,"* CNIL, January 6, 2022; CNIL, *"The Sanctions Issued by the CNIL,"* CNIL, December 1, 2021.

[165] CNIL, *"Cookies: sanction de 60 millions d'euros à l'encontre de FACEBOOK IRELAND LIMITED,"* CNIL, January 6, 2022.

[166] This is primarily because their European headquarters are in Ireland. A fifth of all complaints referred between Data Protection Authorities are referred to the Irish DPC, more than for any other. Johnny Ryan and Alan Toner, "Europe's Enforcement Paralysis: ICCL's 2021 Report on the Enforcement Capacity of Data Protection Authorities" (ICCL, 2021).

[167] Data Protection Commission, *"Data Protection Commission Statement on Funding in 2021 Budget,"* Data Protection Commission, October 13, 2020; Barry O'Halloran, *"Data Protection Commission to Receive €2 Million Extra Funding,"* The Irish Times, October 13, 2020.

[168] The Report of the Irish Council for Civil Liberties conclude: "Almost all (98%) major GDPR cases referred to Ireland remain unresolved." Ryan and Toner, "Europe's Enforcement Paralysis: ICCL's 2021 Report on the Enforcement Capacity of Data Protection Authorities." See also this 2020 critique from Dr Eoin O'Dell: Irish Legal News, *"Data Protection Watchdog Continues to Suffer 'indefensible' Underfunding,"* Irish Legal News, octuber 14 2020

[169] Derek B. Larson and Sara R. Jordan, *"Playing It Safe: Toy Safety and Conformity Assessment in Europe and the United States,"* International Review of Administrative Sciences 85, no. 4 (December 1, 2019): 763–79.

[170] AI Act, art. 71.

If the infringer is a public body, penalties are chosen by the member states.[171]

However, there are reasons one might worry the AI Act will not be sufficiently strictly enforced. Except for biometric identification systems and those systems already covered by existing product safety regulation, every high-risk AI system can get the CE (European Conformity) label through internal self-assessments without involving external certifying bodies, causing some to worry that compliance will not be sufficiently high.[172] On the other hand, many CE-markings do not require input from external certification bodies – often called "notified bodies". The AI Act would simply put in place bodies charged with monitoring compliance across the industry.

The main enforcement bodies of the AI Act are the market surveillance authorities (MSA) in every member state, a common approach in EU product law.[173] MSAs are public bodies with wide-ranging powers to obtain information, apply penalties, withdraw products, and oblige intermediaries to cease offering certain products. It is compulsory for providers of high-risk AI systems to inform an MSA of new risks and malfunctions and for providers to inform the MSA of risks found in their post-marketing monitoring.[174] In the GDPR, users have a right to lodge a complaint, and not-for-profit bodies or associations can also do so on their behalf. That means that if one suspects that a data-processing company acted unlawfully and the company does not react, one can file a report to the MSAs, which then have to take further actions.[175] In contrast, while the proposed MSAs connected to the AI Act may receive complaints from citizens, the MSA is not required to investigate them, something which has drawn criticism from e.g. the Ada Lovelace Institute,[176] the Future of Life Institute,[177] and the Irish Council for Civil Liberties.[178] For the enforcement of the AI Act, the Commission estimates that 1–25 extra staffers will be hired per member state,[179] likely growing over time.[180] Some have argued that this number "is far too small."[181] It remains unclear whether the sanctioning capacity will prove sufficient.

To conclude, the AI Act includes high levels of sanctioning authority, similar to that of the GDPR, whereas its accompanying sanctioning capacity is more uncertain. It might be that infringements are much harder to detect for the AI Act than for the GDPR or that the MSAs will not be staffed with sufficient expertise. It could also be that the lack of ability for citizens to submit complaints to the MSAs will lead to insufficient enforcement. It is difficult to know before the legislation and details of accompanying sanctioning authorities have been finalised.

**2.3.4. First Mover Advantage**

A de facto Brussels Effect for EU AI regulation becomes more likely if the EU is the first jurisdiction to regulate this issue. This is firstly because it reduces the chance that other jurisdictions pass incompatible regulation. Secondly, if the EU is the first mover and another jurisdiction does pass EU-incompatible regulation – i.e. more stringent than the EU regulation in some respect – it is more likely that companies will continue to comply with EU regulation in all other jurisdictions because they have already borne the fixed costs for the EU regulation.

---

[171] AI Act, art. 71(7).
[172] Melissa Heikkilä, *"6 Key Battles Ahead for Europe's AI Law,"* POLITICO, April 21, 2021.
[173] Veale and Borgesius, *"Demystifying the Draft EU Artificial Intelligence Act — Analysing the Good, the Bad, and the Unclear Elements of the Proposed Approach."*
[174] AI Act, art. 62.
[175] See Veale and Borgesius.; *GDPR*, art. 77 and 80.
[176] Alexandru Circiumaru, *"Three Proposals to Strengthen the EU Artificial Intelligence Act,"* December 13, 2021.
[177] Future of Life Institute, *"FLI Position Paper on the EU AI Act"* (Future of Life Institute (FLI), August 6, 2021).
[178] Irish Council for Civil Liberties to European Commission DG CNECT A, *"Flaws in Ex-Post Enforcement in the AI Act,"* February 15, 2022.
[179] Note that the larger data protection authorities have hundreds of staff. European Commission, *"Commission Staff Working Document Impact Assessment Accompanying the Proposal for a Regulation of the European Parliament and of the Council Laying Down Harmonised Rules on Artificial Intelligence (Artificial Intelligence Act) and Amending Certain Union Legislative Acts SWD/2021/84 Final"*, annex III, p. 25; European Data Protection Board, "First Overview on the Implementation of the GDPR and the Roles and Means of the National Supervisory Authorities" (EDPB, February 26, 2019).
[180] AI Act, page 14 mentions that the capacities of the notified bodies have to "be ramped up over time".
[181] Irish Council for Civil Liberties to European Commission DG CNECT A, *"Flaws in Ex-Post Enforcement in the AI Act,"* February 15, 2022.

The EU seems likely to be the first major jurisdiction to pass comprehensive AI regulation. However, given the slow pace of EU regulatory processes and delays in negotiations of the AI Act,[182] we may see smaller jurisdictions[183] adopt comprehensive AI regulation before the EU does, and we are already seeing large jurisdictions adopting regulation for parts of the AI ecosystem. In September 2021, Brazil's lower parliamentary house, the Chamber of Deputies, agreed on a proposed law outlining how AI would be regulated and the role of existing regulators.[184] In April 2022, the Brazilian Senate tasked a commission with proposing a bill on AI regulation, taking into account e.g. the bill proposed by the lower house.[185] Similarly, China has put in place regulation of recommender systems, and it proposed regulation for content-generation systems in early 2022.[186]

However, the EU may still benefit from a first mover advantage via its having published e.g. the AI Act draft and various documents leading up to its drafting.[187] In doing so, other jurisdictions are e.g. more likely to ensure their regulation remains compatible with the EU approach.

## 2.4. Inelasticity within and outside the EU

Demand and supply within and outside the EU must be relatively inelastic; that is, the decrease in the AI product market for any given increase in compliance costs or decrease in product quality as a result of new EU regulation on AI must be small. Positive elasticity, where e.g. demand increases in response to the regulation, would contribute even more to a de facto effect. However, we use the term "inelasticity" for convenience and because negative elasticity seems more plausible. In section 1.1.2, we discussed the regulatory costs of EU regulation for firms, which could be up to 17% of the investment in high-risk AI systems. The higher the elasticity — that is, the more substantially consumption changes for a given increase in regulatory cost or reduction in product quality — the (i) more EU AI spending goes down and firms are less likely to invest in the EU, and (ii) the less likely firms decide to sell EU-compliant products abroad, as the revenue from selling EU-compliant products outside the EU would be smaller.

We discuss four components of inelasticity. First, if buyers have a preference for compliance and compliant products, they are more willing to pay the compliance costs. End consumers could be more trusting of a regulated AI market and AI products that bear a CE mark. This seems likely, though it is possible that the EU-compliant products will be seen as lower quality outside the EU, e.g. if certain functionality is or is assumed to be lacking.[188] Second, if EU buyers can, without great effort, move their consumption of AI products out of the EU, then the demand is more elastic. Third, the more substitutes or alternatives are available for a comparable price, the greater the likelihood that buyers substitute AI products with alternatives — increasing the elasticity of non-EU and EU demand. Fourth, firms' investment decisions being inelastic further increases the chance of de facto diffusion. We argue that the elasticity of firms in response to new EU AI regulation is higher for a more competitive market and for smaller firms.

Within the EU, we tentatively conclude that immediate effects on EU consumption of AI products are unlikely – EU end consumers are unlikely to e.g. move their consumption of AI products out of the EU – but that the increased regulatory burdens could decrease EU consumption over time via supply-side effects. As innovation and adoption of some AI technologies become slower and more costly, thus increasing barriers to entry, suppliers and developers may delay or avoid introducing products in the EU, e.g. choosing to roll out their AI products in other markets first. These higher barriers to entry are likely to differentially affect smaller actors. The size of these effects will crucially depend on the regulatory cost, which is difficult to estimate before the legislation has been finalised.[189]

---

[182] Engler, *"The EU AI Act Will Have Global Impact, but a Limited Brussels Effect."*
[183] As, say, measured by GDP.
[184] Agência Câmara de Notícias, *"Câmara aprova projeto que regulamenta uso da inteligência artificial,"* Portal da Câmara dos Deputados, September 9, 2021. English translation here.
[185] Agência Senado, *"Brasil poderá ter marco regulatório para a inteligência artificial,"* Senado Federal, March 3, 2022. English translation here.
[186] Ding, *"ChinAI #168: Around the Horn (edition 6)"*; Ding, *"ChinAI #182: China's Regulations on Recommendation Algorithms."*
[187] Such as the AI White Paper and results from the HLEG. European Commission, *"White Paper on Artificial Intelligence - a European Approach"*; European Commission, *"High-Level Expert Group on Artificial Intelligence."*
[188] This could happen if, for example, there is a trade-off between a system's accuracy and its interpretability or if the introduction of human oversight into a product makes it slower.

Outside the EU, we conclude that demand is likely inelastic in the regions in which EU compliance is seen as a quality signal, e.g. if EU norms diffuse to other markets.

### 2.4.1. Preferences for Compliant Products

AI regulation could increase consumption of AI by increasing trust in products and the market, increasing legal certainty, and increasing EU harmonisation. Indeed, the EU Commission seems to rely on this being the case, arguing in its preamble to the draft AI Act that "as a result of higher demand due to higher trust, more available offers due to legal certainty, and the absence of obstacles to cross-border movement of AI systems, the single market for AI will likely flourish."[190] Other jurisdictions have made similar statements on the importance of trust. For example, the White House Office for Management and Budget has stated in guidance that "the continued adoption and acceptance of AI will depend significantly on public trust."[191]

What is the connection between product safety regulation, trust, and the size of the AI market? In short, when consumers struggle to judge the quality of products, product safety regulation can increase a market and/or move it closer to a more socially optimal level of product safety. In such markets, sellers will be incentivised to compete on metrics that consumers can perceive, such as price, in turn potentially leading to consumers losing trust in the market[192] or to providers of more quality goods leaving the market, thereby reducing their consumption.[193] This seems particularly the case for "credence goods", where a consumer is unaware of the quality, including its safety, of a good even after having consumed it, but it also applies in cases where judging the quality of a product would take a lot of effort.[194] It seems likely that some AI systems are credence goods, especially when considering lack of discrimination an aspect of quality. This dynamic can be reverted if consumers are provided with some way to identify product quality. Product safety regulation is one such way,[195] though it can also be addressed by industry-led standards, reputations,[196] and, potentially, consumer rating systems.

There are two other mechanisms by which the market might grow: the AI Act increasing legal certainty and it providing a harmonised market. The EU AI market is not unregulated. Existing regulations apply when using AI systems e.g. for human resources functions. However, it might not always be clear how those regulations apply.

---

[189] Some parts of the AI Act are impracticable or very difficult to achieve. If they remain in the final text, the compliance costs may be significant.

[190] AI Act, preamble 3.3: European Commission, *"Commission Staff Working Document Impact Assessment Accompanying the Proposal for a Regulation of the European Parliament and of the Council Laying Down Harmonised Rules on Artificial Intelligence (Artificial Intelligence Act) and Amending Certain Union Legislative Acts SWD/2021/84 Final."*

[191] Vought to Heads of Executive Departments and Agencies, *"Memorandum for the Heads of Executive Departments and Agencies, Guidance for Regulation of Artificial Intelligence Applications,"* November 17, 2020.

[192] This dynamic is explored e.g. in *OECD, Food Safety and Quality: Trade Considerations* (Paris Cedex, France: Organization for Economic Co-operation and Development (OECD), 1999), 37–39.

[193] This is the classic "Lemons Problem" as discussed in George A. Akerlof, *"The Market for 'Lemons': Quality Uncertainty and the Market Mechanism,"* The Quarterly Journal of Economics 84, no. 3 (August 1, 1970): 488–500.

[194] This distinction was first used in Phillip Nelson, *"Information and Consumer Behavior,"* The Journal of Political Economy 78, no. 2 (1970): 311–29,

[195] See Stephan Marette, Jean-Christophe Bureau, and Estelle Gozlan, *"Product Safety Provision and Consumers' Information,"* Australian Economic Papers 39, no. 4 (December 2000): 426–41; OECD, *Food Safety and Quality: Trade Considerations.*

[196] Marette, Jean-Christophe Bureau, and Gozlan, *"Product Safety Provision and Consumers' Information."*

In putting in place new rules, the EU hopes to increase legal certainty on how AI systems can be used. Further, the AI Act is an attempt to put in place one set of EU-level harmonised rules before national governments implement their own potentially incompatible AI regulation. Thus, the AI Act could significantly reduce the cost of operating in the EU, compared to the counterfactual situation where companies would need to comply with up to different 27 national regulations.

Outside the EU, buyers might consume more of the EU-compliant product (including paying more for it) if they perceive it to have more safety-enhancing features or otherwise believe it to be more trustworthy, increasing the revenue of non-differentiation. The perceived quality of EU-compliant products outside the EU varies widely. For instance, the EU's CE mark serves as a signal of product quality in Australia and New Zealand.[197] At the same time, consumers in other regions may be unwilling to accept the higher price or loss in product features associated with the CE mark's regulatory burden and compliance costs.

Similarly, customers outside the EU might have a preference for companies that comply with the EU rules in their jurisdictions. Customers may criticise companies who choose differentiation for complying with one standard for EU customers and another for other customers. For instance, Nestlé has been criticised for producing and selling more-hazardous products in some developing countries.[198] The AI industry may be particularly vulnerable to criticisms of this kind since the media has a large appetite for criticising the AI companies' business practices, as illustrated by the "techlash of big tech".[199] Moreover, various examples demonstrate the motivation of AI workers at major technology companies to engage in internal corporate activism, which increases those companies' potential to have their reputation harmed by offering products of different standards.[200]

The extent to which trust and legal certainty will be increased by the EU's forthcoming AI regulation remains to be seen and will depend largely on the result of ongoing legislative processes and how the requirements in the AI Act are made more concrete in standardisation efforts.

### 2.4.2. Ability to Leave the Market

If buyers can easily move their consumption of AI products outside of the EU AI regulation's jurisdiction, that would significantly increase the elasticity of the EU market in response to that regulation, thereby reducing the EU AI market and decreasing the chance companies offer their AI products on the EU market.

End consumers are unlikely to move out of the EU in response to new AI regulation. For instance, if a restaurant's price increases in the EU, residents do not move out of the EU to enjoy lower restaurant prices. A regulation's scope, including whether that regulation applies to EU imports, influences the inelasticity of buyers. If imports were out of scope, buyers would find it practically costless to substitute the regulated product. However, this is not the case in the proposed AI Act, in line with product safety regulation practices.

In contrast to end consumers, companies that act as buyers in a B2B exchange might be more willing to leave the EU market. For instance, if EU regulation makes a financial derivative more expensive to buy, a hedge fund may not be prepared to pay a higher price for a financial derivative in the EU and may in-stead move its assets out of the EU market. Similarly, some businesses might be incentivised to move their operations out of the EU if that allows them to avoid potentially onerous obligations imposed by the AI Act. This may be possible in some cases. For example, manufacturing companies may do so, as the AI Act concerns the use of machinery, but not end products of a manufacturing process that do not include AI components. Manufacturing companies may experience costs from the AI Act in their potential use of worker management systems and in the use of machinery (the AI Act terms machinery as a high-risk use of AI and calls for existing product safety requirements for machinery to be made consistent with the AI Act).[201] However, moving such operations is likely to be very costly and only justified by very large compliance costs. Perhaps, therefore, added costs to manufacturing processes are more likely to affect decisions to invest in new manufacturing facilities, rather than causing existing facilities to move.

---

[197] Hopkins and McNeill, *"Exporting Hard Law Through Soft Norms: New Zealand's Reception of European Standards"*; Fini, *"The EU as Force to 'Do Good': The EU's Wider Influence on Environmental Matters."*

[198] Nestlé has been criticised by EU consumers and consumer organisations because they do not follow specific guidelines in their factories in other producing countries, such as the Philippines, even though the goods produced in the factories are not sold on the EU market. Bradford, *The Brussels Effect: How the European Union Rules the World*, 36–37.

[199] We thank Shin-Shin Hua for this point. Darrell M. West, *"Techlash Continues to Batter Technology Sector,"* Brookings, April 2, 2021.

[200] See Newton regarding US tech companies' workers protesting e.g. cooperation with the military. Casey Newton, *"Google's Internal Activism Is Spreading across Tech Companies,"* August 14, 2019.

### 2.4.3. Substitutability

The available substitutes within and outside the EU are likely to be different. Within the EU, the substitute for AI products covered by the AI Act will likely be non-AI-based products or solutions, including human labour. If the substitutability of AI products is high within the EU – if it is easy to find comparable products or solutions at a comparable price – the chance of a de facto Brussels Effect is reduced, as EU customers will likely opt for alternatives, reducing the EU market size.

Over time, if AI systems continue to improve and become deeply embedded in business processes, we should expect it to become increasingly difficult to substitute them with e.g. human labour. Over time, it will become increasingly worth making the investment in AI systems. Even now, it is hard to believe that AI systems such as recommender systems in news feeds or content platforms could be effectively replaced with non-AI systems. Thus, we do not expect substitutability to have a large impact on the chances of a de facto effect.

The availability of substitutes for AI systems could reduce the speed or change the direction of innovation as AI systems are incentivised to meet certain requirements and as the cost of producing AI systems for the EU market increases. For example, some have argued that, due to higher taxes on labour than capital investments, the current US tax code incentivises investments in automation replacing human labour beyond what is socially optimal.[202] Further, some argue that incentives should be introduced to promote the development of AI systems that complement rather than displace human labour.[203] We are not sure how the speed of innovation is likely to be affected. We can further suggest that the direction of innovation will change: the AI Act will produce incentives to increase the performance and lower the production cost of AI systems compliant with the EU rules.

Outside the EU, substitutability is likely to be significantly higher: EU-compliant systems outside the EU will be competing with non-EU-compliant products. Thus, we should expect the extent to which EU-compliant products are bought outside the EU to be significantly more sensitive to the changes in price and quality brought about by compliance with EU rules. The extent to which compliance with EU rules makes a product better or more expensive is therefore crucial to firms' decisions of whether to offer EU-compliant products outside the EU.

It is unclear how the substitutability of EU-compliant systems outside the EU will change over time. The difference in price and performance could decrease over time as investment in developing the regulatory technologies to ensure compliance with the requirements ramps up. On the other hand, it could be that the AI Act's requirements become more onerous over time, e.g. if there are trade-offs between a model's accuracy and features required by the AI Act, if ensuring human oversight becomes harder with increasing speed and complexity of AI systems, or if the regulation fails to keep up with technological developments.

---

[201] AI Act, annexes II and III.
[202] Daron Acemoglu, Andrea Manera, and Pascual Restrepo, *"Does the US Tax Code Favor Automation?,"* Working Paper Series 27052 (National Bureau of Economic Research, April 2020).
[203] Daron Acemoglu, *"Harms of AI,"* Working Paper Series 29247 (Cambridge, MA: National Bureau of Economic Research, September 2021); Anton Korinek and Joseph E. Stiglitz, *"Steering Technological Progress,"* February 2021.

### 2.4.4. Supply-Side Elasticity

Another important factor is the elasticity of the firms supplying AI products. They might change their behaviour in response to the EU regulation or in response to changes in demand. The higher the demand elasticity and the higher the regulatory cost, the lower the profitability of supplying products to the EU market. This might mean that firms, investors, and entrepreneurs move their scarce resources (e.g. capital, human resources) out of the EU market or delay investment in the EU market. For example, they might first develop a product for the non-EU market and only then choose to take on the added compliance costs required to expand into the EU market. This dynamic would reduce absolute and relative EU AI spending – weakening the de facto Brussels Effect (see §2.1.1 for a discussion). In sum, if the sellers' investments respond substantively to regulatory costs, a de facto Brussels Effect is less likely.

We can start by looking at how profitability in the AI industry could be affected by the AI Act. Estimates of the profit margin in the AI industry are difficult and diverse. This might be because profitability among AI firms differs drastically and many AI and technology companies incur losses for several years, even when they are already public.[204] Some venture capitalists estimate that the profit margin of the average AI company is between 50 and 60%.[205] Regulation which increases the costs by 10% could lead to a profit margin of 40–50%. Thus, one might expect investors and entrepreneurs to deploy their scarce resources in other markets if those markets can garner higher returns. Competition among AI firms and investors would dampen this effect: the higher returns outside the EU would attract more capital, driving down its value and ability to gain such high returns.

Given EU consumers' difficulty of moving their AI consumption out of the EU and the potential difficulties in finding substitutes for AI products, companies may be able to raise prices to keep profit margins at a similar level. This could be possible provided the competition on the market is not too high.

Furthermore, the AI Act is likely to increase barriers to entry for the EU AI market, which might mean that the profits of a company that succeeds in the EU are more secure.[206] This could mean that large companies, with significant compliance divisions already well set-up to react to new regulation, will not reduce their investments in the EU market, while small and medium enterprises do. This could in turn reduce the innovativeness of the EU AI market over time. The AI Act does include measures, such as regulatory sandboxes,[207] to reduce burdens on smaller actors, but it is unclear if they will be sufficient.

The GDPR provides weak, inconclusive evidence on whether innovation and SMEs will be stifled. A study based on interviews with German start-ups whose products or business models centre on personal data does not find conclusive evidence as to whether the GDPR has increased or stifled innovation.[208] Others report stifled innovation, as the GDPR advantages large companies reducing competitiveness by increasing barriers to entry.[209]

---

[204] Jeffrey Funk, *"AI and Economic Productivity: Expect Evolution, Not Revolution,"* IEEE Spectrum, December 5, 2019.
[205] Martin Casado and Matt Bornstein, *"The New Business of AI (and How It's Different From Traditional Software),"* Future, February 16, 2020.
[206] This has been discussed, for example, in a classic essay by Michael Porter: Michael E. Porter, *"How Competitive Forces Shape Strategy,"* Harvard Business Review, March 1, 1979.
[207] AI Act, art. 53.
[208] Nicholas Martin et al., *"How Data Protection Regulation Affects Startup Innovation,"* Information Systems Frontiers 21, no. 6 (December 1, 2019): 1307–24,

There could be even larger effects on EU innovativeness in AI if research and development would be directly affected by the AI Act, regardless of whether a product has been deployed on the market. If the regulation would also affect AI R&D, then we would expect a bigger supply-side response to the regulation, as research would likely move out of the EU, reducing the amount of AI talent in the region and weakening innovation clusters.

To conclude, the response to the AI Act among buyers and sellers could be sufficiently inelastic to undergird a de facto Brussels Effect. The inelasticity is contingent on the preferences for compliant products. Non-EU consumers are unresponsive to regulation if customers perceive EU-compliant products to be higher quality and requirements don't make a product less attractive to customers, for example if complying with them produces an inferior product in some way. If buyers outside the EU are less willing to pay for EU-compliant products, firms could be discouraged from selling EU-compliant products outside the EU as it would decrease the revenue from non-differentiation.

## 2.5. Costs of Differentiation

The next crucial determinant of whether there will be a de facto Brussels Effect is the cost of differentiation and how it differs from that of non-differentiation. The higher the relative cost of choosing differentiation, the greater the chance of a de facto effect. As illustrated in Figure 2 above, in choosing non-differentiation, firms avoid paying additional fixed regulatory costs as that cost has already been borne in choosing to stay in the EU market. They also avoid potential duplication costs and might face lower verification costs outside the EU. On the other hand, they will have to pay the variable compliance costs associated with offering an EU-compliant product outside the EU.

Before exploring the costs of non-differentiation versus differentiation in more detail, it is useful to note that whether they choose non-differentiation depends on earlier factors. Firstly, the smaller the non-EU's absolute market size (§2.1.1), the smaller the EU variable compliance cost of non-differentiation compared to the fixed costs involved in differentiation. The more oligopolistic the market structure (§2.1.2), the more likely it is that companies can coordinate their compliance strategies, e.g. choosing to all offer non-differentiated products, meaning they are not put at a disadvantage compared to their competitors. The EU's Code of Conduct on Countering Illegal Hate Speech Online illustrates such oligopolistic coordination.[210] The big tech companies, including Google and Facebook, implemented the Code of Conduct worldwide.[211]

We divide our discussion of the relative cost of differentiation into four sections. We consider (i) the additional cost associated with applying the EU standards globally, (ii) the duplication costs and effects of early forking, (iii) the non-EU compliance costs associated with differentiation, and (iv) the extent to which there is existing product differentiation.

### 2.5.1. Variable Costs of Non-Differentiation

A company choosing to offer an EU-compliant product globally would already have incurred the related fixed costs ensuring EU compliance, but incurs the additional costs associated with offering this product globally. If those costs are low — i.e. it is cheap to ensure all of one's products are EU-compliant once compliance for the products sold in the EU has been secured — a de facto effect is more likely.

There are some reasons to think that these variable costs are relatively small and that the fixed costs will be an important factor. One of

---

the most important features of digital products is that they have high fixed development costs but small variable distribution costs. This, many economists argue, is part of why we might expect to see digital markets and the AI industry as winner-take-most markets.[212] In addition, there are trends towards the increasing capital expenditure required to develop frontier AI models: since around 2010, the amount of computational resources required to train machine learning models that advance the state of the art has doubled approximately every 6 months.[213] GPT-3, a state-of-the-art large language model developed in 2020, is believed to have cost around 4.6 million USD to train.[214] Further, some of the AI systems classed as high-risk in the AI Act are in industries with large upfront capital investment in product development, such as those used in medical devices (discussed further in §2.6.2).

On the other hand, the development of AI systems largely consisting of fixed costs does not necessarily mean the same holds true for compliance with EU regulation. For example, to comply with various regulations and demands from its users, social media companies are increasingly investing in content moderation, employing large numbers of content moderators. One 2021 report suggested that Facebook had between 15,000 and 35,000 content moderators.[215] As long as these content moderation tasks are not possible to automate, we should expect moderator numbers to increase almost proportionally with the size of the customer base.[216]

Concretely, some of the requirements imposed by the AI Act may produce variable compliance costs. This could be the case for requirements that there is human oversight over the system, in addition to risk management and post-market monitoring. Some costs related to these requirements would likely already have been incurred in producing an EU-compliant product for the EU market. For example, the company would already have put in the work of integrating their risk management and post-market monitoring systems into their other business practices, e.g. updating how decisions about product launches are made. In addition, ensuring human oversight might require designing user interfaces for the overseers. It could also require retraining or adjusting of the underlying AI systems such that their outputs are more interpretable to meet the requirement that the human overseer can "fully understand the capacities and limitations of the high-risk AI system."[217] However, these requirements also likely involve some variable costs, as companies would likely need to hire additional staff for risk management, post-monitoring, and human oversight should they adopt these requirements globally rather than only in the EU. The extent of these costs is a crucial factor in whether EU-compliant products will be offered outside the EU, as well as which requirements are more likely to be complied with outside the EU.

### 2.5.2. Duplication Costs and Early Forking

Companies' decisions of whether to offer EU-compliant products outside the EU will largely depend on how fundamental the changes needed to comply with the regulations will be. The more fundamental the changes – the earlier the "fork" in the system – and the costlier it is for the company to maintain two separate products, the more likely they are to choose non-differentiation. In short, early forking often implies high duplication costs which incentivise companies to offer one product globally once they have developed an EU-compliant product.

One can think of the production process of an AI system as starting in the design phase, wherein a company or a researcher decides what AI system they are going to produce.

---

[212] Varian, *"Artificial Intelligence, Economics, and Industrial Organization."*
[213] Jaime Sevilla et al., *"Compute Trends Across Three Eras of Machine Learning,"* arXiv [cs.LG] (February 11, 2022), arXiv
[214] Chuan Li, *"OpenAI's GPT-3 Language Model: A Technical Overview,"* Lambda, June 3, 2020
[215] Billy Perrigo, *"'I Sold My Soul.' WhatsApp Content Moderators Review the Worst Material on the Internet. Now They're Alleging Pay Discrimination,"* Time, Originally published: July 15 2021.
[216] Though there are likely some economies of scale as e.g. Facebook has more resources to develop efficient processes and the like.
[217] AI Act, art. 14 §4a.

Next, data is selected, collected, or generated, which the model is subsequently trained on. In some cases, such as in self-playing reinforcement learning systems or GANs,[218] training and data generation happen simultaneously. In some cases, the model will subsequently be fine-tuned or otherwise adapted to a specific use. After some testing and evaluation, the model may be deployed. Customers will often engage with the system via an API or some other user interface. Once the system has been deployed, its performance might be regularly evaluated and reviewed.[219]

Depending on the exact details of the regulation and the nature of the industry, compliance can be achieved by separating – forking – the system at different stages in the process. Some requirements and systems may require early changes in the process. Requirements that an AI system is robust to external threats or new unseen scenarios (e.g. distributional shifts) could require training an entirely new system on new data or using more robust algorithms. Requirements that high-risk AI models are neither biased nor discriminatory (AI Act Recital (44) and Art. 15(3)[220]) could be fulfilled in different ways. For example, high-risk products are required to use less biased and more representative datasets with an aim to reduce the resulting system's bias.[221] Meeting such a requirement would require early forking, in the data-collection process, perhaps requiring systems to be retrained if their original training data did not meet the AI Act's requirements. In contrast, if AI companies could meet requirements by e.g. fine-tuning models or otherwise adjusting them after they have been trained, the duplication costs could be much lower.

In some cases, producers can maintain two separate products cheaply, e.g., by turning off a feature or by making superficial changes to the system. This is particularly common when the change can be made via adjustments at the top of the "technology stack." For instance, Tesla reduced the functionalities of their autopilot for the EU market via a software update in order to comply with a revision of driver assistance systems regulations in 2018 while leaving their cars in other jurisdictions unchanged.[222] Similarly, the AI Act proposes requirements that people be informed when they are engaging with e.g. a chatbot. This requirement could likely be met with a superficial change – by a late forking of the system – by changing the user interface, e.g. by adding a prominent statement saying that an AI system is providing the outputs or by starting any interaction by the chatbot introducing itself as such.

There are several reasons why early forking may produce duplication costs. A core reason is that it may substantially weaken the economies of scale for a product. AI companies usually have large economies of scale.[223] As more people use an AI product, more data becomes available, improving the company's product. So-called foundation models,[224] such as BERT,[225] GPT-3,[226] CLIP,[227] and Gopher,[228] are large deep learning models which can be used in a very wide range of systems – sometimes because they can perform a wide range of tasks and other times because the task they can do is useful in a large number of systems. Once the foundation model is trained, it can be used in many downstream models or specific applications, for example after some fine-tuning. After training, the cost of bringing it to more customers is comparatively small. Such potential variable costs might be computing costs, customer service, and sales. Hence, if complying with EU regulation requires changes to the training and modelling process of the foundation model,[229] differentiation could come with substantial duplication costs. There are also economies of scale regarding computational power and talent. Producing an additional product unit tends to become cheaper the more units are already produced. Therefore, differentiating a product into a compliant and a non-compliant product may lead to higher production costs of differentiation, especially if the forking happens early on, since the firm's production process loses some economies of scale.

---

[218] A generative adversarial network (GAN) is a machine learning framework in which two neural nets contest with each other as a learning process, where e.g. one system attempts to create an image indistinguishable from a photo and another tries to distinguish between the photo and the generated image.
[219] In the real world, many of these steps are not as neat as described. They may happen in tandem, companies might skip steps, or go back a step. Furthermore, models are often updated after deployment as new training data is found or generated.
[220] AI Act.
[221] AI Act, art. 10.
[222] Fred Lambert, *"Tesla Nerfs Autopilot in Europe due to New Regulations,"* May 17, 2019.
[223] Varian, *"Artificial Intelligence, Economics, and Industrial Organization"*; Avi Goldfarb and Daniel Trefler, *"Artificial Intelligence and International Trade,"* in The Economics of Artificial Intelligence: An Agenda, ed. Ajay Agrawal, Joshua Gans, and Avi Goldfarb (University of Chicago Press, 2019), 463–92.
[224] Rishi Bommasani et al., *"On the Opportunities and Risks of Foundation Models,"* arXiv (2021).
[225] Jacob Devlin et al., *"BERT: Pre-Training of Deep Bidirectional Transformers for Language Understanding,"* arXiv [cs.CL] (October 11, 2018), arXiv.
[226] Tom Brown et al., *"Language Models Are Few-Shot Learners,"* in Advances in Neural Information Processing Systems 33 (NeurIPS 2020) (34th Conference on Neural Information Processing Systems, Curran Associates, Inc., 2020), 1877–1901.
[227] Alec Radford et al., *"Learning Transferable Visual Models From Natural Language,"* in Proceedings of the 38th International Conference on Machine Learning, ed. Meila Marina And Tong, vol. 139, Proceedings of Machine Learning Research (PMLR, 2021), 8748–63.
[228] Jack W. Rae et al., *"Scaling Language Models: Methods, Analysis & Insights from Training Gopher,"* arXiv [cs.CL] (December 8, 2021), arXiv.

Engler argues that this dynamic means platforms whose algorithms are considered high-risk (e.g. LinkedIn's algorithms for placing job advertisements and job candidate recommendations) are particularly likely to choose non-differentiation.[230] This seems true insofar as foundational changes are required to the system, which seems to be the case for many requirements. However, certain requirements could be met via shallow changes to the product that could just be implemented in the EU. Such requirements could include having sufficient human oversight or those that can be met via fine-tuning or filtering a model. We discuss these dynamics further in section 2.6.2.2.

### 2.5.3. Non-EU Compliance Costs of Differentiation

If a company chooses to differentiate – offering non-EU-compliant products outside the EU – they incur some other additional costs. Firstly, they'll need to be able to identify what customers should be offered which product. Secondly, they'll need to comply with the regulation of the other jurisdictions.

Unlike companies only offering one EU-compliant product worldwide, companies choosing to differentiate their products need to identify which products are available to which customers – the companies incur an additional *identification cost.* The identification cost consists of determining what jurisdiction applies to the transaction by e.g. checking the customer's IP address or asking the customer to state where they are based. Such identification costs depend not only on how costly it is to get to a certain level of accuracy in identification but also on the cost of misidentifying a customer. Suppose enforcement is stringent and likely, and hence, the cost of falsely identifying an EU consumer as a non-EU consumer is high. In that case, a company finds it optimal to pay for a higher accuracy in identification, which increases the costs of differentiation. In sum, the identification cost will largely depend on the details of the final AI legislation and how liability is distributed among customers, distributors, and producers.

Overall, we expect identification costs to be low and mostly fixed, not requiring that companies do much more than make a good faith effort to check whether EU law applies to a particular transaction. For example, we expect companies to have fulfilled their duty if they e.g. only offer their product on an EU app store or to EU IP addresses. However, it remains to be seen whether this will be the case.

A company choosing non-differentiation would also need to ensure compliance with the requirements of other jurisdictions. We expect the compliance costs of other jurisdictions to be lower than that of the EU, as the EU seems likely to impose some of the most stringent requirements, at least at the time they come into force, and because other jurisdictions are likely to ensure a reasonably high level of compatibility with EU regulation so as to not disadvantage their firms' trading with the EU.

Firms may in particular experience additional verification costs in choosing to differentiate their products. In deploying a different product outside the EU, they would be less able to re-use documentation and other assets used to ensure EU compliance than if they had chosen non-differentiation. This is one reason jurisdictions with significant trade with the EU may be incentivised to establish unilateral recognition schemes with the EU, allowing CE-marked products on their markets without additional regulatory approval or inspection.

---

[229] Such potential regulatory responses are discussed in chapter 5, especially 5.4 of Bommasani et al., *"On the Opportunities and Risks of Foundation Models."*
[230] Engler discusses the case of LinkedIn. Engler, *"The EU AI Act Will Have Global Impact, but a Limited Brussels Effect."*

### 2.5.4. Existing Product Differentiation

If a company has already differentiated products between two different markets, they are more likely to continue down that path in response to the new EU regulation.[231] For illustration, suppose EU AI regulation would require companies to use a specific quality management system (QMS),[232] parts of which differ from the industry's current practices. If a company does not already have differentiated products between EU and non-EU markets, then the new stricter requirements from the EU market have them face the choice of whether to upgrade their global QMS or choose to have two separate ones. If the company has already differentiated their products, on the other hand, and already has two separate QMSs, then the choice is between upgrading both systems or just one system, presumably incurring a larger fixed cost for compliance. Relatedly, this means that industries where there is more product churn, i.e. products are replaced more often, are more likely to see companies choose non-differentiation.

Further, if a company already has differentiated their products, that indicates that non-differentiation is particularly costly or that differentiation is necessary for their product. For example, if two jurisdictions have sufficiently dissimilar non-overlapping legal requirements, differentiating one's products might become a necessity. This seems particularly common in the financial industry where companies already spend significant resources adapting their products to different jurisdictions' legal requirements.

Sometimes, legal requirements can effectively enforce some amount of differentiation, making a de facto Brussels Effect less likely. Data localisation laws, also called data residency laws, are an example of this. They require that data about a nation's citizen or resident must be processed and/or stored inside the country. China, India, and Indonesia have such laws. A dozen others have discussed or implemented them.[233] If the EU or member states adopt such data localisation laws, which some considered in 2013,[234] it would diminish the attractiveness of product non-differentiation as parts of the processes for non-EU data and EU data must be separated anyway.

Similarly, requirements that AI systems used in the EU are trained on EU data could undermine a de facto effect. For example, in the proposed AI Act, high-risk systems are required to "take into account, to the extent required by the intended purpose, … the specific geographical … setting in which the high-risk system is intended to be used."[235] This requirement could undermine a de facto Brussels Effect, if interpreted sufficiently strictly, e.g. such that supervised learning systems deployed in the EU must be trained solely on EU data. This would particularly be the case if other jurisdictions implement similar requirements or if companies are reluctant to offer a product trained solely on EU data in other jurisdictions. Less strict interpretations of the requirement, allowing e.g. fine-tuning of the system on EU data, would have a smaller effect.

---

[231] *Engler*.
[232] AI Act, art. 17.
[233] Anupam Chander and Uyên P. Lê, *"Data Nationalism,"* Emory Law Journal 64, no. 3 (2015): 677
[234] Data localisation discussions also happened in the EU. In 2013, both France and Germany considered such rules. *Chander and Lê, 690ff*.
[235] AI Act, art. 10 §4.

## 2.6. Likelihood of a De Facto Brussels Effect for Different Industries and Regulatory Requirements

The above sections suggest that the AI industry as a whole may have many of the features that make a de facto Brussels Effect more likely. However, what holds for AI in general might not hold for the specific industries and AI systems that the EU AI regulation will apply to. Though it is difficult to make predictions on how these will interact – especially before the legislation has been finalised – this section offers some tentative predictions. The most common reasons we find that a Brussels Effect might not occur is if (i) the industry or compliance within it is already regionalised (as discussed in §2.1 and parts of §2.5), (ii) compliance with the requirements does not require early forking (as discussed in §2.5), (iii) the additional cost of compliance with EU regulation abroad is large even once compliance for EU products has already been secured, and (iv) compliance with EU regulation does not increase perceived product quality outside the EU, making up for the additional compliance costs.

We focus on the AI Act and updates to the product liability regime. However, both the DSA and the DMA could substantially influence the AI industry, and we encourage other researchers to investigate their likelihood of de facto diffusion.

This section proceeds by looking at particular requirements that will be introduced, what industries and systems these requirements may affect, and discussing whether the factors described above make a de facto Brussels Effect likely or not. Section 2.6.1 explores the chance of a de facto effect in the realm of limited-risk systems. Section 2.6.2 focuses on high-risk systems, which will receive the most detailed discussion. Section 2.6.3 discusses prohibitions of certain limited-risk systems. Finally, section 2.6.4 concerns updates to the EU liability regime.

### 2.6.1. Transparency Obligations for Some Lower-Risk AI Systems

The Commission's proposed AI Act includes provisions requiring deployers to inform users if their system (i) interacts with humans and (iia) is used to detect emotions or determine association with (social) categories based on biometric data, or (iib) generates or manipulates content, e.g. deep fakes or chatbots.[236] As the costs of differentiation as well as the regulatory costs for such systems are likely to be low, we argue that a de facto Brussels Effect is plausible insofar as norms shift such that customers come to see disclosure as a sign of a company or product being trustworthy.[237]

The differentiation costs and the compliance costs associated with these transparency requirements are likely low. A company using chatbots on their website can comply with this requirement by adding a small text box telling the customer they are engaging with an AI system or starting the conversation with the chatbot identifying itself as such. The differentiation costs are similarly low. Companies could identify whether a user is covered by EU law or not via their IP address and make a slight change to the user interface, e.g. by adding a disclosure note.

The revenue from non-differentiation depends on the preferences of non-EU consumers. A disclosed chatbot might be less effective in providing customer service and satisfaction. At the same time, norms around the disclosure of AI systems could provide a reputational boost from non-differentiation, by e.g. notifying non-EU customers if they are engaging with a chatbot. Consumers might be distrustful of a company that has a reputation for not disclosing AI systems. In this case, non-EU consumers would be unresponsive (§2.4) because they prefer disclosure — increasing the revenue from non-differentiation and making a de facto Brussels Effect more likely. Another factor is the extent to which non-EU customers would punish actors for holding a "double standard", having different transparency policies in the EU and elsewhere.

---

[236] However, Veale and Borgesius criticise the transparency obligation for category (iii) as being unenforceable. How can a market surveillance authority find the undisclosed deep fakes? Veale and Borgesius, *"Demystifying the Draft EU Artificial Intelligence Act — Analysing the Good, the Bad, and the Unclear Elements of the Proposed Approach."*. See also AI Act, Title IV.

[237] This is in contrast to Engler, who thinks it is more likely. Engler, *"The EU AI Act Will Have Global Impact, but a Limited Brussels Effect."*

We could get a sense of the chance of a Brussels Effect of such transparency obligations by considering California's 2018 Bot Disclosure Act.[238] This law requires some companies interacting with Californian consumers, including importers to California, to highlight when a user interacts with a bot. "Any public-facing internet website, web application, or digital application" with more than ten million unique monthly American visitors, i.e., the 80 most popular websites,[239] must disclose bots.[240] Some commentators expected the Bot Disclosure Act to exhibit a California Effect.[241] Unfortunately, to date, no impact assessment or similar evaluation has been published to evaluate the California Effect of the BOT Act.

**2.6.2. Conformity Assessments for High-Risk AI Systems**

What parts of the EU's regulation of high-risk AI regulation are most likely to see a de facto Brussels Effect? To answer this question, we first look at what high-risk uses of AI (including in which industries) and what requirements from the draft AI Act are most prone to seeing de facto diffusion. For a recap of what systems are classified as high-risk and what requirements are imposed on them, please refer back to section 1.1.2 and Table 1.

**2.6.2.1. What High-Risk Uses of AI Are Most Likely to See a De Facto Effect?**

We believe that we're most likely to see de facto regulatory diffusion in the use of AI in the following domains: (i) many of the products already covered by existing product safety regulation under the New Legislative Approach, most notably medical devices; (ii) worker management, including hiring, firing, and task allocation; (iii) some general AI systems or foundation models used across a wide range of uses and industries; and (iv) less confidently, the use of AI in the legal sector and the use of biometric identification and categorisation of natural persons.[242] We argue that most other uses considered high-risk in the proposed AI Act will not see a strong de facto Brussels Effect, as the market structure or the product differentiation is already regionalised (see §§2.1.2 and 2.5). This is partly because many of the uses deemed high-risk in the AI Act are government uses of AI.

The majority of the high-risk uses of AI outlined in the AI Act's Annex III concern government uses of AI, which naturally pushes in favour of a regional market structure. These uses include management and operation of certain critical infrastructure (e.g. road traffic); admission and grading within educational settings; decisions regarding granting or revoking access to public benefits; various uses by law enforcement; uses in migration, asylum, and border control management; and AI systems to assist judicial authorities (e.g. courts) in their work.[243]

---

[238] California Senate, *"An Act to Add Chapter 6 (commencing with Section 17940) to Part 3 of Division 7 of the Business and Professions Code, Relating to Bots,"* Pub. L. No. 1001, CHAPTER 892 (2018), http://bcn.cl/2b6q3. It is also known as the BOT ("Bolstering Online Transparency") Act or California Senate Bill 1001.
[239] Quantcast, *"Audience Measurement & Analytics Platform,"* Quantcast (Quantcast Inc, August 30, 2020).
[240] California Senate, *An act to add Chapter 6 (commencing with Section 17940) to Part 3 of Division 7 of the Business and Professions Code*, relating to bots.
[241] CITRIS Policy Lab, *"Fair, Reliable, and Safe: California Can Lead the Way on AI Policy to Ensure Benefits for All,"* Medium, May 28, 2019. The regulatory diffusion of California legislation is often compared to the EU's regulatory diffusion. Historical examples for a "California Effect" include data privacy, food safety, and vehicle regulation. Bradford, *The Brussels Effect: How the European Union Rules the World*. California is frequently among the earliest US states to adopt new legislation to strengthen democratic ideals, consumer rights, and individual freedom or rights.
[242] The first two of these four groups are also discussed in Engler, *"The EU AI Act Will Have Global Impact, but a Limited Brussels Effect."*
[243] AI Act, annex III.

Beyond these government uses of AI, additional high-risk uses outlined in Annex III appear to either have a regional market structure or already have products differentiated along regional lines. One such example is the management and operation of critical infrastructure such as the supply of water, gas, heating, and electricity.[244] Though the market in these industries tends to be slightly more globalised earlier in the supply chain (e.g. in producing and trading electricity), the markets for supplying these commodities tend to be fairly regionalised, making a de facto Brussels Effect significantly less likely.

High-risk uses of AI in the financial sector also seem unlikely to us to see a de facto Brussels Effect, due to a regionalised market structure and having products differentiated along regional lines. Though the financial sector as a whole is reasonably globalised, especially with regard to financial services for corporations and with regard to investment, the particular financial-sector uses picked out by the AI Act see less globalisation: assessments of creditworthiness or credit scores of natural persons.[245] Such assessments tend to be carried out by national or regional companies, partly because of differences in rules and regulations between jurisdictions.

There is still some chance that we will see a de facto effect in these regionalised domains. The most plausible mechanism by which this would happen is if the *provision* of AI systems for these uses is globalised — that is, if e.g. governments procure systems for the high-risk uses and those vendors are global actors — and/or if compliance with EU requirements becomes seen as a quality signal. The latter could end up being the case, especially since these uses of AI (e.g. the use of AI for admission decisions), are likely to be controversial. Whether this effect ends up strong enough to produce a de facto effect remains to be seen.

Moving on to the uses of AI we think are more likely to see a de facto Brussels Effect, many of the high-risk uses of AI in domains already covered by other product safety regulations appear likely to see a de facto Brussels Effect. Such products (listed in the AI Act's Annex II) notably include medical devices and in vitro diagnostic medical devices, but also another ten domains already covered by product safety regulation including machinery, personal protective equipment, radio equipment, and toys.[246]

Medical devices seem likely to see a de facto Brussels Effect if new requirements are introduced, as medical device companies tend to produce one product for the global market and are unlikely to leave the EU market. The medical device industry is large and dominated by US- and EU-based companies. The EU being one of the regions with the highest consumption of medical devices,[247] it seems unlikely that companies will exit the EU market. Further, companies tend to offer one product globally, seeking to ensure it is compliant with both EU and US requirements (such compliance will as a rule allow the product to enter many other markets too).[248] This is partly because EU and US requirements tend to be the most strict[249] and because product differentiation tends to be costly. A large part of the development cost for medical devices is running studies to prove their safety and efficacy,[250] and so any changes that would require re-running those studies would likely cause huge increases in costs. As such, if the AI Act introduces new requirements, it seems likely that companies will attempt to remain compliant with the EU regulation globally.

---

[244] AI Act, annex III, §2.
[245] AI Act, annex III, §4b.
[246] *Engler.*
[247] Observatory of Economic Complexity (OEC), *"Medical Instruments,"* OEC, accessed july, 09 2022.
[248] Industry & Analysis, "2016 Top Markets Report Medical Devices: A Market Assessment Tool for U.S. Exporters" ( International Trade Administration, U.S. Department of Commerce, May 2016).
[249] Christa Altenstetter and Govin Permanand, *"EU Regulation of Medical Devices and Pharmaceuticals in Comparative Perspective,"* The Review of Policy Research 24, no. 5 (September 2007): 385–405; EMERGO, *"Europe Medical Devices Regulation (MDR) CE Marking Regulatory Process,"* EMERGO, August 23, 2017
[250] Aylin Sertkaya, Amber Jessup, and Rebecca DeVries, "Cost of Developing a Therapeutic Complex Medical Device for the U.S. Market," 2019.

A deeper investigation into whether de facto diffusion is likely for machinery covered by existing EU product safety regulation appears interesting, as AI systems are particularly likely to have a large impact in manufacturing and because the consumption of machinery plausibly has higher responsiveness than other goods covered by existing product safety regulation. Much machinery will be purchased and used by companies engaged in manufacturing, who could move their operations to countries with less stringent requirements on autonomous manufacturing equipment should the AI Act prove too onerous.

There are a number of other industries also covered by EU-wide product safety regulation under the so-called "Old Approach", including automotive and aviation. Though these product safety regulations are not directly affected by the proposed AI Act – the AI Act specifically says that only Article 84, which concerns the Commission's duties to evaluate the AI Act's effects and implementation, will apply to Old Approach product safety regulation – the recitals of the AI Act state that the ex ante requirements for high-risk systems "will have to be taken into account when adopting relevant implementing or delegated legislation under those acts". Aviation and automotive typically involve large fixed development and production costs, incentivising companies to produce one product for the global market, as illustrated e.g. in Vogel's early study of the California Effect.[251] As such, if the AI Act's requirements for high-risk systems end up being applied to Old Approach product safety regulation, it seems likely that it would produce a de facto effect.

What about general AI systems or "foundation models"[252] that are used across a wide range of applications? Examples of such systems include general purpose visual recognition systems that could be used for a wide range of tasks such as identifying whether a video includes a certain branded product. There are four routes by which these systems end up complying with the requirements for high-risk systems, potentially creating a Brussels Effect. Providers of general AI systems may (i) have legal requirements imposed on them in the AI Act, (ii) have contractual duties to ensure compliance with some requirements, (iii) see reputational benefits from compliance, or (iv) want to directly apply the system to high-risk domains, therefore incurring the relevant duties.

Whether providers of general purpose AI systems will have requirements imposed on them by the AI Act is a matter of ongoing discussions between the EU Council, Parliament, and Commission. While the originally proposed AI Act did not include any language about general purpose systems, proposed updates to the act do. In November 2021, the EU Council's Slovenian presidency proposed amendments to the AI Act in a compromise text, according to which general purpose AI systems would not be considered high-risk unless they are explicitly intended for high-risk uses.[253] In contrast, a May 2022 compromise text by the subsequent French presidency of the Council would impose duties on general purpose AI systems that may be used by other actors in high-risk applications.[254] Such systems would need to comply with a subset of the requirements for high-risk systems, concerning e.g. risk management, data and data governance, post-market monitoring, accuracy, and robustness.[255] The pro-posal includes exemptions for small and medium enterprises as well as general systems where the provider has explicitly excluded any high-risk uses in the instructions of use for the system.[256]

---

[251] See e.g. Vogel, *Trading Up: Consumer and Environmental Regulation in a Global Economy*.

[252] Bommasani et al., *"On the Opportunities and Risks of Foundation Models."*

[253] European Parliament, "Regulation (EC) No 1907/2006 of the European Parliament and of the Council of 18 December 2006 Concerning the Registration, Evaluation, Authorisation and Restriction of Chemicals (REACH), Establishing a European Chemicals Agency, Amending Directive 1999/45/EC and Repealing Council Regulation (EEC) No 793/93 and Commission Regulation (EC) No 1488/94 as Well as Council Directive 76/769/EEC and Commission Directives 91/155/EEC, 93/67/EEC, 93/105/EC and 2000/21/EC" article 52a.

[254] La Présidence Française du Conseil de l'Union européenne, "Proposition de Règlement Du Parlement Européen et Du Conseil établissant Des Règles Harmonisées Concernant L'intelligence Artificielle (législation Sur L'intelligence Artificielle) et Modifiant Certains Actes Législatifs de l'Union - Texte de Compromis de La Présidence - Article 3, Paragraphe 1 Ter, Articles 4 Bis à 4 Quater, Annexe VI (3) et (4), Considérant 12 Bis Bis," May 13, 2022.

[255] Specifically, they would comply with requirements from the AI Act regarding having a risk management system (Art. 9), data and data governance (Art. 10), technical documentation (Art. 11), providing users with explicit instructions for use (Art. 13(2) and (13)(3)(a) to (e)), and requirements surrounding accuracy, robustness, and cybersecurity (Art. 15). They would also need to comply with additional requirements regarding e.g. providing their customers with information needed for their compliance, conducting a lighter conformity assessment, and conducting post-market monitoring. *La Présidence Française du Conseil de l'Union européenne* Art. 4b.

It seems plausible that there would be a Brussels Effect for some general purpose systems offered by large corporates if something similar to the French presidency's proposal becomes law. This effect could be dampened if companies choose to not allow their system to be used for high-risk uses in an effort to avoid potential controversial uses of their systems. For example, OpenAI's usage guidelines for their natural language model GPT-3 explicitly disallows some uses – for example applications that "help determine eligibility for credit, employment, housing, or similar essential services" – that the AI Act would classify as high-risk.[257] Companies' postures could change as the market in these areas grows and as it becomes clearer that negative effects in high-risk domains can be avoided.

Incentives to have general AI systems fulfil AI Act requirements on high-risk systems could also come from customers adapting the AI system to high-risk uses. For example, a company might wish to use a large language model such as GPT-3 or Gopher to summarise candidates' answers to questions in a hiring process. It may be difficult to build a compliant high-risk system using a non-compliant general purpose system, in which case ensuring the general system's compliance with some of the AI Act's requirements could become a contractual obligation. Such contracts could allow the provider of the general system to charge a premium and would likely benefit both parties, assuming it would be cheaper for the provider of the general system to comply with the relevant requirements than for the purchaser to do so.

Some requirements might fairly straightforwardly create such demand, such as requirements that the training procedure of the relevant AI system be included in a high-risk system's technical documentation.[258] Other requirements concern the behaviour of the model, such as its accuracy in the target setting. Whether this ends up requiring adjustments to a general model adapted for a high-risk use is largely an empirical question of how much the behaviour of an AI system can be shaped by e.g. fine-tuning it – that is, training the system on some additional data more relevant to a desired task – or filtering its outputs. If it is easier to ensure compliance by making changes further down in the technology stack, no changes to the general system may be necessary. There is ongoing work on this question in e.g. the domain of natural language processing,[259] and it remains to be seen if such tools will be sufficient or whether it is advantageous to instead train the foundation model itself with compliance in mind.

Providers of general models may also be incentivised to ensure compliance with requirements for high-risk systems if it provides a boost to their reputation. Compliance with the high-risk requirements seems likely to be a strong quality signal. Indeed, it is plausible that some general models widely available today via APIs from companies such as Google, Hugging Face, Microsoft, and OpenAI are already compliant with many of the AI Act's requirements for high-risk systems.

The AI Act classifies a number of worker management tools as high-risk, including those that assist with or make decisions about access to employment or self-employment opportunities and task allocation. There has been significant growth in such tools over the past few years, in particular those used to assist with hiring decisions and to allocate tasks on a micro level. The latter category includes gig economy companies' (e.g. Uber's) use of AI systems to allocate jobs to different workers and the use of such systems in warehouse management[260] and by fast food companies.[261]

---

[256] Though the exemption would not hold if there were sufficient reason to believe the system would be misused. La Présidence Française du Conseil de l'Union européenne Art. 4c and 55a.
[257] OpenAI, *"Usage Guidelines (responsible Use): App Review,"* OpenAI's API, accessed July 13, 2022.
[258] See AI Act, annex IV §2.
[259] See e.g. Irene Solaiman and Christy Dennison, *"Process for Adapting Language Models to Society (PALMS) with Values-Targeted Datasets,"* Advances in Neural Information Processing Systems 34 (2021).

We are unsure whether the AI Act will produce a de facto Brussels Effect for worker management systems. On the one hand, uses of AI in these domains may be controversial – e.g. as evidenced by the negative press Amazon received when news broke that the CV screening system they used was biased in favour of male applicants[262] – pushing companies to voluntarily comply with stricter standards. On the other hand, it could be that meeting the EU requirements has a significant negative impact on the performance of the system. There might also be other pressures towards differentiation: the same traits may not be indicators of a successful employee across regions, encouraging companies to train or fine-tune their AI systems for different jurisdictions. If worker management systems are provided on international platforms, such as LinkedIn for job applications, a de facto Brussels Effect is likely because the costs of differentiation are higher.[263]

There is also a possibility of a de facto Brussels Effect in domains where companies have particularly large needs to build trust in their products. For example, this may be the case in use cases that are seen as controversial or sensitive, such as technologies for legal work or in using biometric data to categorise or identify individuals. In the former case, even though the software is used by lawyers at private firms rather than by the judicial authorities and would therefore not be covered by the regulation, compliance with the strictest possible standards is likely to be important for customers.

In addition to the above, the AI Act's requirements for high-risk systems might come to be seen as the gold standard for responsible AI development and deployment. If so, these requirements could produce a de facto effect for systems that the AI Act does not consider high-risk. This effect could be further bolstered if influential voluntary codes of conduct that the AI Act encourages are set up and include requirements similar to those for high-risk systems.[264] Whether such diffusion of high-risk requirements to non-high-risk systems, not only inside the EU but also outside it, will take place is difficult to tell.

In summary, we believe that a de facto effect is particularly likely for any changes to requirements in existing product safety regulation (e.g. for medical devices) and that there may be a de facto effect for general AI systems, worker management systems, and other domains where compliance with the AI Act is likely to be a strong quality signal. A de facto effect connected to a number of high-risk uses of AI, such as the use of AI in law enforcement or in the financial sector, is made unlikely by the regionalised compliance or market structure. Lastly, we may see de facto diffusion beyond high-risk systems if the AI Act's requirements on high-risk uses of AI comes to be seen as the gold standard for responsible AI development and deployment.

**2.6.2.2. What Requirements for High-Risk AI Systems Are Most Likely to Produce a De Facto Effect?**

The chance of a de facto effect differs not only between high-risk uses of AI but also between the requirements imposed on such systems. Which requirements are most likely to produce a de facto effect depends on a complex interaction of e.g. the factors outlined in sections 2.4 and 2.5. The devil will be in the details. Below, we will explore these dynamics for a subset of the requirements imposed by the AI Act: those pertaining to data and data governance, risk management systems and post-market monitoring, and technical documentation, as well as accuracy, robustness, and cybersecurity.

---

[260] Arthur Cole, *"AI Technology Modernizes Warehouse Management,"* November 1, 2021.
[261] Alex Glenn, *"Spanish Startup Reduced McDonald's Waiting Time,"* August 26, 2021.
[262] Jeffrey Dastin, *"Amazon Scraps Secret AI Recruiting Tool That Showed Bias against Women,"* REUTERS (Reuters, October 10, 2018).
[263] Engler, *"The EU AI Act Will Have Global Impact, but a Limited Brussels Effect."*
[264] AI Act, Title IX.

First, the AI Act would introduce requirements on the data used to train, validate, and test high-risk AI systems. The data should e.g. be "relevant, representative" and "have the appropriate statistical properties". At first glance, these requirements seem likely to produce a de facto effect. They would often have effects early on in the system's life cycle, thus requiring early forking and introducing high costs to differentiation. Once the costs for adaptation of the data-collection process are paid, using the same compliant data for non-EU products might not be costly — provided there are no steep trade-offs between less biased and more accurate AI models.[265]

However, the requirements on data could undermine a de facto effect of the AI Act if they require training on local data; that is, if they require or encourage companies to use EU data for AI systems deployed in the EU. This would undermine a de facto effect as it would effectively force companies to differentiate their products between different jurisdictions. Such differentiation in turn lowers the chance of a de facto effect as the cost of maintaining two different AI systems has already been taken on (for details, see §2.5). Parts of the requirements could be read as encouraging training on local data: data is meant to be "relevant [and] representative", and datasets should take into account "the characteristics or elements that are particular to the specific geographical ... setting within which the high-risk AI system is intended to be used."

Second, the AI Act imposes requirements regarding internal company processes, such as requiring there be adequate risk management systems and post-market monitoring. Such requirements could produce a de facto effect by causing more companies to set up new processes and apply them globally, or if their existing processes are brought up to the AI Act's standards globally. If a company does set up a new risk management system as a result of the AI Act, it seems plausible to us that such systems will often be applied worldwide, as companies commonly have risk management functions and the cost in setting up such a system might be primarily fixed.[266]

The risk management and post-market monitoring requirements could also produce a de facto effect indirectly. Even if the AI Act's requirements do not diffuse outside the EU, such enhanced risk management procedures could identify risks and issues within the EU that companies may feel compelled to address globally. This would particularly be the case if, for instance in the eyes of US courts, companies would have good reason to believe that the problem identified in the EU would also exist in the US. Speculatively, this could push some companies away from de facto diffusion if they wish to ensure that faults or risks identified for their EU products are not applicable to the rest of the world, causing them to differentiate their products and risk management teams.

Third, the AI Act introduces requirements on documentation of companies' AI systems, to be shared with regulators[267] and users.[268] Similar to "model cards,"[269] an AI system should be accompanied by information about its intended purpose, accuracy, performance across different groups and contexts, likely failure modes, and so on. Once such documentation has been created, it will likely be advantageous to provide it to the market outside the EU, insofar as the documentation is applicable to those systems. It is a service which some customers might appreciate and few will object to.

Fourth, requirements on accuracy, robustness, and cybersecurity of AI systems (Art. 15) are likely to exhibit a de facto Brussels

---

[265] Bommasani et al., "On the Opportunities and Risks of Foundation Models," sec. 5.4.
[266] "setting up a new QMS may cost EUR 193,000–330,000 upfront plus EUR 71,400 yearly maintenance cost." Renda et al., *"Study to Support an Impact Assessment of Regulatory Requirements for Artificial Intelligence in Europe Final Report (D5),"* 12.
[267] AI Act, art. 11.
[268] AI Act, art. 13.
[269] Margaret Mitchell et al., *"Model Cards for Model Reporting,"* arXiv [cs.LG] (October 5, 2018), arXiv.

Effect insofar as they (i) lead to updates in the relevant underlying models, (ii) will not substantially reduce product quality for non-EU consumers, and (iii) mainly consist of fixed costs. A 2021 study for the Commission suggests that the requirement could be implemented through technical solutions, e.g. tests against adversarial examples, model flaws, controlled studies in real-world conditions, and brainstorming of possible external threats.[270] This knowledge produced will then likely also affect the robustness of the products sold outside the EU.

To conclude, based on our cursory assessment, requirements regarding accuracy, robustness, cybersecurity, and documentation seem reasonably likely to produce a de facto effect. We may see a de facto effect with regard to the data requirements, so long as they do not introduce strong requirements to train on local data.

### 2.6.3. Prohibited AI Practices

The proposed AI Act bans certain AI applications, including (i) certain "real-time" biometric identification systems for law enforcement, (ii) AI-based social scoring, and (iii) AI systems used for subliminal manipulation. For the most part, we should expect bans to not produce a de facto effect, as companies would simply not offer prohibited products on the EU market.

However, there are two mechanisms by which prohibitions could contribute to a de facto effect. Firstly, some products that engage in prohibited uses can be adjusted to steer clear of prohibited uses. In such cases, there is a possibility of a de facto effect if it is advantageous to remain in the EU market, make the necessary changes, and apply those changes globally. Secondly, prohibitions in the EU could change consumer preferences abroad, making it more likely that companies could see reputational risks by offering EU-prohibited applications outside the EU.

The first mechanism may be important with regard to the proposed prohibition of "subliminal techniques … to materially distort a person's behaviour in a manner that causes … physical or psychological harm".[271] Depending on the interpretation of such a ban, many AI systems, e.g. those used for content moderation, could run the risk of engaging in prohibited uses. Should language similar to this make it into the final text — though it seems far from likely that it will[272] — companies are likely to put a lot of effort into ensuring that their systems are not considered manipulative, likely making changes early in the production process, potentially causing a de facto Brussels Effect.

The second mechanism could play a role in the prohibitions on social scoring and "real-time" biometric identification systems used by law enforcement. We might expect the latter to have some effect on multinational AI companies, seeing as many of them have already made commitments not to offer remote biometric identification to law enforcement. In 2020, Microsoft, Amazon, and IBM all announced that they would not offer facial recognition technology to US police departments or for their use on body camera footage.[273]

A de jure Brussels Effect is more likely for these prohibitions. We consider such a de jure Brussels Effect specifically in section 3.2.

### 2.6.4. Liability of AI Systems

In addition to the AI Act, the Commission is considering updates to the EU's liability rules with regard to AI systems.[274] Three factors negatively affect our ability to as-sess whether these changes will produce a de facto effect. First, it is difficult to evaluate whether firms change decisions in response to any liability regulation. Scholars have struggled to find such effects. Second, for corporate actors to not simply take on the liability but also to change trade-offs and decisions because of the liability rules, the liable actor along the supply chain must be the one who can reduce the relevant risk. Producers of AI systems, especially of increasingly generalised AI systems, could be far removed from the end-users, potentially limiting the number of cost-effective interventions they can undertake to reduce liability claims. If liability was placed on producers, they may be incentivised to simply take on the liability risk or to transfer the related costs to actors further along the AI supply chain. Third, due to the invisibility of compliance, it is difficult to evaluate whether a de facto Brussels Effect of product liability has occurred in the past. Due to these uncertainties, no conclusive statement about a de facto Brussels Effect of AI liability rules is possible. However, we can conclude that such regulatory diffusion is more likely if a firm's changes in response to the liability rules are either at the beginning of the technology stack or entail mostly fixed costs, such as post-market monitoring. However, this does not necessarily mean that such interventions would be the most cost-effective way of increasing the trustworthiness of AI products while avoiding undue regulatory burdens.

---

[270] Renda et al., *"Study to Support an Impact Assessment of Regulatory Requirements for Artificial Intelligence in Europe Final Report (D5),"* 132ff.
[271] AI Act, Title II, art. 5, §1a.
[272] We should expect forceful lobbying against there being ambiguity on this point in the final regulation. Facebook e.g. raised these concerns in their submission to the AI Act. Facebook, *"Response to the European Commission's Proposed AI Act."*
[273] Though Microsoft noted that they may offer such products, if appropriate federal legislation is put in place.

As of our writing, the Commission is actively developing changes to the EU liability regime concerning AI and other emerging technologies – by either changing the Product Liability Directive (PLD) or harmonising aspects of national civil liability law regarding the liability of certain AI systems.[275] The latter could include adopting strict liability for AI operators or the adaptation of the burden of proof.[276] Hence, liability affects all categories of risks discussed in this report. If a company conforms to all product safety standards, it is still liable for possible defects.[277] Liability regulation, complemented by product safety rules, determines who compensates users for damages incurred. Therefore, it has an ex ante deterrence effect by encouraging companies to adopt different internal procedures that result in safer products.[278] Whether the liability of AI exhibits de facto regulatory diffusion is subject to several significant uncertainties, which we'll discuss below.

First, as the Commission has not yet published draft AI liability rules, it is difficult to estimate its effect on the AI industry. For instance, the extent to which the rules exhibit regulatory stringency is unknown.

Second, a firm's response to liability regulation is difficult to observe. In theory, one would expect that liability regulation changes the firm's trade-off between profits and risks of potential defects – altering their decisions.[279] But how do we measure whether and how this happens? There is some evidence that firms do change behaviour.[280] However, because the Product Liability Directive (PLD) has not led to many court cases,[281] one might suspect firms are not strongly incentivised to change decisions because the cost of causing defects has not increased.

---

[274] European Commission, *"Commission Collects Views on Making Liability Rules Fit for the Digital Age, Artificial Intelligence and Circular Economy,"* Internal Market, Industry, Entrepreneurship and SMEs, October, 20 2021.

[275] See e.g. Public consultation in October 2021: European Commission; European Commission, *"Inception Impact Assessment: Proposal for a Directive Adapting Liability Rules to the Digital Age and Artificial Intelligence,"* June 6, 2021; European Commission, *"Civil Liability – Adapting Liability Rules to the Digital Age and Artificial Intelligence,"* European Commission, 2021.

[276] European Commission and Directorate-General for Justice and Consumers, Liability for Artificial Intelligence and Other Emerging Digital Technologies (Publications Office of the European Union, 2019); European Commission, *"Inception Impact Assessment: Proposal for a Directive Adapting Liability Rules to the Digital Age and Artificial Intelligence."*

[277] "The EU Product Liability Directive (PLD), that governs the responsibility for such defects, should be applied 'without prejudice' to the product safety regime" European Parliament, *"Directive 2001/95/EC of the European Parliament and of the Council of 3 December 2001 on General Product Safety (Text with EEA Relevance),"* CELEX number: 32001L0095, Official Journal of the European Union L 11, January 15, 2002, 4–17, art. 17.

[278] Andrea Bertolini, *Artificial Intelligence and Civil Liability*, PE 621.926 (European Parliament, 2020)

[279] John Prather Brown, *"Toward an Economic Theory of Liability,"* The Journal of Legal Studies 2, no. 2 (June 1, 1973): 323–49.

[280] Benjamin van Rooij, Megan Brownlee, and D. Daniel Sokol, *"Does Tort Deter? Inconclusive Empirical Evidence about the Effect of Liability in Preventing Harmful Behaviour,"* in The Cambridge Handbook of Compliance (Cambridge University Press, 2021), 311–25.

In addition, the liability ought to fall on the actor along the production process who can reduce the likelihood and severity of a defect. Some developers of AI systems, especially of increasingly generalised AI systems, are far removed from the end-users, limiting the number of cost-effective interventions they can undertake to reduce liability claims. In this case, a profit-maximising producer would conceivably buy liability insurance, accept the risk of liability claims, charge higher prices for their system, or transfer some of the risks to users of their system via contractual means. And if firms are not reacting to liability changes, the reaction cannot diffuse to other world regions.

For EU AI liability law, developers likely hold some liability.[282] In the PLD, "all producers involved in the production process should be made liable".[283] Hence, theoretically, all actors along the supply chain – who can reduce the likelihood and severity of AI risks – also have liability.

Third, the de facto regulatory diffusion of AI liability law is difficult to estimate. We find no evidence on whether the PLD has exhibited a de facto Brussels Effect. As noted above, firms' response to liability regulation is not easily observable. It is even more difficult to evaluate whether the response not only occurred but also diffused to other world regions.

Taken together, this should reduce our credence in de facto diffusion of the corporate responses to AI liability rules.

Nonetheless, we now turn our attention towards non-differentiation for responses to liability regulation. We can conclude some general trends. Suppose the liability incentivises actors to undertake more post-market monitoring than required for high-risk AI systems by the EU AI Act. In that case, a de facto Brussels Effect is likely because of the low costs of non-differentiation – monitoring may be mostly a fixed cost. Besides, if the most cost-effective interventions to reduce defects are early in the technology stack, the costs of differentiation are much higher – increasing the likelihood of a de facto Brussels Effect. Hence, if the liable actors are the producers of foundation models, they are less likely to divide compliance[284] and produce two different products. However, suppose instead that only downstream deployers respond to the liability or that the API access and interface of the foundation model will be changed. In that case, a de facto Brussels Effect is less likely because the costs of differentiation are lower. It is a completely different question as to whether compliance early or later on the stack is the most desirable, i.e. most cost-efficient, in achieving its regulatory aims.

To conclude, it is unclear whether changes in liability rules for AI systems will produce a de facto effect. This is because there is little evidence on whether and how liability rules in the EU have changed company behaviour, there is even more uncertainty about whether such changes have had impacts outside the EU, and it is not yet clear how the liability of AI systems will be distributed across the AI supply chain. If liability requires changes early on in the technology stack, e.g. changes to the training of a foundation model, or requires changes that once made are cheap to apply outside the EU, a de facto effect seems more likely.

---

## 2.7. De Facto Brussels Effect Conclusion

In this section, we explored the dynamics of the de facto Brussels Effect and applied those to the context of AI and the EU AI Act. We conclude that the AI industry as a whole has many of the features needed to produce de facto regulatory diffusion. We also conclude that some parts of the EU's proposed AI Act are likely to produce a de facto effect.

First, we outlined five factors which determine the likelihood of a de facto Brussels Effect, building and expanding on Anu Bradford's work[285] and arguing that these factors looked reasonably favourable to a de facto effect in the AI industry. The current and prospective market for AI products in the EU is large. The EU accounts for 5 to 20% of worldwide AI spending. Moreover, multinational and oligopolistic firms dominate the AI industry – making non-differentiation more attractive (see §2.1). The EU's regulatory capacity is strong, including the expertise, ability, and interest to sanction non-compliance (see §2.3).

Further, EU AI regulation is expected to be more stringent than other jurisdictions' regulation because of the EU's regulation-friendly public opinion and regulatory culture (see §2.2). In addition, the regulatory process is ahead of other major jurisdictions, potentially providing the EU with a first mover advantage (see §2.3.4). However, some worry that the proposed AI Act will place excessive demands on AI companies, potentially leading to reduced consumption of and investment into AI products in the EU (see §2.4). If some requirements are redesigned and hence less costly, there could be significantly smaller effects on the EU AI industry. EU consumers are unlikely to start consuming non-EU products(e.g. by moving out of the EU), though some might use non-EU VPNs to access AI systems online. Next, we explored the dynamics of non-differentiation, arguing that early forking, high perceived product quality as a result of compliance, and low variable costs to complying beyond the EU once EU compliance is secured are important factors in making a de facto Brussels Effect more likely.

Second, we applied the above framework to the specific requirements set out in the AI Act on prohibited, high-risk, and limited-risk uses of AI (see §2.6). Narrowing down on these particular requirements, the EU AI Act appears less likely to produce a de facto effect than the previous sections might indicate, e.g. as the act focuses heavily on government and regional uses of AI.

In the Commission's proposed AI Act, certain AI applications, such as chatbots and deepfakes, have transparency requirements — it must be disclosed that they are AI products. In this case, the costs of differentiation are low because only the interface has to be changed. Hence, a de facto Brussels Effect would only occur if the non-EU consumers value disclosure or if the reputational costs of differentiation are substantial.

High-risk AI practices are subject to product safety requirements under the proposed AI Act. De facto diffusion in cases where the product is regionalised, e.g. because it is used by governments or because the industry already differentiates products between jurisdictions (such as in the financial sector or with regard to some critical infrastructure), is less likely. It is only likely to happen insofar as the AI Act's high-risk requirements come to be seen as the gold standard of responsible use of AI or if provision of these products is globalised, though the use is regional.

We believe that we're most likely to see de facto regulatory diffusion in the high-risk use of AI in the following domains: (i) many of the products already covered by existing product safety regulation under the New Legislative Approach, notably medical devices; (ii) worker management, including hiring, firing, and task allocation; (iii) some general AI systems or foundation models used across a wide range of uses and industries; and (iv) less confidently, the use of AI in the legal sector and the use of biometric identification and categorisation of natural persons. There could also be diffusion of these standards outside high-risk uses of AI if the requirements become seen as the gold standard of responsible AI development and deployment. We also explore which requirements on high-risk AI systems are most likely to see a Brussels Effect, highlighting requirements regarding data, risk management, documentation, and the accuracy, robustness, and cybersecurity of high-risk AI systems as reasonably likely to see diffusion.

---

[285] Bradford, *"The Brussels Effect"*; Bradford, *The Brussels Effect: How the European Union Rules the World.*

Depending on the interpretation of the outright prohibitions in the final AI Act, many AI systems, particularly those used for content moderation, risk being banned. Should such strict language make it into the final legislation, companies will likely invest heavily in ensuring that their systems are for instance not considered manipulative. This may involve making changes early in the production process, potentially causing a de facto Brussels Effect because of substantial differentiation costs.

For AI liability updates,[286] the plausibility of de facto regulatory diffusion is uncertain. First, it is difficult to evaluate whether firms change decisions in response to liability regulation because such changes are barely visible. Second, for corporate actors to not simply accept the liability but to change trade-offs and decisions because of the liability rules, the liable actor along the supply chain must be the one who can improve the source code or the data-collection process or who makes deployment decisions. However, we can conclude that regulatory diffusion is more likely if a firm's changes in response to the liability rules are either at the beginning of the stack or entail mostly fixed costs, such as post-market monitoring.[287]

---

# 3. Determinants of the De Jure Brussels Effect

Foreign jurisdictions may also adopt regulation that resembles EU regulation; a phenomenon termed the de jure Brussels Effect.[288] Below, we describe four channels that may produce a de jure Brussels Effect, building on Bradford, Young, and Schimmelfennig & Sedelmeier.[289] First, foreign jurisdictions could adopt the blueprint voluntarily. Second, the EU may promote their blueprint through multilateral agreements or mutual recognition agreements. Third, a de facto Brussels Effect can help cause a de jure effect. For example, multinational companies may lobby their governments to adopt regulations similar to the EU because, otherwise, their national competitors may benefit from less stringent requirements. Fourth, conditionality describes the situations in which another jurisdiction incorporates the EU blueprint because external incentives provided by the EU, such as trade requirements or treaties, encourage it. The GDPR, as an instance of and an analogy for AI regulation, caused regulatory diffusion partly through significant conditionality.

We argue that a de jure Brussels Effect with respect to AI is plausible for at least parts of the EU AI regulatory regime, though it is far from certain. The Blueprint Channel will likely be the most influential, seeing as the EU is one of the first movers in regulating AI and is responding to regulatory pressures also felt by other jurisdictions. We are reasonably likely to see diffusion of the risk-based approach and the operationalisation of what "trustworthy AI" entails. There is a decent chance that other jurisdictions, in particular liberal democracies, will prohibit some of the same systems as the EU. It seems likely that other jurisdictions will introduce transparency requirements for e.g. chatbots and deepfakes, though that would more accurately be termed a "California Effect," as California was first to introduce such requirements with the BOT Disclosure Act passed in 2018.[290] We are unsure how influential the list of high-risk systems in Annex III will be, i.e. those high-risk systems not already covered by safety regulation, though it seems likely that AI systems' use in domains like hiring and loan decisions will be viewed as controversial across the globe, as evidenced e.g. by current White House proposals for an AI Bill of Rights.

What about the other channels? A de jure Brussels Effect via the Multilateralism Channel seems most likely for those parts of EU regulation that could feed into international standard setting processes, in particular, the requirements put on high-risk uses of AI. If a de facto Brussels Effect of AI occurs, multinational companies are likely to lobby for EU-like AI regulation abroad, attempting to create a de jure effect (the De facto Channel). However, it is uncertain how successful such efforts would be. We remain unsure about the extent to which the Conditionality Channel will cause regulatory diffusion.

---

[288] See Damro, *"Market Power Europe"*; Bradford, *"The Brussels Effect"*; Bradford, *The Brussels Effect: How the European Union Rules the World*. Note that the term is sometimes used specifically for what we term the De Facto Channel. Dempsey et al., *"Transnational Digital Governance and Its Impact on Artificial Intelligence."*

[289] Bradford discusses the channels we cover in §§3.1 and 3.3 Bradford, *"The Brussels Effect"*; Bradford, *The Brussels Effect: How the European Union Rules the World*. Schimmelfennig and Sedelmeier argue that one should distinguish between the channels discussed in §§3.1 and 3.4. Schimmelfennig and Sedelmeier, *"Governance by Conditionality: EU Rule Transfer to the Candidate Countries of Central and Eastern Europe."* From Young, we added the channel in §3.3 to the framework. Young, *"Europe as a Global Regulator? The Limits of EU Influence in International Food Safety Standards."*

[290] California Senate, *An act to add Chapter 6 (commencing with Section 17940) to Part 3 of Division 7 of the Business and Professions Code,* relating to bots.. It is also known as the BOT ("Bolstering Online Transparency") Act or California Senate bill 1001.

Before proceeding, it is worth noting that a jurisdiction adopting EU-like regulation is not sufficient to establish that there has been a de jure effect. The EU's actions must also have played a causal role. This is because the EU and the other jurisdictions might have independently adopted the regulation for the same reason, e.g. because they are responding to the same regulatory issues. This might seem particularly likely in the case of AI since it is a new regulatory domain where many jurisdictions are facing similar regulatory challenges, meaning some are likely to reach for similar regulatory solutions. One relevant factor in assessing whether there has been a causal link is time – who adopted the regulation first – though it is important to note that regulations can have a global impact even before they are adopted, as visible in the AI Act's being discussed abroad. Another way to assess the causal link is to trace specific causal pathways by which the EU might affect regulation abroad. Below, we outline some of these pathways and speculate on how likely they are to have an effect in the EU case. We encourage other authors to trace these pathways as the EU's regulatory regime is being designed and implemented.

## 3.1. Blueprint Adoption Channel

Foreign jurisdictions often copy EU regulations believing this approach might meet their regulatory goals. This Blueprint Adoption Channel is more likely if (i) the issue is on the political agenda of other countries because of similar concerns and interests, (ii) the EU is the first mover for regulation, and (iii) the EU advertises and promotes its regulation including via networks and multilateral institutions. Such adoption might be the result of what Shipan & Volden[291] call "learning" – adopting similar regulation after it has been adopted and proved successful – or "imitation" – adoption before data on the success of the regulation is available. We argue that (ii) and (iii) are relatively likely for AI regulation. Then, we describe the diffusion of EU AI policy principles in recent years, arguing that this provides some indication in favour of the EU's issue framings spreading internationally. However, we also note that other jurisdictions adopting EU-like regulation does not necessarily suggest that the EU caused this adoption. It may be that the EU and the other jurisdictions are responding to the same regulatory pressures.

First, the Blueprint Channel is more likely if the issue regulated in the EU is also on the political agenda of other countries and is so out of similar concerns. Artificial intelligence is on many policymakers' agendas, though their reasons differ. Some emphasise AI's importance for national competitiveness and economic growth, while others put more emphasis on the potential harms AI systems might cause. Relative to other jurisdictions, EU policymakers seem more focused on the harms of AI. They also tend to place greater weight on the claim that a thriving AI industry requires public trust and that public trust relies on regulation.[292]

What parts of the AI Act seem most likely to meet regulatory needs faced by other jurisdictions? Firstly, though jurisdictions may differ in which AI systems they find worthy of additional regulatory burdens, they will all need to decide what rules such systems should comply with. Thus, we expect the EU's requirements for high-risk systems to end up being influential abroad. Secondly, many populations in liberal democracies worry about the use of AI systems by the government, which might suggest the EU's list of prohibited uses of AI potentially could become influential.

---

[291] Charles R. Shipan and Craig Volden, *"The Mechanisms of Policy Diffusion,"* American Journal of Political Science 52, no. 4 (October 2008): 840–57.
[292] See e.g. *European Commission, "AI Act"* recital 3.3.

Thirdly, as AI systems become more prevalent in society and it becomes more difficult to distinguish AI-generated speech, text, and art from that generated by humans, it seems likely that policymakers will feel a need to have citizens be informed of the origin of the content they are engaging with. Therefore, one could expect the EU's regulation on transparency requirements for certain AI systems to influence how these other jurisdictions meet that regulatory challenge. However, if this happens, the term "California Effect" would be more apt, as California was the first to introduce such requirements with the BOT Disclosure Act passed in 2018.

If the EU publishes the first regulation on some issue, it is more likely to be copied. More jurisdictions will have an opportunity to use the blueprint, and the EU could be seen as the leader concerning the topic at hand and its standards as the gold standard of responsible AI development. In section 2.3.4, we argue that it is likely that the EU is the first mover among large jurisdictions proposing comprehensive AI regulation.

Moreover, the EU's promotion of its regulatory blueprint makes international adoption more likely. The EU can promote a global narrative that this blueprint solves an important problem, which makes copying more likely. For instance, the worldwide promotion of the CE marking was partly responsible for its significant de jure Brussels Effect.[293]

Taking inspiration from the success of de jure diffusion of the GDPR, the Commission plans to promote its regulatory regime on AI. Many smaller nations have adopted regulation that is GDPR-adequate – where the GDPR allows transfer of personal data to a jurisdiction outside of the EU without specific authorization, as the jurisdiction's data protections are deemed similar enough to the EU's[294] – which the Commission also states as one achievement in a 2019 assessment of the GDPR.[295]

In the proposed AI Act, the Commission states that "[s]pearheading the ethics agenda, while fostering innovation, has the potential to become a competitive advantage for European businesses on the global marketplace."[296] It does so with an awareness of the global effects the GDPR has had: "Many countries around the world have aligned their legislation with the EU's strong data protection regime. Mirroring this success, the EU should actively promote its model of a safe and open global Internet."[297] As an example, the EU's ban and conformity assessments of different biometric identification systems could increase the international condemnation of such systems, limiting deployment even outside the EU's borders.

Further, the narrative diffusion of EU AI thinking since approximately 2018 provides valuable information regarding the international susceptibility to adopting EU thinking on AI and the future Blueprint Adoption Channel. A 2020 report from Access Now suggests that the European "Trustworthy AI" approach, outlined in the EU's High-level expert group's Ethics Guidelines[298] has had a significant global effect, e.g. via the OECD. Many of the concepts and related principles were included in the OECD principles,[299] which were signed by 42 countries and heavily influenced a subsequent G20 declaration.[300] EU member states partially adopted the EU's focus on AI trustworthiness and human rights. After the HLEG Ethics Guidelines were published and before the Commission published its 2020 AI White Paper, 17 national strategies were published by EU member states, of which five countries explicitly mentioned "Trustworthy AI".[301] One member state, Malta, fully integrated the seven requirements of the EU Ethics Guidelines for Trustworthy AI.

---

[293] Hanson, *CE Marking, Product Standards and World Trade* p. 193.
[294] GDPR, art. 45.
[295] European Commission, *"General Data Protection Regulation Shows Results, but Work Needs to Continue,"* European Commission, July 24, 2019.
[296] European Commission, *"Annex to the Communication from the Commission to the European Parliament, the European Council, the Council, the European Economic and Social Committee and the Committee of the Regions,"* December 7, 2018, 18.
[297] European Commission, *"Communication from the Commission to the European Parliament, the Council, the European Economic and Social Committee and the Committee of the Regions Shaping Europe's Digital Future,"* CELEX number: 52020DC0067, February 19, 2020, 14 "A strategy for standardisation, which will allow for the deployment of interoperable technologies respecting Europe's rules, and promote Europe's approach and interests on the global stage (Q3 2020)."; *European Commission and Directorate-General for Communications Networks, Content and Technology, "Shaping Europe's Digital Future"* (Publications Office, 2020).
[298] European Commission, *Ethics Guidelines for Trustworthy AI*.
[299] OECD AI Policy Observatory, *"The OECD Artificial Intelligence (AI) Principles,"* OECD AI Policy Observatory, accessed July 14, 2022.
[300] Leufer and Lemoine, *"Europe's Approach to Artificial Intelligence: How AI Strategy Is Evolving."*

Several other countries have incorporated EU language into their national AI strategies, including Singapore and New Zealand. For instance, New Zealand has taken up the EU language and principles in its Aotearoa AI Principles.[302] They explicitly mention that they drew upon the European Commission's Ethics Guidelines for Trustworthy AI[303] and also used a human rights approach.[304] However, it remains to be seen whether this narrative diffusion will lead to similar regulation.

Some jurisdictions are further along in their regulatory processes. In April 2022, the Brazilian Senate tasked a commission with proposing a bill on AI regulation, taking into account e.g. a bill proposed by the lower house of the Brazilian National Congress.[305] Though the Brazilian approach may diverge from the EU's regulatory regime – the lower chamber's proposal comprised a significantly more sectoral approach, not introducing new AI-specific regulators or regulation – the forthcoming EU regime seems to be given significant attention. The rapporteur of the lower chamber said the EU's efforts were the main inspiration for the proposed changes and that the Senate commission will explicitly consider the EU regime.[306]

Policy discussions in the US are also starting to concern issues addressed by the EU's AI Act, though it is unclear whether there is a causal relationship. For instance, in October 2021, Eric Lander and Alondra Nelson from the White House Office for Science and Technology Policy published an opinion piece in *Wired*, arguing that the US needed an AI Bill of Rights.[307] They encouraged debate about what such updated rights in light of AI technologies might be. They encouraged discussion of e.g. rights to not be subject to biased or unaudited algorithms, to know if and how an AI system is influencing decisions important to one's civil liberties, and to not be subject to pervasive and discriminatory surveillance, many of which the AI Act seeks to protect. Their piece was accompanied by a Request for Information on the use of AI for biometric technologies, including their use for "inference of attributes including individual mental and emotional states", signalling interest in proposing concrete regulation. The extent to which such efforts will make their way into regulation is hard to predict. It would depend on the amount of bipartisan support for such efforts and the extent to which the Biden administration can introduce measures via executive powers.

History suggests that a strong de jure Brussels Effect through the Blueprint Adoption Channel reaching the US seems unlikely, while China regularly takes inspiration from EU regulation. In contrast to countries in the Asia-Pacific, Latin America, or Eastern Europe, only a few cases of a de jure Brussels Effect have been observed in the United States.[308] One such case is the EU chemical regulation REACH that led to both a de facto and de jure Brussels Effects in the United States.[309] It regulates chemical products un-der the so-called New Approach of product safety, just as the conformity assessments for high-risk AI systems in the proposed AI Act.[310] EU legislation, in particular REACH, was cited in state-level reforms in the US states of California, Massachusetts, and Maine, and in American federal-level reforms, including the "Kid-Safe Chemicals Act."[311] Moreover, EU chemical regulation influenced the behaviour and thinking of American producers, consumers, and the public. For instance, it led to American consumers demanding more information about the safety of chemicals and NGOs lobbying for improvements to American chemical regulations.[312]

---

[301] *Leufer and Lemoine.*

[302] *Leufer and Lemoine.*

[303] The AI Forum of New Zealand, *"Trustworthy AI in Aotearoa: AI Principles"* (The AI Forum of New Zealand, March 2020), 2.

[304] Leufer and Lemoine, *"Europe's Approach to Artificial Intelligence: How AI Strategy Is Evolving,"* 10

[305] Agência Senado, *"Brasil poderá ter marco regulatório para a inteligência artificial"*; Agência Câmara de Notícias, *"Câmara aprova projeto que regulamenta uso da inteligência artificial."* English translations here.

[306] Agência Senado, *"Brasil poderá ter marco regulatório para a inteligência artificial"*; Agência Câmara de Notícias, *"Câmara aprova projeto que regulamenta uso da inteligência artificial."* English translations here.

[307] Eric Lander and Alondra Nelson, *"Americans Need a Bill of Rights for an AI-Powered World,"* Wired, October 8, 2021.

[308] Scott, *"Extraterritoriality and Territorial Extension in EU Law"*; Bradford, *The Brussels Effect: How the European Union Rules the World.*

[309] Joanne Scott, *"From Brussels with Love: The Transatlantic Travels of European Law and the Chemistry of Regulatory Attraction,"* The American Journal of Comparative Law 57, no. 4 (October 1, 2009): 897–942.

Despite the low base rate of a de jure Brussels Effect for the US, AI could be different. AI is a fairly new regulatory domain, where policymakers on both sides of the Atlantic may be facing somewhat similar regulatory pressures. The citizens of both jurisdictions are, for example, worried that the government could use AI technologies for repression. As a result of responding to the same regulatory pressures, and the EU identifying appropriate mechanisms for AI regulation first, the US might adopt regulation inspired by the EU.

This would most likely happen via US states first passing regulation, which then diffuses to the federal level. Since about the 1990s, many US states have adopted more stringent risk regulation than the federal government, often inspired by EU regulation.[313] We are already seeing this in the case of AI, where Oregon, New Hampshire, and California have banned the use of facial recognition software on body cam footage.[314] This more stringent regulation could then diffuse to the US federal level, for example via the De Facto Channel.[315]

This Blueprint Channel regularly reaches China – the Chinese government has in the past copied EU rules quickly after EU adoption. Examples include data protection legislation, chemical regulation, toy safety standards, competition rules, merger rules, and genetically modified organism (GMO) labelling.[316] In 2021, China adopted the Personal Information Protection Law, which provides GDPR-like protections for citizens against private corporations.[317] Chinese officials have also publicly stated that they take inspiration from EU regulation. For instance, China's former vice minister of commerce, Ma Xiuhong, said that "China has borrowed many experiences of European Competition Law in various aspects for the enactment of Antimonopoly Law."[318]

However, recent Chinese efforts to regulate AI reduce the chance of de jure diffusion to China with regard to AI regulation. Chinese regulators have charged ahead in some domains, regulating AI sooner and likely more stringently in certain dimensions than the EU will. In March 2022, the Cyberspace Administration of China adopted regulations for recommendation systems, including requirements for providers to protect users' personal information, to allow them to conveniently turn off recommendation services and to be informed about how the recommendation system works, and to ensure algorithms do not "go against public order and good customs, such as by leading users to addiction or high-value consumption".[319] In January that same year, rules on AI-gener-ated content such as deepfakes were proposed, including provisions like requiring consent from the subject of deepfake images, audio, or video, and that the recipient of AI-generated content be made aware of its source. This last provision may be inspired by e.g. California's BOT Disclosure Act, though we have not found evidence on this issue. As is common for Chinese regulation, the proposal is significantly more far-reaching than EU proposals, stating that regulatees "may not produce, reproduce, publish, or disseminate: information inciting subversion of State power or harming national security and social stability; obscenity and pornography; false information; information harming other people's reputation rights, image rights, privacy rights, intellectual property rights, and other lawful rights and interests …". Matt Sheehan argues that these initiatives should not be ignored by western observers: there may be lessons to learn from the success or failure of these initiatives.[320] At the same time, the AI Act aims to be more general and comprehensive than the Chinese regulation. The South China Morning Post discusses what Hong Kong and China can learn from the AI Act.[321]

---

[310] *European Commission, "AI Act."*

[311] Scott, *"From Brussels with Love: The Transatlantic Travels of European Law and the Chemistry of Regulatory Attraction."*

[312] *Scott.*

[313] Vogel, *The Politics of Precaution: Regulating Health, Safety, and Environmental Risks in Europe and the United States.*

[314] Samsel, *"California Becomes Third State to Ban Facial Recognition Software in Police Body Cameras."*

[315] For example, in 2019, Democrat Elissa Slotkin brought the Bot Disclosure and Accountability Act to Congress, co-sponsored by 4 Republicans. The bill subsequently "died in committee." Elissa Slotkin, *"H.R.4536 - 116th Congress (2019-2020): Bot Disclosure and Accountability Act of 2019,"* September 27, 2019.

[316] For a longer description and list of such regulatory diffusion, see Bradford, *The Brussels Effect: How the European Union Rules the World.* For instance: chapter 5, p.153 for data protection legislation; chapter 7 page 225 for the RoHS directive; page 180 for the GMO labelling; pages 201, 203 for the chemical regulation REACH; some toy safety standards page 204; China's 2008 Anti-Monopoly Law (pages 117 and 118); merger rules page 118

[317] Though note that it does not afford any protections against state uses of personal data. Lomas, *"China Passes Data Protection Law."*

[318] Bradford, *The Brussels Effect: How the European Union Rules the World*, 118.

[319] Ding, *"ChinAI #168: Around the Horn (edition 6)"*; Ding, *"ChinAI #182: China's Regulations on Recommendation Algorithms"*; Rogier Creemers and Graham Webster, *"Translation: Internet Information Service Deep Synthesis Management Provisions (Draft for Comment) – Jan. 2022,"* DigiChina, February 4, 2022.

If the Blueprint Channel does reach China, it is unlikely to do so for regulation curtailing the government's ability to use AI technology for surveillance, censorship, and the like. The Chinese Personal Information Protection Law notably does not put any restraints on government use of data.

De jure diffusion via the Blueprint Channel seems particularly likely for the requirements imposed on high-risk systems. Firstly, many jurisdictions beyond the EU may establish more sectoral and piecemeal regulatory regimes for AI, where significant responsibility is given to existing regulators and new domains in need of regulation as a result of advances in AI are dealt with one-by-one. For example, the UK National AI Strategy suggested it would take a largely sectoral approach[322] and the same seems likely for Brazil.[323] We are already seeing e.g. China and US states producing regulation aimed at specific regulatory challenges produced by AI. Relatedly, the AI Act has been criticised for taking an overly product safety–focused lens on AI regulation.[324] This makes diffusion of the structure of the EU regulatory regime less likely. Secondly, for major EU trading partners, the main source of trade friction with the EU would stem from imposing requirements incompatible with the EU on companies trading with the EU. Whether the company is regulated by a sectoral regulator outside the EU and an AI-specific regulator within the EU has a less significant impact. This is one mechanism by which the EU's requirements for high-risk AI systems might become a gold standard or a crucial starting point for other jurisdictions and actors attempting to make concrete responsible AI development and deployment practices.

## 3.2. Multilateralism Channel

A de jure Brussels Effect can also be caused by international standards being influenced by EU norms.[325] For instance, this has been the case for International Organization for Standardization (ISO) standards. We argue that this channel could work through international standard setting organisations (such as the ISO, IEEE, and ITU) in which the EU has historically had significant influence.[326] Through such standard setting bodies, the EU is likely to be able to spread its conception of what responsible deployment of AI systems entails (e.g. its list of requirements for high-risk systems). In addition, there are also informal diplomatic routes through which the EU norms influence other jurisdictions.

---

[320] Matt Sheehan, *"China's New AI Governance Initiatives Shouldn't Be Ignored,"* Carnegie Endowment for International Peace, January 4, 2022.

[321] Andy Chun, *"Europe's AI Regulation Seeks a Balance between Innovation and Risk. Is Hong Kong Ready?,"* South China Morning Post, March 18, 2022.

[322] Office for Artificial Intelligence, Department for Digital, Culture, Media & Sport, and Department for Business, Energy & Industrial Strategy, *"National AI Strategy"* (HM Government, September 22, 2021).

[323] Agência Câmara de Notícias, *"Câmara aprova projeto que regulamenta uso da inteligência artificial."*, English translation here.

[324] See e.g. Ada Lovelace Institute, *"People, Risk and the Unique Requirements of AI: 18 Recommendations to Strengthen the EU AI Act"* (Ada Lovelace Institute , March 31, 2022).

[325] Young calls this regulatory diffusion through competition. Alasdair R. Young, *"Liberalizing Trade, Not Exporting Rules: The Limits to Regulatory Co-Ordination in the EU's 'new Generation' Preferential Trade Agreements,"* Journal of European Public Policy 22, no. 9 (October 21, 2015): 1253–75.

[326] Engler suggests that CEN and ISO's efforts for convergence should make this more likely. Engler, *"The EU AI Act Will Have Global Impact, but a Limited Brussels Effect"*; CEN-CENELEC, *"ISO and IEC,"* CEN-CENELEC, accessed July 14, 2022.

Several multinational institutions are involved in the global AI policy dialogue. Among them is the subcommittee of the ISO that is developing international standards for the AI industry.[327] Generally, the EU has significant influence in ISO negotiations.[328] For instance, the US's decentralised regulatory process for developing product standards, compared to the more hierarchical structure in the EU, puts the US at a disadvantage globally at the ISO.[329]

However, the EU might have less influence in standard setting for AI than other standard setting processes, because the US and China increasingly see AI and standard setting for AI as important for national security, leading to concerted efforts by both countries to exert influence. Chinese engagement has significantly increased since 2012.[330] Since the release of the China AI White Paper, the China Electronics Standardization Institute has been actively engaged in developing relevant international standards, including as an active member of the ISO subcommittee focused on AI standards.[331] In the US, there have been recent calls, for example from the National Institute on Standards and Technology (NIST),[332] to increase US engagement in standard setting processes for AI. Moreover, in international negotiations, the EU is one of the most stringent regimes. This outlying preference could put the EU at a disadvantage if the standard setting process is very structured and majority rule decisions are made.[333]

Moreover, there can be international or bilateral mutual recognition agreements (MRAs). If the EU and the US sign an MRA, EU-compliant products can be sold in the US and vice versa. Historically, the more stringent jurisdictions have been advantaged in the negotiation processes of the MRAs.[334] For more details on the MRA for product safety, we refer to the appendix of this report.

This Multilateralism Channel, however, encompasses much more than the formal international institutions and agreements, such as the ISO and MRAs, and can also occur on the basis of informal bilateral negotiations. For AI, the EU-US bilateral efforts illustrate such an informal channel. In June 2021, the EU and US launched the Trade and Technology Council to "lead digital transformation."[335] Its goals include (i) cooperating on developing compatible international standards and (ii) facilitating cooperation on regulatory policy and enforcement. For both goals, working groups were set up. The G7 countries also pledged to support their respective "effective standard-setting" of AI systems.[336]

## 3.3. De Facto Effect Channel

A de facto Brussels Effect can lead to a de jure effect. Suppose there is a strong de facto Brussels Effect. In that case, foreign companies who use the EU blueprint as their international policy already bear the compliance cost and so will lobby other governments to adopt the EU regulation, whereas domestic competitors who do not operate in or export to the EU do not.[337] Furthermore, countries may be more inclined to implement regulation that has seen a de facto Brussels Effect in their country, as passing such regulation comes with smaller regulatory costs for their companies. For instance, when European corporate actors were subjected to an EU regulation requiring Eco-Management and Auditing Scheme (EMAS) standards on public disclosure of the results of company policies,[338] they started an alliance with the green movement to support diffusion of the standard. Consequently, the ISO 14001 standard was adopted in 1996, copied from the EU regulation.[339]

---

[327] See the ISO website on AI Standards and their ongoing work: ISO, *"ISO/IEC JTC 1/SC 42."*
[328] Hairston, "Hunting for Harmony in Pharmaceutical Standards."
[329] Walter Mattli and Tim Büthe, *"Setting International Standards: Technological Rationality or Primacy of Power?,"* World Politics 56, no. 1 (October 2003): 1–42; Young, *"Europe as a Global Regulator? The Limits of EU Influence in International Food Safety Standards."*
[330] Mark Montgomery and Natalie Thompson, *"What the U.S. Competition and Innovation Act Gets Right About Standards,"* Lawfare, August 13, 2021.
[331] The Big Data Security Standards Special Working Group of the National Information Security Standardization Technical Committee, *"Artificial Intelligence Security Standardization White Paper (2019 Edition),"* trans. Etcetera Language Group, Inc. (Center for Security and Emerging Technology, 2019); Peter Cihon, *"Standards for AI Governance: International Standards to Enable Global Coordination in AI Research & Development"* (Center for the Governance of AI Future of Humanity Institute, University of Oxford, April 2019).
[332] National Science and Technology Council, *"U.S. Leadership in AI: A Plan for Federal Engagement in Developing Technical Standards and Related Tools"* (National Science and Technology Council, August 9, 2019).
[333] Cihon, *"Standards for AI Governance: International Standards to Enable Global Coordination in AI Research & Development."*
[334] Young, *"Europe as a Global Regulator? The Limits of EU Influence in International Food Safety Standards"*; Young, *"Liberalizing Trade, Not Exporting Rules: The Limits to Regulatory Co-Ordination in the EU's 'new Generation' Preferential Trade Agreements."*
[335] European Commission, *"EU-US Launch Trade and Technology Council to Lead Values-Based Global Digital Transformation."*
[336] The White House, *"Carbis Bay G7 Summit Communiqué,"* The White House, June 13, 2021.

Market structure influences the strength of this channel. Lobbying can be seen as a coordination problem: the bigger and the more oligopolistic the firms, the stronger the lobbying, as firms can more easily cooperate in providing the common pool resource. Moreover, the greater the total market size, the more likely firms will have enough political power to achieve their aim.[340]

However, though it seems plausible that companies would engage in this lobbying if they were subject to a de facto effect, there is little reported evidence for this channel leading to a de jure Brussels Effect to date. The Eco-Management and Auditing Scheme (EMAS) standard is the only clear reported example we found. There is some evidence that companies will start lobbying governments if they are subject to greater regulatory burdens from the EU. Privacy regulation offers one such example. Since the GDPR took effect, CEOs of Alphabet,[341] Apple,[342] and Microsoft[343] have called for the US to pass similar regulation. However, it is difficult to tell the extent to which such public statements translate into on-the-ground lobbying efforts by the companies and, if they do, whether such lobbying would be successful.

The De Facto Channel is seemingly more common for the California Effect than the Brussels Effect. US automobile emission standards, capital market regulation,[344] pollution standards, and other environmental standards have been diffused from the US state level to the US federal level[345] and to other countries through such a De Facto Channel.[346] As such, we believe the De Facto Channel is most likely to have an effect on US federal policy via states adopting stringent EU-like regulation of AI. Though US companies would likely strongly oppose such state-level regulation if it risks their profitability, once passed we predict that many of those same companies would push for similar regulation at the federal level. This might be the most plausible route by which EU-like regulation is eventually passed in the US.

---

In summary, contingent on a de facto Brussels Effect, we expect that multinational AI firms will lobby other jurisdictions to pass similar AI regulation, as the AI industry is relatively big and has an oligopolistic structure. This is particularly likely to happen if some US states, notably California, pass EU-like regulation. We are unsure how successful such efforts would be. Further, we should also expect legislators to be, on the margin, more inclined to implement EU-compatible legislation as it will introduce smaller compliance costs. It is unclear how successful this would be.

## 3.4 Conditionality Channel

Conditionality and external incentives can also lead to the adoption of EU blueprints abroad. This Conditionality Channel requires equivalence clauses and/or extraterritoriality, which are both unlikely for upcoming AI regulation.

Equivalency clauses, especially common in EU financial regulation,[347] are the clearest example of conditionality. These clauses condition ease of market access on the demonstration of equivalent rules in home markets. Countries that adopt EU-like rules can trade more easily with the EU. For example, the Commission undertakes adequacy decisions for data protection regulation, concluding whether a third country, one of its sectors, or an international organisation have equivalent data protection levels. Such decisions permit cross-border data transfer with diminished regulatory burdens. Hence, foreign jurisdictions, including the United States and Japan, experienced external incentives to adopt stronger data protection regulation.[348] After Japan increased its data privacy standards, it received an adequacy decision from the EU, improving data trade and transmission.[349]

A high degree of extraterritoriality can also put pressure on other countries to implement EU-equivalent regulation.[350] The EU has been shifting towards more extraterritoriality, expanding beyond the inclusion of EU imports.[351] Extraterritoriality is a feature of European aviation law, competition law, and data privacy law. It was a significant cause of the GDPR's de facto and de jure Brussels Effect (see appendix). This Conditionality Channel is significant if data privacy regulation is interpreted as an instance of AI regulation. AI product safety standards will likely not exhibit the same degree of extraterritoriality. Whereas the GDPR applies to any entity that handles any data from EU citizens, the AI Act would only apply to companies that put products on the EU market.

Taken together, a de jure Brussels Effect of AI is plausible. However, it might predominantly reach jurisdictions that have less geopolitical power. A de jure Brussels Effect is more likely to reach China than the US (see §3.1). The Multilateralism and Blueprint Channels are the most likely channels. If a de facto Brussels Effect occurs, it is plausible that multinational companies will lobby other jurisdictions, though it is unclear whether this will lead to success. While there are several examples of this channel as a California Effect, there is only one reported instance of such a de jure Brussels Effect.

---

[347] Jerome Deslandes, Magnus Marcel, and Cristina Pacheco Dias, *"Third Country Equivalence in EU Banking and Financial Regulation"* (European Parliament, August 2019).

[348] Ivy Yihui Hu, *"The Global Diffusion of the 'General Data Protection Regulation' (GDPR),"* ed. K. H. Stapelbroek and S. Grand (Erasmus School of Social and Behavioural Sciences, 2019)

[349] European Commission, *"Adequacy Decisions: How the EU Determines If a Non-EU Country Has an Adequate Level of Data Protection,"* European Commission, accessed July 14, 2022,

[350] Raphael Bossong and Helena Carrapico, eds., *EU Borders and Shifting Internal Security: Technology, Externalization and Accountability* (Springer International Publishing, 2016).

[351] For a recent clear example, see *"Developments in the Law: Extraterritoriality,"* Harvard Law Review 124, no. 5 (2011): 1226–1304. Much of the discussion focuses on EU competition (antitrust) law. See. e.g., Berkeley Electronic Press, *"Flying Too High? Extraterritoriality and the EU Emissions Trading Scheme: The Air Transport Association of America Judgment,"* Eutopia Law, 2012; Barbara Crutchfield George, Lynn V. Dymally, and Kathleen A. Lacey, *"Increasing Extraterritorial Intrusion of European Union Authority into U.S. Business Mergers and Competition Practices: U.S. Multinational Businesses Underestimate the Strength of the European Commission from G.E.-Honeywell to Microsoft,"* Connecticut Journal of International Law 19, no. 3 (2004): 571–616.

# 4. Appendix: Case Studies

We consider the history and causes of regulatory diffusion for (i) EU data protection legislation, (ii) the EU Product Liability Directive, and (iii) the product safety framework and CE marking.

## 4.1. Data Protection

The EU Data Protection Directive (DPD), a potential analogy to AI regulation, led to regulatory diffusion via a de jure Brussels Effect.[352] The 2018 General Data Protection Regulation (GDPR) exhibited a strong de facto Brussels Effect. Despite the recentness, the GDPR has led to a de jure Brussels Effect in more than five countries.

One can learn about the AI Brussels Effect from this case study as the market partly overlaps; AI and privacy regulation affect many of the same systems and products. We tentatively conclude that the unique features of data protection regulation were responsible for substantial parts of its de facto Brussels Effect. For data regulation, the forking happens earlier on, and the wide extraterritorial claims increased the market size to which the regulation applied. The de jure Brussels Effect appears to have been mostly caused by the attraction of foreign jurisdictions to the EU data protection blueprint, a process that has been ongoing since the Council of Europe's 1981 Convention 108 and the EU's 1995 Data Protection Directive.[353]

Assessing the potential of an AI Brussels Effect requires careful consideration of China and the United States since these countries are home to the largest number of world-leading AI companies. The history of data protection regulatory diffusion indicates that China experienced a de facto and de jure Brussels Effect of limited scope, while the US saw a limited de facto effect and a de jure effect with regard to some states.

**4.1.1. The Analogy between Data Protection and AI Regulation**

Scholars and politicians frequently refer to data protection regulation as an analogue for EU AI regulation.[354] The analogue is fitting in that (i) both laws apply to similar companies, including Amazon, Facebook, Google, IBM, and Microsoft; (ii) they both regulate the technology B2C market; (iii) the regulatory target is similar; and (iv) collected data is often used in machine learning algorithms, one prominent AI technique.

Data protection regulation can be considered an instance of AI regulation as, for example, the GDPR regulates aspects of AI development and deployment.[355] For example, Article 5(1)(a) of the GDPR requires data processing to be fair and transparent. This includes information fairness, i.e. providing data subjects with information on how their data is used, and substantive fairness, which means that the content of an automated inference or decision must be fair.[356] This requirement challenges the application of AI models that are biased or trained on biased data. Moreover, principles from data protection regulation might shape the AI landscape. The GDPR includes what some have called a right to explanation,[357] stating that data subjects have a right to receive "meaningful information about the logic involved" in automated decisions, which would often be made by AI systems.

---

[352] Steven R. Salbu, "The European Union Data Privacy Directive and International Relations" (William Davidson Institute, December 2001).
[353] Lee A. Bygrave, *"The 'Strasbourg Effect' on Data Protection in Light of the 'Brussels Effect': Logic, Mechanics and Prospects,"* Computer Law & Security Review 40 (April 1, 2021): 105460.
[354] However, the analogy might also be politically motivated (see the van der Leyen speech in the European Parliament) to make the plans on AI regulation look more impressive. Directorate-General for Neighbourhood and Enlargement Negotiations, *"Speech by President-Elect von Der Leyen in the European Parliament Plenary on the Occasion of the Presentation of Her College of Commissioners and Their Programme"*; EPIC, *"At G-20, Merkel Calls for Comprehensive AI Regulation."*
[355] There are more ways through which the GDPR affected the AI industry. This includes data minimisation (5(1)(c)), accuracy (5(1)(d)), consent (which might affect whether data from the internet can be scraped to train AI models), and repurposing of data. One might also wonder how the right to erasure should be applied to an AI model which is already trained with one's data. Giovanni Sartor and Francesca Lagioia, *"The Impact of the General Data Protection Regulation (GDPR) on Artificial Intelligence"* (European Parliamentary Research Service, June 2020).

However, while the GDPR might be an instance of AI regulation, there are also reasons to believe that the GDPR analogy is not very informative when forecasting a Brussels Effect for other regulations of AI.

First, if a firm has two internal data protection policies or data management processes, one EU-compliant and one non-compliant, the costs of differentiation (see §2.5) may be high, and compliance is mostly a fixed cost, making non-differentiation more attractive. Second, suppose data protection rules require you to treat the input data for AI systems differently. In that case, this might have high costs of differentiation because the data-collection and management processes are one of the first steps in the production pipeline – forking happens early on. Both make non-differentiation and a de facto Brussels Effect more likely than it is for other AI regulation (see §2.5). On the other hand, there are examples of cases in which the higher EU data privacy standards for social media companies were not diffused to other jurisdictions – suggesting non-differentiation is not the profit-maximising choice in all scenarios. One illustration might be the Facebook-owned messaging app WhatsApp, which initiated a compulsory data privacy update in January 2021. Some of the most widely criticised parts of the regulation were only implemented for users outside of Europe.[358]

Moreover, the GDPR applies to all data subjects that are physically in the EU. Unless a website is intentionally not making itself available to EU IP addresses, they have to have a GDPR-compliant version.[359] Hence, EU data privacy regulation has significant extraterritorial jurisdictional claims, i.e. it governs activities occurring outside the jurisdiction's border.[360] In the DPD, GDPR's non-harmonized predecessor, the definition of an establishment (Article 4), i.e. the territorial scope, was left to the individual member states, which resulted in different national laws having differing degrees of extraterritoriality.[361] The GDPR directly applies to all member states and thus reduces legal uncertainty. While the DPD also had a de jure Brussels Effect, the GDPR led to a strong de facto Brussels Effect. The GDPR's extraterritoriality could have contributed to its de facto regulatory diffusion as it increased the effective market to which the regulation applies.

In addition, the lower regulatory burden of moving data to jurisdictions the EU Commission considers to provide an adequate data protection level has further bolstered a de jure Brussels Effect. Concretely, these GDPR requirements mean that the Commission determines whether a country outside the EU offers an adequate level of data protection. Only when a jurisdiction has been determined to provide adequate protection is personal data allowed to flow from the EU (and Norway, Liechtenstein, and Iceland) to that country without requiring any further safeguards. Twelve countries are on this whitelist, including Israel, Uruguay, and Japan.[362] The United States got a partial and temporary exemption. The Commission reviews the data protection levels of these whitelisted countries every four years. Japan went further, developing its own whitelist.[363] These rules provide economic incentives for non-EU countries to adopt an EU-equivalent data protection level to ensure the free flow of data. It is unclear whether there will be an analogous rule for AI products, as the AI Act does not include any provisions on adequacy assessments.[364]

---

[356] See also GDPR, recital 71; *Sartor and Lagioia*.

[357] Though this is contended by Watcher et al. with a compelling response in Selbst and Powles. Sandra Wachter, Brent Mittelstadt, and Luciano Floridi, *"Why a Right to Explanation of Automated Decision-Making Does Not Exist in the General Data Protection Regulation,"* International Data Privacy Law 7, no. 2 (May 1, 2017): 76–99, https://doi.org/10.1093/idpl/ipx005; Andrew D. Selbst and Julia Powles, *"Meaningful Information and the Right to Explanation,"* International Data Privacy Law 7, no. 4 (November 1, 2017): 233–42.

[358] Jenny Darmody, *"Explainer: What You Need to Know about the WhatsApp Update,"* Siliconrepublic, January 14, 2021.

[359] GDPR, art. 3.

[360] Deborah Senz and Hilary Charlesworth, *"Building Blocks: Australia's Response to Foreign Extraterritorial Legislation,"* Melbourne Journal of International Law 2, no. 1 (June 1, 2001): 69–121. Dan Jerker B. Svantesson, *"The Extraterritoriality of EU Data Privacy Law – Its Theoretical Justification and Its Practical Effect on U.S. Businesses,"* Stanford Journal of International Law 50, no. 1 (2014): 53–102. Argument for significant extraterritoriality: Benjamin Greze, *"The Extra-Territorial Enforcement of the GDPR: A Genuine Issue and the Quest for Alternatives,"* International Data Privacy Law 9, no. 2 (April 21, 2019): 109–28.

[361] Svantesson, *"The Extraterritoriality of EU Data Privacy Law – Its Theoretical Justification and Its Practical Effect on U.S. Businesses."*

### 4.1.2. Regulatory Diffusion

The EU DPD led to regulatory diffusion via a de jure Brussels Effect.[365] The 2018 GDPR exhibited a strong de facto Brussels Effect. It has also led to a de jure Brussels Effect in more than five countries despite its recentness.

In 1980 and 1981, international data privacy regulation efforts were initiated with two international agreements, the OECD's nonbinding privacy principles and the binding Convention 108 of the Council of Europe (CoE).[366] In 1995, the EU DPD followed,[367] which resembles its successor, the 2018 GDPR, in its vast scope. The regulatory targets are data-processing activities conducted by organisations established in the EU, activities offering goods or services (even if for free) to data subjects situated in the EU (not restricted to EU citizens), and the monitoring of such data subjects. For instance, the US company Clearview AI falls under the GDPR.[368] It offers US law enforcement agencies a service where they can search for all photos[369] of an individual and, for instance, identify them in CCTV footage.

Due in part to the Council of Europe Convention 108's preceding and inspiring the EU DPD, some have argued that the spread of European data protection should not be ascribed to the EU. As the Council of Europe,[370] headquartered in Strasbourg, is separate from the EU and includes more countries, some have argued that the spread of these norms should perhaps be termed a "Strasbourg Effect".[371] We will not discuss this question in detail, as most commentators seem to agree that the EU's regulatory efforts played a significant role in the diffusion of European data protection norms, regardless of its role in originating these norms.

**The Data Protection Directive**

The DPD led to a de jure Brussels Effect, partly due to its unprecedented extraterritorial jurisdictional claims.[372] These extraterritorial demands were reasonable from the perspective of European policymakers because they are required to provide adequate protection for European citizens.[373]

Comprehensive data privacy laws that apply to all types of personal data have been adopted by 145 countries, including India, Japan, Malay-

---

[362] European Commission, *"Adequacy Decisions: How the EU Determines If a Non-EU Country Has an Adequate Level of Data Protection."*

[363] Paul M. Schwartz, *"Global Data Privacy: The EU Way,"* New York University Law Review 94, no. 4 (October 2019): 771–818.

[364] AI Act.

[365] Salbu, "The European Union Data Privacy Directive and International Relations."

[366] Council of Europe, *"Details of Treaty No.108,"* Council of Europe, accessed July 14, 2022. Note that the Council of Europe is not an institution of the EU.

[367] European Parliament, *"Directive 95/46/EC of the European Parliament and of the Council of 24 October 1995 on the Protection of Individuals with Regard to the Processing of Personal Data and on the Free Movement of Such Data,"* CELEX number: 31995L0046, Official Journal of the European Communities L 281 31 (November 23, 1995), https://eur-lex.europa.eu/legal-content/EN/TXT/?uri=CELEX:31995L0046. (in the following: data protection directive).

[368] The Hamburger DPA deemed their behaviour illegal but only issued a narrow request rather than a pan-European order. NOYB, *"Clearview AI Deemed Illegal in the EU, but Only Partial Deletion Ordered,"* noyb.eu, January 28, 2021.

[369] These pictures and the metadata were scraped from Facebook, YouTube, Venmo, etc.

[370] Not to be confused with the European Council, which is a part of the EU.

[371] Bygrave, *"The 'Strasbourg Effect' on Data Protection in Light of the 'Brussels Effect': Logic, Mechanics and Prospects."*

[372] Svantesson, *"The Extraterritoriality of EU Data Privacy Law – Its Theoretical Justification and Its Practical Effect on U.S. Businesses,"* 53–102; European Parliament, *"Directive 95/46/EC of the European Parliament and of the Council of 24 October 1995 on the Protection of Individuals with Regard to the Processing of Personal Data and on the Free Movement of Such Data"*, art. 25 and 26.

[373] In the case of AI regulation, such extraterritoriality is most likely not necessary to protect the safety and interest of EU consumers.

[374] Graham Greenleaf, *"Global Data Privacy Laws 2021: Uncertain Paths for International Standards,"* Privacy Laws & Business International Report 169 (Privacy Laws & Business, 2021).

sia, South Korea, Taiwan, South Africa, the Economic Community of West African States, and some Latin American countries.[374] The privacy laws of these countries are not only influenced by the earlier OECD guidelines or the Council of Europe Convention, but they also incorporate unique parts of the EU DPD. A 2012 study considered 33 of the 39 non-European national data protection laws and found that 19 out of 33 national privacy laws contain at least 7 of the 10 elements which were added to the DPD but were not present in either the OECD and the CoE document.[375] Thirteen out of the 33 contain at least nine of these ten features.[376] All 75 non-European data privacy laws enacted at least 7 out of 10 of the 1995 EU directive principles. The EU Data Protection Directive exhibited a strong de jure Brussels Effect.

Because of the absence of studies on the potential de facto Brussels Effect of the DPD, we do not further analyse whether there was such an effect. However, a de facto effect was undermined by the lack of harmonisation of the DPD. Due to its nature, the directive is less centralised in its implementation, reducing the internal cohesion. This reduces the market size and thus weakens the de facto Brussels Effect.[377]

**The Council of Europe 108 Convention**

The Council of Europe 108 Convention also exhibited regulatory diffusion and was adopted by countries which were not members of the CoE.[378] All 126 privacy laws worldwide share the ten core elements from the CoE Convention 108.[379] This might be relevant for actors in the AI policy space as the CoE is also developing AI regulation.[380]

**The General Data Protection Regulation**

In 2018, the GDPR replaced the DPD to (i) achieve more harmonisation, (ii) adapt the law to the new technology landscape, and (iii) better govern international data transfers. As a regulation rather than a directive, the GDPR leads to greater regulatory consistency between EU member states, reducing regulatory and other overhead costs.[381] Moreover, the GDPR improved the legal enforcement system, stressed the importance of individual rights, and changed the consent definition.[382] Companies outside of Europe are adopting GDPR compliance for their operations worldwide.[383]

Further, the regulation also affects business-to-business interactions. One example of this phenomenon is Microsoft, which requires all of its suppliers to be GDPR-compliant.[384] Hence, for instance, Australian businesses serving the business-to-business market, which do not themselves have customers in the EU, have been pressured by their multinational clients to ensure that their software products will be GDPR-compliant.[385] The GDPR has exhibited a strong de facto Brussels Effect.

---

[375] Greenleaf, *"The Influence of European Data Privacy Standards Outside Europe: Implications for Globalization of Convention 108."*

[376] *Greenleaf.*

[377] Greenleaf, *"Global Data Privacy Laws 2021: Uncertain Paths for International Standards"*; Graham Greenleaf, *"Global Data Privacy Laws 2021: Despite COVID Delays, 145 Laws Show GDPR Dominance,"* Privacy Laws & Business International Report 169 (Privacy Laws & Business, 2021), https://doi.org/10.2139/ssrn.3836348.

[378] Greenleaf, *"The Influence of European Data Privacy Standards Outside Europe: Implications for Globalization of Convention 108."*

[379] Graham Greenleaf, *"Global Convergence of Data Privacy Standards and Laws: Speaking Notes for the European Commission Events on the Launch of the General Data Protection Regulation (GDPR) in Brussels & New Delhi, 25 May 2018,"* University of New South Wales Law Research Series 56 (University of New South Wales, May 25, 2018).

[380] Council of Europe, "CAHAI - Ad Hoc Committee on Artificial Intelligence," Artificial Intelligence, accessed July 14, 2022.

[381] In contrast to a regulation, a directive can vary from member state to member state. Thus, a multinational company has to slightly adapt its compliance to different national jurisdictions. Besides, a directive requires more regulatory costs, as not only the European institutions but also national institutions have to work on the law.

[382] This includes the right to be informed, the right of access, the right of rectification, the right to erasure, the right to restrict processing, the right to data portability, the right to object, and rights related to automated decision-making and profiling. Griffin Drake, *"Navigating the Atlantic: Understanding EU Data Privacy Compliance amidst a Sea of Uncertainty,"* Southern California Law Review 91, no. 1 (November 2017).

[383] Greenleaf, *"Global Convergence of Data Privacy Standards and Laws: Speaking Notes for the European Commission Events on the Launch of the General Data Protection Regulation (GDPR) in Brussels & New Delhi, 25 May 2018."*

[384] *Greenleaf.*

[385] Graham Greenleaf, *"'GDPR Creep' for Australian Businesses But Gap in Laws Widens,"* University of New South Wales Law Research Series 54 (University of New South Wales, June 6, 2018).

It is still too early to evaluate the final extent of the de jure Brussels Effect of the GDPR. However, some evidence exists for de jure regulatory diffusion.[386] In 2019 and 2020, 13 new countries adopted data privacy legislation and 13 other countries updated existing laws, of which the GDPR influenced almost all.[387] The countries held adequate under the 1995 DPD can renew their status until 2022.[388] To date, the EU has made adequacy decisions approving 14 jurisdictions, including Argentina, Canada, Israel, South Korea, Japan, Switzerland, the United Kingdom, and New Zealand.[389]

In addition, some US states have or are in the process of adopting regulation with elements from the GDPR. In 2018, California adopted the California Consumer Privacy Act,[390] originally introduced as a ballot proposition, with many similarities to the GDPR.[391] In 2021, the Consumer Data Protection Act[392] was signed into law in Virginia, with many similarities with the GDPR and the California Consumer Privacy Act.[393] Furthermore, there have been repeated attempts to pass a similar law in Washington State.[394]

In addition, the European institutions have shaped the global narrative surrounding data privacy through their regulatory efforts. While personal data might also be considered a commodity in the United States, data privacy is regarded as a human right in the EU. Importantly, this European narrative appears to have influenced the positions of American technology companies. For instance, the president of Microsoft tweeted, "We believe privacy is a fundamental human right".[395] Similarly, the CEO of Apple told CNN that "privacy is a fundamental human right".[396] In the same vein, the European narrative on AI – such as the concept of "trustworthy AI" – may influence the positions and actions of non-European AI companies.

Moreover, the regulation might have also strengthened certain industries. The GDPR has provided a strong business case for privacy-enhancing technologies (PET). One would expect that the GDPR will increase the development and deployment of these techniques. However, since PETs are not mature enough to be widely employed, it is currently difficult to evaluate such diffusion.

**China**

EU data protection rules have also influenced China. The 2017 Cyber Security Law includes explicit consent from the users and the requirement that the data used for processing be adequate and not excessive.[397] This Chinese policy process also received funding from the Commission,[398] which also set up policy dialogues between the two jurisdictions.[399] In August 2021, the Chinese government passed the Private Information Act.[400] The Cyberspace Administration of China, which has been encouraged in recent years to fiercely enforce regulation in the technology industry, will enforce the act. At the same time, not all EU aims of privacy legislation

---

[386] "Early examples just from Asia include Malaysia (data portability); Korea (4% administrative fines); Indonesia ("right to be forgotten"); and mandatory data breach notification (DBN) in six countries." Greenleaf, *"Global Convergence of Data Privacy Standards and Laws: Speaking Notes for the European Commission Events on the Launch of the General Data Protection Regulation (GDPR) in Brussels & New Delhi, 25 May 2018."* See also Greenleaf, *"Global Data Privacy Laws 2021: Uncertain Paths for International Standards."*

[387] See for instance: Greenleaf, *"Global Data Privacy Laws 2021: Uncertain Paths for International Standards."*

[388] European Commission, *"Adequacy Decisions: How the EU Determines If a Non-EU Country Has an Adequate Level of Data Protection."*

[389] *European Commission.*

[390] California State Legislature, *"Bill Text - AB-375 Privacy: Personal Information: Businesses,"* June 29, 2018.

[391] Francesca Lucarini, *"The Differences between the California Consumer Privacy Act and the GDPR,"* April 13, 2020.

[392] Virginia's Legislative Information System, *"2021 Special Session I: HB 2307 Consumer Data Protection Act; Personal Data Rights of Consumer, Etc,"* LIS, accessed July 14, 2022.

[393] Sarah Rippy, *"Virginia Passes the Consumer Data Protection Act,"* International Association of Privacy Professionals, March 3, 2021.

[394] Jim Halpert et al., *"The Washington Privacy Act Goes 0 for 3,"* International Association of Privacy Professionals, April 26, 2021.

[395] Schwartz, *"Global Data Privacy: The EU Way."*

[396] *Schwartz.*

[397] Bradford, *The Brussels Effect: How the European Union Rules the World*, chap. 5.

[398] Bradford, *chap. 5*; Zhang Xinbao, *"Status Quo Of, Prospects for Legislation on Protection of Personal Data in China,"* 北大法宝V6官网, 2007; Zhang Xinbao and Liao Zhenyun, "中国个人数据保护立法的现状与展望," 中国法律：中英文版, 2007.

[399] Bradford, *The Brussels Effect: How the European Union Rules the World*, chap. 5.

[400] Lomas, *"China Passes Data Protection Law."*

have been achieved in China. The Chinese public sector is completely exempt. Digital authoritarianism, the blocking and filtering of online content, the social credit system, and facial recognition techniques all clash with the aims and values of EU data protection rules and are not curtailed by the regulation adopted by the Chinese Communist Party.[401]

**Extraterritoriality and the United States**

The extraterritorial reach of the GDPR contributed to the de facto and de jure GDPR Brussels Effect.[402] At the same time, the extraterritoriality also illustrates how powerful and economically advanced countries, especially the US, try to resist the Brussels Effect.

While the US experienced de facto Brussels Effects for various EU legislative efforts, including the Code of Conduct regarding hate speech, parts of the DPD, and the GDPR,[403] it can also resist selectively, sometimes successfully and sometimes not.[404] The United States partially circumvented the extraterritorial claims of the DPD and GDPR, particularly the requirements for international data transfers. The EU and the US adopted two data transmission agreements, the Safe Harbor agreement in 2000 and the Privacy Shield in 2015, allowing unhindered data transmission between the United States and the EU. Both agreements were adopted even though the US data privacy standards were not equivalent to the EU, which is a requirement for data transmission agreements in both the DPD and GDPR.[405] Consequently, both data transmission agreements were declared invalid by the European Court of Justice (ECJ) in 2015 and 2020, respectively.[406] Since 2020, the US and the Commission have stated their intention to negotiate a new agreement. In June 2021, both sides asserted their commitment to find a successor to the Privacy Shield.[407] At the same time, *Politico* reported that Facebook, for instance, continues to send data across the Atlantic.[408]

In addition, despite the EU's data protection concerns, the United States and the EU have signed bilateral passenger name record (PNR) agreements for flights.[409] These agreements allow for the exchange of both the information provided by passengers when they book tickets and when checking in for flights, and the exchange of data collected by air carriers for commercial purposes. Both agreements are examples of the United States resisting European pressure for regulatory convergence. The US might have leveraged its considerable regulatory capacity in customs policy and homeland security, market size, or geopolitical power.

Despite the resistance of the United States, both data transmission agreements potentially led to a Brussels Effect: they brought the US closer to the EU privacy standards.[410] The 2000 Safe Harbor Agreement encouraged company self-regulation. By 2015, 4,500 US companies had publicly affirmed the Safe Harbor principles. Consequently, the Safe Harbor agreement has indirectly led to a Brussels Effect in the United States.

The United States has no omnibus privacy laws — suggesting the absence of a de jure Brussels Effect. Nevertheless, the EU discourse and regulation around data privacy has significantly influenced the United States. The narrative around data privacy in the United States appears to have increasingly moved away from consumer safety and towards fundamental rights.[411] Despite the temporary special treatment received by the United States in the DPD and GDPR, many American companies have followed the Safe Harbor principles and adopted stricter data protection practices than required by the US government. The Safe Harbor agreement called for self-regulatory efforts by companies, which led to further agreements between companies and more significant regulatory actions.[412]

---

[401] On these grounds and the demands of the Chinese government, companies like Google withdrew from China. Matt Sheehan, "How Google Took on China—and Lost," MIT Technology Review, December 19, 2018.

[402] See §2.1.3 of this report or Bradford, *The Brussels Effect: How the European Union Rules the World*.

[403] For more, see *Bradford*.

[404] Bach and Newman, "The European Regulatory State and Global Public Policy: Micro-Institutions, Macro-Influence."

[405] The directive called for an "essential equivalent". This is not given in the case of the US privacy regulation. Schwartz, "Global Data Privacy: The EU Way."

[406] Court of Justice of the European Union, "Judgment in Case C-362/14 Maximillian Schrems v Data Protection Commissioner: The Court of Justice Declares That the Commission's US Safe Harbour Decision Is Invalid," Press Release 117/15 (Court of Justice of the European Union, October 6, 2015); European Commission, "EU-US Data Transfers: How Personal Data Transferred between the EU and US Is Protected," European Commission, accessed July 14, 2022. New discussions were initiated in August 2020.

[407] Kenneth Propp, "Progress on Transatlantic Data Transfers? The Picture After the US-EU Summit," Lawfare, June 25, 2021.

[408] Vincent Manancourt, "Despite EU Court Rulings, Facebook Says US Is Safe to Receive Europeans' Data," POLITICO, December 19, 2021.

[409] Javier Argomaniz, "When the EU Is the 'Norm-taker': The Passenger Name Records Agreement and the EU's Internalization of US Border Security Norms," Journal of European Integration 31, no. 1 (January 1, 2009): 119–36.

[410] Jennifer Daskal, "Borders and Bits," Vanderbilt Law Review 71, no. 1 (2018): 179.

[411] For a discussion see Schwartz, "Global Data Privacy: The EU Way." E.g. The president of Microsoft, tweeted, "We believe privacy is a fundamental human right." In a similar fashion, Tim Cook, the CEO of Apple, told CNN that "privacy is a fundamental human right."

### 4.1.3. Conclusion

The EU data protection regime has exhibited a strong de jure Brussels Effect. This was partly mediated via the international spread of the EU-supported narrative of data privacy as a human right, a unique feature of European data protection regulation. For instance, non-EU countries passed stronger data protection clauses that were not required in order to trade with the EU. In a 2012 study, 28 out of 33 examined data privacy laws also have border control data export limitations.[413] Similar to the EU, Japan created a whitelist of countries to which Japanese data can flow.[414]

The European regulation also exhibited a de facto Brussels Effect. However, it is unclear whether this offers transferable insights for (other) AI regulation since the de facto Brussels Effect of data protection regulation may have been due to unique features of this regulatory target and design. First, the narrative change increased the revenue from non-differentiation. Second, the extraterritorial claims also made non-differentiation more attractive. Third, the regulation required early forking, which increased the costs of differentiation – making non-differentiation more likely.

## 4.2. Product Liability Directive

As of January 2022, the Commission is preparing to propose AI-specific changes to the EU liability regime – by either changing the Product Liability Directive (PLD) or by harmonising aspects of national civil liability law regarding the liability of certain AI systems.[415] The regulatory diffusion of the PLD can inform us about the likelihood of a future Brussels Effect of AI liability rules. For instance, if another jurisdiction has liability regulation strongly influenced by the PLD, then the EU AI liability becomes more attractive and feasible as a blueprint. In addition, the PLD and the AI liability update might share several features.

The PLD influenced the product liability legislation of many countries – evidence for a future de jure effect of AI liability updates. US liability law has much higher economic costs and is less easy to copy than the PLD. This made the de jure Brussels Effect more likely as other jurisdictions were less likely to take the US regulation as a blueprint. As discussed in section 2.6.4, whether liability law exhibited de facto regulatory diffusion is extremely difficult to study. Hence, one should be less confident that future AI liability law will lead to a de facto Brussels Effect.

The Commission's legislative efforts may include the adoption of strict liability for AI operators or the adaptation of the burden of proof.[416] The EU AI White Paper 2020 and the Inception Impact Report 2021[417] propose, among other things, to include software in the definition of a product and to shift the burden of proof more towards the AI companies. In doing so, companies would be given the responsibility to demonstrate the safety of their AI products,

---

rather than requiring consumers to prove in court that the AI product was defective.

### 4.2.1. Regulatory Diffusion

The PLD has become an internationally leading blueprint, having been copied in more than a dozen countries. Iceland, Liechtenstein, Malta, Norway, and Switzerland voluntarily adopted it simultaneously with the EU – these countries regularly opt for EU legislation because the EU is their main trading partner.[418] Countries in Asia-Pacific, among them Australia, China, Taiwan, Japan, Malaysia, Indonesia, and South Korea, adopted it as a blueprint 7 to 15 years after the EU adoption.[419] Russia, Israel, and Quebec also used the PLD as a blueprint.[420]

China is another example of a country to which the European PLD diffused. The structure of China's Tort Liability Law broadly follows the German civil code law, which implements the PLD. Moreover, China used the PLD as a blueprint for the liability along the supply chain, the burden of proof in liability cases, and defining a defect.[421]

The European liability model has become dominant on a global level such that "the American approach has become almost an outsider".[422] There are two explanations for why the European rather than American liability model served as a blueprint. First, the number of liability claims in the US, their awards, and their publicity are significantly higher than anywhere else in the world – involving substantial economic costs.[423] This may have made the PLD, under which liability claims are harder to win and awards are smaller, relatively more attractive.[424] Second, the PLD is more concise and easier to understand than its American counterpart.[425] Taken together, this de jure Brussels Effect might show that EU-crafted liability law is attractive for other jurisdictions.

### 4.2.2. Impacts of EU-style Liability Law

Despite the strong de jure Brussels Effect of EU liability law, it is difficult to assess whether there have been any flow-through effects on company behaviour. The regulation has caused only minor detectable changes in EU courtrooms – conceivably suggesting the absence of any actual effect as companies do not have enough pressure to change behaviour.[426] When EU consumers sue because of product damages, they rarely rely on the PLD but rather on pre-existing national law.[427] There is only scarce evidence for litigation in the countries using the EU blueprint of the PLD, likely because the PLD is too restrictive and only "supplemented pre-existing na-tional liability regimes".[428] It is generally difficult to measure compliance with liability law since compliance can look differently for every company.

---

[418] "Consolidated Text: Council Directive 85/374/EEC of 25 July 1985 on the Approximation of the Laws, Regulations and Administrative Provisions of the Member States Concerning Liability for Defective Products."

[419] Luke R. Nottage and Jocelyn Kellam, "Europeanisation of Product Liability in the Asia-Pacific Region: A Preliminary Empirical Benchmark," Legal Studies Research Paper, No. 07/30 (Sydney Law School, May 1, 2007), https://doi.org/10.2139/ssrn.986530. The adoption happened in the following years: 1992. Australia; 1993, People's Republic of China; 1994, Taiwan; 1995, Japan; 1999, Malaysia and Indonesia; 2000, Korea.

[420] Reimann, "Product Liability in a Global Context: The Hollow Victory of the European Model," European Review of Private Law 11, no. 2 (2003): 128–54, See also William Boger, "The Harmonization of European Products Liability Law," Fordham International Law Journal 7, no. 1 (1983): 1–60; Cheon-Soo Kim, "Theories and Legislation of Products Liability in the Southeast Asian Countries," Journal of Social Studies Research 55 (1999). For China, see Claudius Hans Taschner and Karola Taschner, 10 Jahre EG-Richtlinie Zur Produkthaftung : Rückblick, Umschau, Ausblick, vol. 15, Schriftenreihe Deutscher Jura-Studenten in Genf (Genève: Unité de droit allemand, Faculté de droit, 1996., 1996), 13–14.

[421] "Overall, it would appear that China has chosen to follow the EC Directive rather than the US Third Restatement." Kristie Thomas, "The Product Liability System in China: Recent Changes and Prospects," The International and Comparative Law Quarterly 63, no. 3 (July 2014): 755–75. The same conclusion was reached by: Reimann, "Product Liability in a Global Context: The Hollow Victory of the European Model."

[422] Reimann, "Product Liability in a Global Context: The Hollow Victory of the European Model"; Alfred E. Mottur, "The European Product Liability Directive: A Comparison with U.S. Law. An Analysis of Its Impact on Trade and a Recommendation for Reform so as to Accomplish Harmonisation and Consumer Protection," Law and Policy in International Business 25 (1993-1994).

[423] Mathias Reimann, "Product Liability," in Comparative Tort Law: Global Perspectives, ed. Mauro Bussani and Anthony J. Sebok, Research Handbooks in Comparative Law (Edward Elgar Publishing Limited, 2021), 236–63.

[424] Reimann, "Product Liability in a Global Context: The Hollow Victory of the European Model."

[425] Reimann; European Parliament, "Directive 95/46/EC of the European Parliament and of the Council of 24 October 1995 on the Protection of Individuals with Regard to the Processing of Personal Data and on the Free Movement of Such Data", art. 1-13. Reimann discusses the complexities of US liability law, which differs for every state. In contrast, the EU PLD has already been translated into 20 languages. In sum, other countries might have not even understood the US Law "Foreign drafters might have just adopted whatever they understood."

[426] Reimann, "Product Liability in a Global Context: The Hollow Victory of the European Model." Although the reports of the European commission see the limited number of court cases as a sign of success of the PLD, as summarised in Bertolini, Artificial Intelligence and Civil Liability, 55ff.

[427] In its 2001 Report, the Commission mentioned barely a hundred court decisions under the new regime for the last fifteen years in all the member states combined.

[428] This might be because the directive and other laws merely supplemented national law and the literature is critical how much the directive has actually harmonised European product liability law at all. See: Mathias Reimann, "Liability for Defective Products at the Beginning of the Twenty-First Century: Emergence of a Worldwide Standard?," The American Journal of Comparative Law 51, no. 4 (October 1, 2003): 751–838; Jane Stapleton, "Product

In general, there is only weak evidence for companies adopting changes in response to any liability law. For example, a 2021 meta-study finds limited but inconclusive evidence that firms reduce risks, internalise externalities, and add safety precautions after liability law was passed.[429] Therefore, it remains unclear whether multinational firms become more cautious in response to liability law updates for AI products and services.

While it is difficult to evaluate whether liability law has domestic effects on company behaviour, as discussed in the previous paragraph, it is even more difficult to assess whether there has been a de facto Brussels Effect of liability law: whether multinational companies have changed their practices outside the EU in response to the PLD.[430]

### 4.2.3. Conclusion

We conclude tentatively that an EU liability law update for certain AI systems is likely to cause a de jure Brussels Effect. Countries that already use the PLD as a blueprint will find it easiest to copy the European approach to regulating liability for AI and other emerging technologies. At the same time, however, it is difficult to measure to what extent (i) firms change in response to liability legislation; (ii) that response is global, i.e. a de facto Brussels Effect occurs; and (iii) EU law was causally responsible for the adoption of similar legislation abroad, a de jure Brussels Effect.

## 4.3. Product Safety and CE Marking

The proposed EU AI Act would largely be part of the EU product safety regulatory regime. The act outlines that high-risk AI systems, those applied to specific use cases, should first be self-assessed in conformity assessments before being sold on the common market – though biometric identification systems must be assessed by a conformity assessment body. Products which are regulated in the "New Legislative Framework", the EU product safety regime, and include AI systems must also abide by the product safety rules of the AI Act.[431] The AI requirements are tested by the product-specific conformity assessment body. The AI product safety requirements also apply to the other harmonisation regulation (see Annex II, Section B). The New Approach for product safety, i.e. recent EU product safety rules, has historically caused both a de jure and de facto Brussels Effect. Thus, upcoming AI product safety regulation might also lead to regulatory diffusion. EU product safety regulations apply to all EU imports but exclude EU exports. In general, EU product safety legislation exhibited a strong de jure and de facto Brussels Effect, making a future Brussels Effect more likely for AI-specific product safety rules.

### 4.3.1 The EU Product Safety Framework

The EU uses the New Approach to product safety, which originated in the 1985 Council resolution on a New Approach to Technical Harmonization and Standardization.[432] This so-called New Legislative Framework consists of 29 mostly product-specific directives. The PSD establishes the legal framework that implements the New Approach to product safety. The conformity assessments apply to all EU imports but not EU exports. For these products, such as electronics and children's toys, regulat-

---

ory bodies and internal and external industry experts develop safety goals and conformity assessments.[433] Instead of requiring firms to implement specific measures, firms have to reach safety targets. To this end, the firms that are responsible for proving that their products are safe are free to use any means. They can follow voluntary standards by the European Committee for Standardization (CEN) and the European Committee for Electrotechnical Standardization (CENELEC) or have the safety of their products verified independently. In practice, most companies follow the CEN and CENELEC standards.[434]

The EU AI Act says that approved non-governmental bodies need to conduct conformity assessments for biometric identification systems. In all other cases, firms conduct a self-assessment, potentially relying on the CEN or CENELEC standards, or verify the safety with an approved non-governmental body.[435] Afterwards, the product gets a CE (Conformité Européenne) mark and can be sold on the EU market.

**4.3.2. Regulatory Diffusion**

The safety targets and specific guidelines developed for particular CE marks have exhibited strong de jure and de facto Brussels Effects, making a Brussels Effect for the upcoming CE marking of high-risk AI applications likely. Several prominent examples of the Brussels Effect, such as the chemical regulation REACH, are part of this New Approach to product safety. Other examples of the Brussels Effect include the directives on the Safety of Toys (88/378/EEC),[436] Machinery (89/3321/EEC), Medical Devices (93/42/EEC), Pressure Equipment (97/23/EC), Telecommunication Terminal Equipment (98/392/EEC), and pharmaceuticals.[437]

National standards, for products that are not covered under the EU product safety legislation, are less likely to exhibit regulatory diffusion, plausibly because they lack — in contrast to the EU — the regulatory coherence (§2.3.2) and market size (§2.1.1) necessary to influence the international market and foreign jurisdictions. Regulation on the European level makes compliance more worthwhile for multinational firms.[438]

The EU conformity marking also caused a de jure Brussels Effect. For example, the Chinese conformity marking, "CCC", developed in 2003, is similar to the EU system, the "CE" mark.[439] The international influence of the EU conformity marking is in part due to its stringency.[440] Local regulation agencies seek to comply with key trading partners. Since the European regulation is the most stringent and non-EU regulators aim to maintain access to the EU market, these regulators effectively comply with EU regulation. Several countries have incorporated "CE" marks into their national legislation to support their export industry. For instance, New Zealand incorporated all EU conformity marking standards in its national law, especially in industries with significant exports to the EU market.[441] The United States, the United Kingdom, and other countries are converging towards the European standard level of conformity marking. However, this development is much slower for the United States.[442]

In addition to the de jure Brussels Effect for EU conformity marking, New Zealand and Australia have also experienced a de facto Brussels Effect. For example, wine regulation is weaker in both countries than the EU product safety rules for wine. Nevertheless, Australian wine producers and exporters decided to comply with the EU export requirements, exemplifying a de facto Brussels Effect[443] Moreover, the CE mark has become a prominent product quality signal in New Zealand since the country lacks a national product safety mark and the market is dominated by Asian imports, which consumers trust less.[444] This means that the revenue from non-differentiation is higher.

---

[434] Veale and Borgesius, *"Demystifying the Draft EU Artificial Intelligence Act — Analysing the Good, the Bad, and the Unclear Elements of the Proposed Approach."*
[435] The second option is not common. *Veale and Borgesius.*
[436] European Parliament, *"Directive 2009/48/EC of the European Parliament and of the Council of 18 June 2009 on the Safety of Toys,"* CELEX number: 32009L0048, Official Journal of the European Union L 170 1 (June 18, 2009): 1–37.
[437] Marco de Morpurgo, *"The European Union as a Global Producer of Transnational Law of Risk Regulation: A Case Study on Chemical Regulation,"* European Law Journal 19, no. 6 (November 2013): 779–98.
[438] Notably, the costs of leaving a market regulatory stringency, such as the EU, increase with the size of the market at hand.
[439] Hanson, *CE Marking, Product Standards and World Trade.*
[440] For more, see "the trade to the top" Bradford, *The Brussels Effect: How the European Union Rules the World.*
[441] Hopkins and McNeill, *"Exporting Hard Law Through Soft Norms: New Zealand's Reception of European Standards."*

The Commission also uses free trade agreements as a channel to promote the regulatory diffusion of CE marking. For instance, the free trade agreements with Mexico and Mercosur in 2019 include a commitment to the local adoption of CE marking.[445]

Moreover, Canada, the United States, Australia, Switzerland, and New Zealand have Mutual Recognition Agreements on Conformity Assessment (MRA) with the EU.[446] These agreements entail the reciprocal acceptance of conformity assessments for two jurisdictions with similar product safety levels and equivalent assessment authorities for particular product families. If a country raises its standards to a level on par with the CE mark and establishes an MRA with the EU, the national industry avoids costs when trading with the EU or expanding to the EU market. Hence, the possibility of MRAs makes the copying of EU-like regulation more attractive and a de jure Brussels Effect more likely.[447]

There are three more explanations for the de jure Brussels Effect exhibited by the EU product safety regulations. First, the EU actively promoted its product safety regulations worldwide.[448] Second, the EU wields substantial influence in international standard setting bodies, which have adopted aspects of European product safety regulations and serve as a channel for the de jure Brussels Effect.[449] Third, corporate interest groups have a strong interest in the convergence of product safety standards for all globalised markets. For this reason, medical technology companies lobbied for the international convergence of medical devices.[450] However, whether an internationally harmonised standard setting procedure increases or decreases the de jure Brussels Effect of CE marking remains unclear. Convergence on standard setting could lead to international standards being adopted that are lower than the EU rules, therefore weakening the de jure and de facto Brussels Effects.[451] On the other hand, some International Organization for Standardization (ISO) standards are the same as the EU standards. For example, the European standard EN 1050 (risk assessment for machinery) became ISO 14120, and EN 292 (machinery safety) became ISO 1200-1.[452] See also the discussion in section 3.2.

### 4.3.3. Conclusion

European product safety regulation and the CE mark led to substantial global regulatory diffusion. The EU's strategy to regulate only safety targets rather than specific safety precautions appears effective. This strategy ensures that product safety does not hinder innovation and supports the regulation of rapidly developing technologies and products, such as AI systems.[453] The CE mark is considered a sign of product quality, which increases the revenue from non-differentiation and contributes positively to the de facto Brussels Effect.

---

[443] Fini, *"The EU as Force to 'Do Good': The EU's Wider Influence on Environmental Matters."*

[444] Hopkins and McNeill, *"Exporting Hard Law Through Soft Norms: New Zealand's Reception of European Standards."*

[445] For the 2019 agreement, see European Commission, "Trade Part of the EU-Mercosur Association Agreement Without Prejudice," 2019, and for the general strategy; Hanson, *CE Marking, Product Standards and World Trade,* 190.

[446] For a list of MRAs, see: European Commission, *"Mutual Recognition Agreements,"* Internal Market, Industry, Entrepreneurship and SMEs, accessed July 14, 2022.

[447] See Björkdahl et al., *Importing EU Norms Conceptual Framework and Empirical Findings,* vol. 8, chap. 8.

[448] Hanson, *CE Marking, Product Standards and World Trade,* 19.

[449] Hairston, "Hunting for Harmony in Pharmaceutical Standards."

[450] In the past in the GHTF and after 2012 in the IMRDF International Medical Device Regulators Forum, *"About IMDRF,"* International Medical Device Regulators Forum, accessed July 14, 2022.

[451] For a discussion, see Peter Cihon's FHI report: Cihon, *"Standards for AI Governance: International Standards to Enable Global Coordination in AI Research & Development."* More specifically, international standards weaken the Brussels Effect if the standards are weaker than national laws passed by non-EU countries in response to EU rules in the absence of international rules.

[452] Hopkins and McNeill, "Exporting Hard Law Through Soft Norms: New Zealand's Reception of European Standards."

[453] For discussion on for instance privacy by design: Eric Lachaud, *"Could the CE Marking Be Relevant to Enforce Privacy by Design in the Internet of Things?,"* in Data Protection on the Move: Current Developments in ICT and Privacy/Data Protection, ed. Serge Gutwirth, Ronald Leenes, and Paul De Hert (Dordrecht: Springer Netherlands, 2016), 135–62.

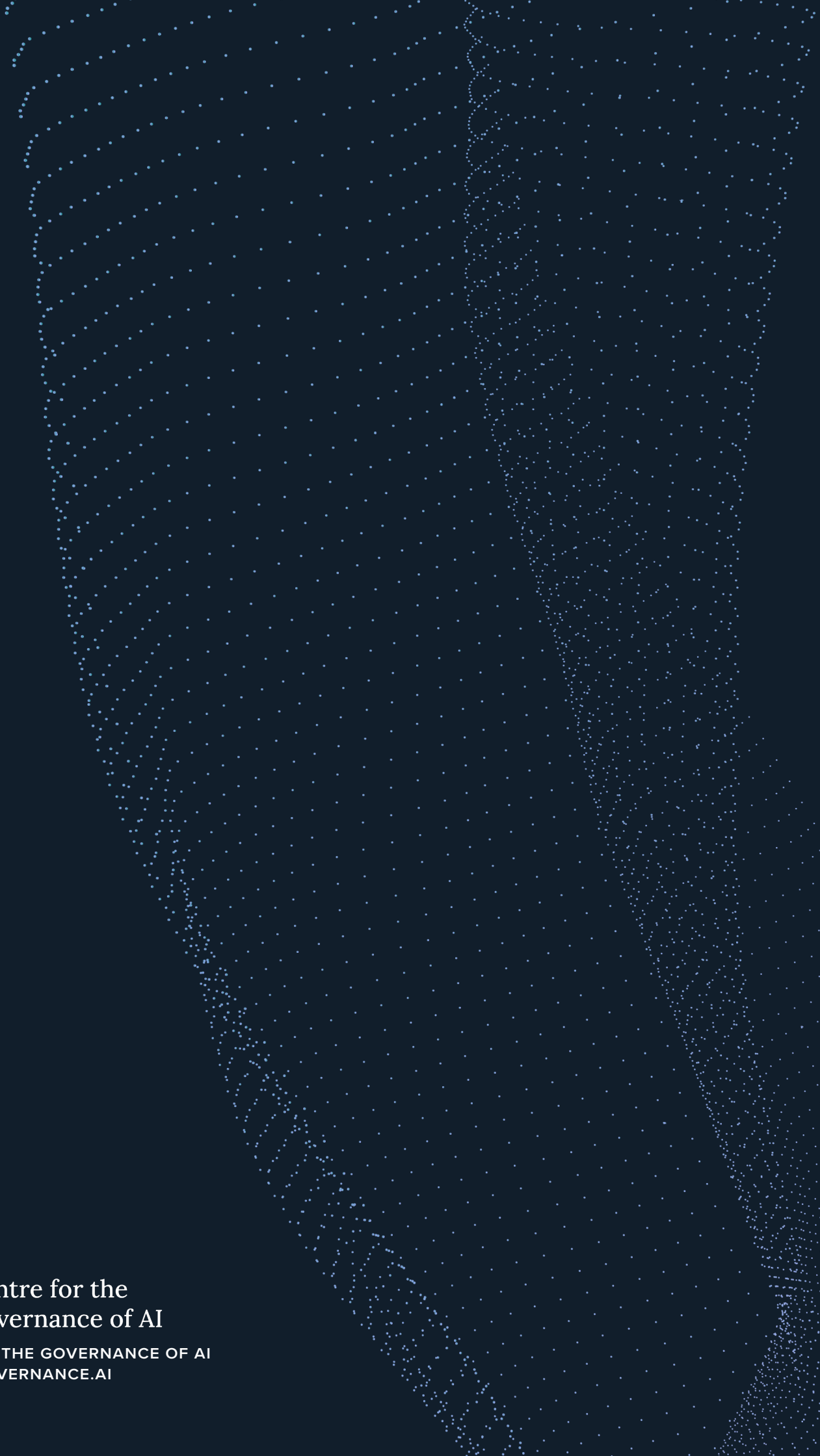
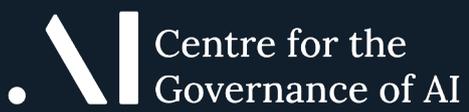